\documentclass{article}

\usepackage{arxiv}

\usepackage[utf8]{inputenc} 
\usepackage[T1]{fontenc}    
\usepackage{hyperref}       
\usepackage{url}            
\usepackage{booktabs}       
\usepackage{amsfonts}       
\usepackage{nicefrac}       
\usepackage{microtype}      
\usepackage{lipsum}		
\usepackage{graphicx}
\usepackage{natbib}
\usepackage{doi}

\usepackage{amsmath}
\usepackage{amssymb}
\usepackage{cool}
\usepackage{colortbl}%

\newcommand{\va}{{\mathbf a}}
\newcommand{\vb}{{\mathbf b}}

\newcommand{\vc}{{\mathbf c}}

\newcommand{\ve}{{\mathbf e}}

\newcommand{\vo}{{\mathbf o}}

\newcommand{\vr}{{\mathbf r}}

\newcommand{\vu}{{\mathbf u}}

\newcommand{\vv}{{\mathbf v}}
\newcommand{\vw}{{\mathbf w}}

\newcommand{\vx}{{\mathbf x}}
\newcommand{\xx}{{\mathbf x}}
\newcommand{\vy}{{\mathbf y}}

\newcommand{\vNull}{{\mathbf 0}}
\newcommand{\vxi}{{\boldsymbol \xi}}

\newcommand{\mA}{{\mathbf A}}
\newcommand{\mB}{{\mathbf B}}

\newcommand{\mE}{{\mathbf E}}

\newcommand{\mH}{{\mathbf H}}
\newcommand{\mI}{{\mathbf I}}
\newcommand{\mJ}{{\mathbf J}}

\newcommand{\mM}{{\mathbf M}}

\newcommand{\mQ}{{\mathbf Q}}
\newcommand{\mR}{{\mathbf R}}
\newcommand{\mS}{{\mathbf S}}
\newcommand{\mT}{{\mathbf T}}
\newcommand{\mU}{{\mathbf U}}
\newcommand{\mV}{{\mathbf V}}
\newcommand{\mW}{{\mathbf W}}
\newcommand{\mX}{{\mathbf X}}
\newcommand{\mY}{{\mathbf Y}}

\newcommand{\Transp}{{{\mathrm T}}}

\Style{IdentityMatrixSymb=\mI,%
       DSymb={\mathrm d},%
       IntegrateDifferentialDSymb={\mathrm d}}

\newtheorem{theorem}{Theorem}

\usepackage{tikz}
\usetikzlibrary{positioning}

\title{Objective Flow Measures based on Few Trajectories}

\author{ \href{https://orcid.org/0000-0002-8009-7070}{\includegraphics[scale=0.06]{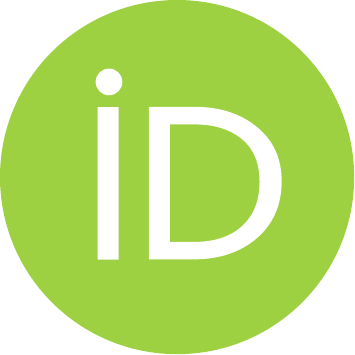}\hspace{1mm}Holger Theisel}
		\\
	Department of Computer Science\\
	University of Magdeburg\\
	Magdeburg, Germany\\
	\texttt{theisel@ovgu.de} \\
	\And
	\href{https://orcid.org/0000-0002-5469-1324}{\includegraphics[scale=0.06]{orcid.pdf}\hspace{1mm}Anke Friederici} \\
	Department of Computer Science\\
	University of Magdeburg\\
	Magdeburg, Germany\\
	\texttt{anke@isg.cs.uni-magdeburg.de} \\
	\And
	\href{https://orcid.org/0000-0002-3020-0930}{\includegraphics[scale=0.06]{orcid.pdf}\hspace{1mm}Tobias G{\"u}nther} \\
	Department of Computer Science\\
	Friedrich-Alexander-University Erlangen-N{\"u}rnberg\\
	Erlangen, Germany \\
	\texttt{tobias.guenther@fau.de} \\
}

\renewcommand{\headeright}{}
\renewcommand{\undertitle}{}

\begin{document}
\maketitle

\begin{abstract}
Sparse trajectory data consist of a low number of trajectories such that the reconstruction of an underlying velocity field is not possible. Recently, approaches have been introduced to analyze flow behavior based on a single trajectory only: trajectory stretching exponent (TSE) to   detect hyperbolic (stretching) behavior, and trajectory angular velocity (TRA) to detect   elliptic (rotation) behavior. In this paper, we analyze these approaches and in particular show that they are -- contrary to what is claimed in the literature -- not objective in the extended phase space. Furthermore, we introduce the first objective measure of rotation behavior that is based on only few trajectories: at least 3 in 2D, and at least 4 in 3D. For this measure -- called trajectory vorticity (TRV) -- we show that it is objective and that it can be introduced in two independent ways: by approaches for unsteadiness minimization and by considering the relative spin tensor. We apply TRV to a number of constructed and real trajectory data sets, including drifting buoys in the Atlantic, midge swarm tracking data, and a simulated vortex street.
\end{abstract}


\section{Introduction}
In flow visualization, time-dependent velocity fields, obtained by simulation or measurement, are of high interest, as they describe many natural phenomena. In fact, the data behind most techniques for flow visualization and flow analysis are continuous time-dependent velocity fields~\citep{McLoughlin09,Edmunds12:SurfaceSTAR,Bujack20:UnsteadyTopoSTAR}. Alternatively, sets of trajectories became popular as another representation of flows~\citep{Bujack15:LagrangianRepresentation,Sane18:InSituLagrangian}. Usually, sets of trajectories are assumed to be sufficiently dense. 
In this paper, we are interested in flows where only a very low number of trajectories is known. 
Examples are the analysis of observational drifter data~\citep{GDP:buoys}, balloon data, particle tracks from particle tracking velocimetry (PTV),  or swarms of tracked animals~\citep{sinhuber2019three} or robots.
For such sparse sets of trajectories the question arises: can we infer information about the behavior of the underlying flow if only a very few trajectories are available? We are concerned with the question:
can we get information about hyperbolic (stretching) or elliptic (rotation) behavior in this case? In particular, we consider information that is invariant under different choices of moving reference frames, i.e., is objective.
While objectivity of flow measures is a common and obvious demand, it is in fact a rather strong condition, especially when checking the rotation/swirling behavior of moving particles. Objective measures give the same result, no matter whether the observer is at a fixed position, travelling e.g. with one particle, or the observer's coordinate system is in rotating motion itself.  In fact, the most challenging part  for objective measures is to distinguish between swirling around a common center and a rotating movement of the coordinate system.
The first approach to tackle this problem was proposed by \cite{Haller21singletrajecory}, who introduced measures based on a single trajectory only.
For this, the concept of  quasi-objectivity is introduced: Contrary to classical objectivity where a scalar value must be invariant under arbitrary time-dependent Euclidian transformations, for quasi-objectivity a condition (A) is introduced, and invariance is not demanded for all Euclidean transformations but only for those fulfilling (A).  Then, \cite{Haller21singletrajecory}  introduced several measures based on a single trajectory: {\em extended trajectory stretching exponents} $\mbox{TSE}$ and 
$\overline{\mbox{TSE}}$, and {\em extended trajectory angular velocity} $\mbox{TRA}$,    $\overline{\mbox{TRA}}$. 
\cite{Haller21singletrajecory}  claimed that  $\mbox{TSE}$ and 
$\overline{\mbox{TSE}}$ are objective in the extended phase space, and that $\overline{\mbox{TRA}}$ is quasi-objective in the extended phase space under a certain condition put to the average vorticity in a certain neighborhood of the trajectory.

\begin{figure}[t]
    \centering
    \includegraphics[width=0.7\linewidth]{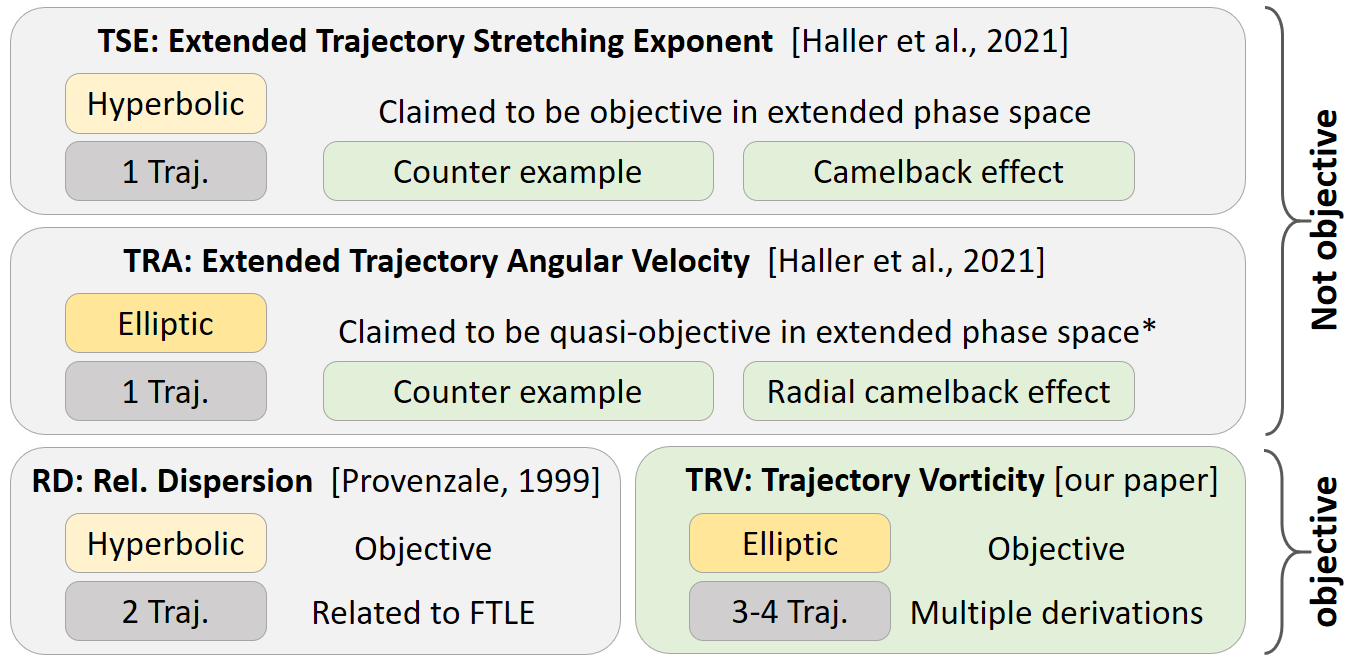}\\
    \centering
    {*under avg. vorticity condition}
    \caption{Summary of concepts, with our contributions highlighted in green. Using counterexamples, we show that TSE and TRA are not objective and we discuss their camelback effects. Afterwards, we introduce the TRV measure, which is based on multiple trajectories and proven to be objective.}
    \label{fig:overview-methods}
\end{figure}

In this paper, we make the following contributions:
\begin{itemize}
\item
We show that the claims in \citep{Haller21singletrajecory} concerning objectivity of   $\mbox{TSE}$, $\overline{\mbox{TSE}}$, $\overline{\mbox{TRA}}$ are incorrect. In fact, we show that  neither   $\mbox{TSE}$ nor 
$\overline{\mbox{TSE}}$ are objective in the extended phase space. Further, $\overline{\mbox{TRA}}$ is not quasi-objective in the extended phase space  under an averaged-vorticity-based condition.
\item
We present a further analysis of $\overline{\mbox{TSE}}$ and $\overline{\mbox{TRA}}$ showing a "camelback effect" that limits the usefulness of $\overline{\mbox{TSE}}$ and $\overline{\mbox{TRA}}$.
\item
We introduce a new flow measure  $\mbox{TRV}$ ({\em trajectory vorticity}), which measures rotational behavior based on at least three trajectories (in 2D) or four trajectories (in 3D), respectively.
\item
We show that $\mbox{TRV}$ can be derived in two independent ways: by  approaches for unsteadiness minimization, and by considering the relative spin tensor.
\item
We prove that $\mbox{TRV}$ is objective.
\item
We apply the new measure $\mbox{TRV}$ to a number of sparse trajectory data sets, including drifting buoys in the Atlantic, midge tracking data, and trajectories in a simulated vortex street.
\end{itemize}
Figure~\ref{fig:overview-methods} summarizes the main concepts in our paper. The green components are our novel contributions.

\section{Basic Concepts and Related Work}

Objectivity, a concept from continuum mechanics, refers to the invariance of a measure under a moving reference system. Let $s(\vx,t)$, $\vw(\vx,t)$, $\mT(\vx,t)$ be time-dependent scalar-, vector-, and tensor fields, respectively. Further, let $\tilde{s}(\tilde{\vx},t)$, $\tilde{\vw}(\tilde{\vx},t)$, 
 $\tilde{\mT}(\tilde{\vx},t)$ be their observations under the Euclidean frame change
 \begin{equation}
\label{eq_define_movingsystem}
\vx = \mQ(t) \; \widetilde{\vx} + \vb(t)
\end{equation}
where $\mQ=\mQ(t)$ is a time-dependent rotation tensor and  $\vb(t)$ is a time-dependent translation vector. Then $s, \vw, \mT$ are \emph{objective} if the following conditions hold, cf.~\cite{truesdell_book}:
 \begin{eqnarray}
\tilde{s}(\tilde{\vx},t)=s(\vx,t) \;,\; \tilde{\vw}(\tilde{\vx},t) = \mQ^\Transp \, \vw(\vx,t) \;,\;
\tilde{\mT}(\tilde{\vx},t) =  \mQ^\Transp \, \mT(\vx,t) \,   \mQ.
\end{eqnarray} 
Since its introduction to flow analysis \citep{Astarita79}, objectivity became a common demand for newly-introduced flow measures\citep{haller2005}. In fact, there are a variety of objective flow measures focusing on hyperbolic (stretching) properties, such as FTLE \citep{Shadden2005}. Also, objective flow measures focusing on elliptic (rotational) behavior have been introduced and can roughly be divided into three classes: (1) {\em Replacing the spin tensor by the relative spin tensor} 
 \citep{Drouot76,Astarita79}: These approaches use the fact that the rate-of-strain-tensor is objective 
 and consider the spin tensor (vorticity) in the local frame given by the rate-of-strain tensor.  (2) {\em Replacing the spin tensor by the spin deviation tensor} 
  \citep{haller2016,Liu19:SpinDeviation}, where the fact is used that the difference of two spin tensors at different locations but the same time is objective. 
 (3) {\em Finding optimal reference frames minimizing the unsteadiness of the observed flow}: introduced by    \cite{Guenther:2017:Siggraph}, this created a number of follow-up work 
 \citep{Guenther:2019:VIS,Hadwiger19,Rojo20,Guenther:2020:TVCG,Rautek21,Zhang2021KillingObserverInteraction}. Recently, objectivity of unsteadiness minimization approaches has been questioned  \citep{Haller20:can} but confirmed \citep{Theisel:2021:PhysFluids}.

All approaches mentioned so far have in common that they rely on an underlying velocity field and its derivatives. For our problem where only a few trajectories are available they are not applicable. There are, however, a few flow measures based on only few trajectories. 
The {\em relative dispersion}
was introduced by \cite{Provenzale_1999_review} and was further analyzed by \cite{Haller2000,Haller21singletrajecory}.
Given are two distinct $C^1$ continuous trajectories $\vx_1(t), \vx_2(t)$ along with their derivatives $\dot{\vx}_1(t), \dot{\vx}_2(t)$ . Defining the local relative dispersion
\begin{equation}
\label{eq_RD1}
\mbox{rd} = \mbox{rd}_{\vx_1(t),\vx_2(t)}(t) = \frac{ (\vx_2-\vx_1)^\Transp \, (\dot{\vx}_2-\dot{\vx}_1) }{   (\vx_2-\vx_1)^\Transp \, (\vx_2-\vx_1)},
\end{equation}
one gets the relative dispersion by integrating $\mbox{rd}$ along trajectories:
\begin{equation}
\label{eq_RD2}
\mbox{RD}_{\vx_1(t),\vx_2(t)}^{t_0,t_N} = \int_{t_0}^{t_N} \mbox{rd} \; dt = \ln
\frac{| \vx_2(t_N) - \vx_1(t_N) |}{| \vx_2(t_0) - \vx_1(t_0) |}.
\end{equation}
Note that $\mbox{RD}$ is objective  \citep{Haller21singletrajecory}.

\cite{Haller21singletrajecory} introduced measures for stretching and rotation that are based on single trajectories only: 
 {\em Extended trajectory stretching exponents} $\mbox{TSE}$ , $\overline{\mbox{TSE}}$, and {\em extended trajectory angular velocity} $\mbox{TRA}$,    
 $\overline{\mbox{TRA}}$. 
Given is a $C^2$ continuous trajectory $\vx(t)$ for $t \in [t_0,t_N]$, its first and second derivatives $\dot{\vx}(t), \ddot{\vx}(t)$, and a positive constant $v_0$ accounting for a certain ratio between space and time units to make them non-dimensionalized. Considering $\vx(t)$ in an extended phase space gives for the first and second derivative of a trajectory $\underline{\vx}(t)$:
\begin{equation}
 \dot{\underline{\vx}}(t) = \begin{pmatrix}
  \frac{1}{v_0} \,   \dot{\vx}(t)\\
 1
\end{pmatrix}
\;\;\;,\;\;\;
 \ddot{\underline{\vx}}(t) = \begin{pmatrix}
\frac{1}{v_0} \, \ddot{\vx}(t)\\
 0
\end{pmatrix}.
\end{equation}
Then a local stretching measure can be defined as
\begin{equation}
\mbox{tse} = \mbox{tse}_{\vx(t),v_0}(t) = \frac{ \dot{\underline{\vx}}^\Transp \, \ddot{\underline{\vx}} }{ \dot{\underline{\vx}}^\Transp \, \dot{\underline{\vx}} }
= \frac{ \dot{\vx}^\Transp \, \ddot{\vx} }{ \dot{\vx}^\Transp \, \dot{\vx} + v_0^2}
\end{equation}
from which the Lagrangian measures $\mbox{TSE}$ and $\overline{\mbox{TSE}}$ are computed by integrating $\mbox{tse}$ along the trajectory:
\begin{align}
\label{eq_define_TSE1}
\mbox{TSE}_{\vx(t),v_0}^{t_0,t_N} &= \frac{1}{\Delta t} \int_{t_0}^{t_N} \mbox{tse} \; dt 
\;=\;
 \frac{1}{\Delta t} \ln \sqrt{  \frac{ |  \dot{\vx}(t_N)   |^2 + v_0^2  }{ |  \dot{\vx}(t_0)   |^2 + v_0^2} }
\\
\overline{\mbox{TSE}}_{\vx(t),v_0}^{t_0,t_N} &= \frac{1}{\Delta t} \int_{t_0}^{t_N} | \mbox{tse}| \, dt
\label{eq_appr_TSEbar}
\; \approx\;
 \frac{1}{\Delta t}
 \sum_{i=0}^{N-1} 
 \left|
 \ln \sqrt{  \frac{ |  \dot{\vx}(t_{i+1})   |^2 + v_0^2  }{ |  \dot{\vx}(t_i)   |^2 + v_0^2} }
 \right|
\end{align}
with $\Delta t = t_N-t_0$. The discretization in Eq.~\eqref{eq_appr_TSEbar} samples $\vx(t)$ at $N+1$ time steps $t_0 < t_1 < ... < t_N$.

For defining $\mbox{TRA}$, the $(n+1)$-dimensional matrix function
\begin{equation}
\mathbf{tra} = \mathbf{tra}_{\vx(t),v_0}(t) =
\frac{ \dot{\underline{\vx}} \,  \ddot{\underline{\vx}}^\Transp -  \ddot{\underline{\vx}} \,  \dot{\underline{\vx}}^\Transp }{  \dot{\underline{\vx}}^\Transp \,  \dot{\underline{\vx}} }
\end{equation}
can be introduced that describes the local angular velocity. Note that $\mathbf{tra}$ is an anti-symmetric matrix, from  
which one gets by integration along the trajectory Lagrangian measures
\begin{align}
\mbox{TRA}_{\vx(t),v_0}^{t_0,t_N} &= \frac{1}{\Delta t} \frac{\sqrt{2}}{2}
\left|
\int_{t_0}^{t_N}  \mathbf{tra} \; dt
\right|_{Fr}
\\
&=
\label{eq_define_TRA1}
\frac{1}{\Delta t}
\cos^{-1}
\frac{ \dot{\vx}(t_0)^\Transp  \, \dot{\vx}(t_N) + v_0^2  }{ \sqrt{  | \dot{\vx}(t_0)|^2 + v_0^2 } \sqrt{  | \dot{\vx}(t_N)|^2 + v_0^2 } }
\\
\overline{\mbox{TRA}}_{\vx(t),v_0}^{t_0,t_N} &= \frac{1}{\Delta t}  \frac{\sqrt{2}}{2}  \int_{t_0}^{t_N} \left|  \mathbf{tra} \right|_{Fr}  \, dt
\\
&\approx
\frac{1}{\Delta t}
\sum_{i=0}^{N-1}
\cos^{-1}
\frac{ \dot{\vx}(t_i)^\Transp  \, \dot{\vx}(t_{i+1}) + v_0^2  }{ \sqrt{  | \dot{\vx}(t_i)|^2 + v_0^2 } \sqrt{  | \dot{\vx}(t_{i+1})|^2 + v_0^2 } }
\end{align}
where $_{Fr}$ denotes the Frobenius norm of a matrix. \cite{Haller21singletrajecory}  claimed that  $\mbox{TSE}$ and 
$\overline{\mbox{TSE}}$ are objective in the extended phase space, and that $\mbox{TRA}$ and  $\overline{\mbox{TRA}}$ are quasi-objective in the extended phase space under a certain condition put to the average vorticity in a certain neighborhood of the trajectory.

\section{$\mbox{TSE}$, $\mbox{TRA}$, and objectivity}

Single-trajectory flow measures are attractive because they need minimal information to infer the flow behavior of an underlying field. Obviously, single-trajectory measures cannot be objective in the Euclidean observation space because one may think of a reference system moving with the trajectory, making each trajectory zero \citep{Haller21singletrajecory}. Because of this, \cite{Haller21singletrajecory} considered objectivity in an extended phase space.  In this section, we analyze and correct statements of \cite{Haller21singletrajecory} about objectivity in the extended phase space.

\subsection{Definition of $\mbox{TSE}$ and $\mbox{TRA}$}

We recapitulate the definition of $\mbox{TSE}$ from \cite{Haller21singletrajecory}, keeping their notation as much as possible. We start with a single observed trajectory
$\vx(t)$  in $n$-D ($n=2,3$) for $t \in [t_0, t_N]$ running from $\vx_0 = \vx(t_0)$ to  $\vx_N = \vx(t_N)$. Further, we assume that $\vx(t)$ is a trajectory (path line) of an underlying unsteady velocity field $\vv(\vx,t)$, i.e., $\dot{\vx}(t) = \frac{d \vx}{d t} = \vv(\vx(t),t)$ for all $t \in [t_0, t_N]$.  
Following \cite{Haller21singletrajecory}, $\vv$ is transformed into a non-dimensionalized field $\vu$ by
\begin{equation}
\vy = \frac{\vx}{L}  \;\;\;,\;\;\; \tau = \tau_0 + \frac{t-t_0}{T} \;\;\;,\;\;\; v_0 = \frac{L}{T}
\label{eq:non-dimension-params}
\end{equation}
where $L,T, v_0$ are certain positive constants for a field that need to be determined by additional knowledge about the data. Generally, the scaling factor $v_0$ is non-zero, i.e., $v_0\neq 0$. This transformation rephrases  $\vx(t)$ into
the non-dimensionalized trajectory 
\begin{equation}
\vy(\tau) = \frac{1}{L} \vx(t_0 + T(\tau-\tau_0))
\label{eq:define_y}
\end{equation}
running from $\vy_0=\vy(\tau_0)=\frac{1}{L} \vx_0$ to $\vy_N=\vy(\tau_N)=\frac{1}{L} \vx_N$ with $\tau_N = \tau_0 + \frac{t_N-t_0}{T}$.
Further, it gives the non-dimensionalized vector field
\begin{equation}
\label{eq_define_u}
\vu(  \vy, \tau) =  \frac{1}{v_0} \vv \left( L \vy , t_0 + T (\tau- \tau_0) \right).
\end{equation}
Note that (\ref{eq_define_u}) contains a correction of a missing term $\frac{1}{v_0}$
 in formula (26) in \citep{Haller21singletrajecory}. The error in  formula (26) in \citep{Haller21singletrajecory} can be seen in the following way: suppose $\vv$ is a constant vector field, i.e., $\vv(\vx,t)=\vv_c$. Then formula (26) in \citep{Haller21singletrajecory} would give $\vu(\vy,\tau)=\vv_c$ no matter how $v_0$ is chosen. This would contradict to the formula before (33) in \citep{Haller21singletrajecory}.

Following \citep{Haller21singletrajecory} further, an extended phase space $\mY = 
\begin{pmatrix}
  \vy\\
 z
\end{pmatrix}
$ is introduced. Transformation of $\vy(\tau)$ and $\vu(  \vy, \tau)$ into this extended phase space gives
\begin{equation}
\mY(\tau) = 
\begin{pmatrix}
  \vy(\tau)\\
 \tau
\end{pmatrix}
\;\;\;,\;\;\; 
\mU(\mY) = 
\begin{pmatrix}
  \vu(\vy,z)\\
 1
\end{pmatrix}
\label{eq:define-extended-phase-space}
\end{equation}
where $\mY(\tau)$ is the trajectory in the extended phase space 
running from $\mY_0 = \mY(\tau_0) = 
 \begin{pmatrix}
   \vy_0\\
 \tau_0
\end{pmatrix}$ to  $\mY_N = \mY(\tau_N) = 
 \begin{pmatrix}
   \vy_N\\
 \tau_N
\end{pmatrix}$, and $\mU(\mY)$ is the underlying vector field. The tangent vector of $\mY(\tau)$ is
\begin{equation}
\mY'(\tau) = \frac{ d \, \mY}{ d \, \tau} =
 \begin{pmatrix}
   \vy'(\tau)\\
1
\end{pmatrix}
=
\begin{pmatrix}
 \frac{1}{v_0}  \dot{\vx}( t_0 + T(\tau-\tau_0 ) )\\
1
\end{pmatrix}.
\end{equation}
Note that $\mU(\mY)$ is an autonomous dynamical system now: $\mU$ is  a steady velocity field in the extended phase space. Then \cite{Haller21singletrajecory} defines  $\mbox{TSE}$ and $\mbox{TRA}$ as
 \begin{eqnarray}
\label{eq_define_TSE2} 
\mbox{TSE}_{t_0}^{t_N}(\vx_0,v_0) &=& 
 \frac{1}{\Delta t} \ln \frac{| \mY'(\tau_N) | }{ | \mY'(\tau_0) | }
= \frac{1}{\Delta t} \ln \frac{| \mU(\mY_N) | }{ | \mU(\mY_0) | }
\\
\label{eq_define_TRA2} 
\mbox{TRA}_{t_0}^{t_N}(\vx_0,v_0) &=& 
 \frac{1}{\Delta t} \cos^{-1} \frac{ \mY'(\tau_0)^\Transp \, \mY'(\tau_N)   }{ | \mY'(\tau_0) | |\mY'(\tau_N)| }
 \\
&=& \frac{1}{\Delta t} \cos^{-1} \frac{\mU(\mY_0)^\Transp \, \mU(\mY_N)  }{ | \mU(\mY_0) | | \mU(\mY_N) | }
\end{eqnarray}
where $\Delta t = t_N-t_0$, (\ref{eq_define_TSE2}) is identical to the right-hand side of (\ref{eq_define_TSE1}),
and (\ref{eq_define_TRA2}) is identical to the right-hand side of (\ref{eq_define_TRA1}).
To show objectivity of $\mbox{TSE}$ in the extended phase space, one has to prove that $\mbox{TSE}$ is invariant under observation in any moving Euclidean reference system in the extended phase space. Analogous to Eq.~\eqref{eq_define_movingsystem}, such moving reference system is defined by
 \begin{equation}
 \label{eq_define_moving_frame_1}
 \mY = 
 \boldsymbol{\mathcal Q(\tau)}  \widetilde{\mY} + \mB(\tau)
 \;,\;
   \boldsymbol{\mathcal Q}(\tau)
 =
 \begin{pmatrix}
\mQ(\tau) &  \vNull \\
 \vNull^\Transp & 1
\end{pmatrix}
\;,\;
\mB(\tau) =
 \begin{pmatrix}
\vb(\tau)  \\
0
\end{pmatrix}
\end{equation}
with $\mQ(\tau) \in SO(n)$ being a rotation matrix, and $\vNull$ being the zero-vector. The observed trajectory $\widetilde{\mY}(\tau)$ and the underlying velocity field  $\widetilde{\mU}(\widetilde{\mY},\tau)$ in the new moving reference system are 
 \begin{align}
\label{eq_trafo_in_extended_reference_system2} 
\widetilde{\mY}(\tau) &=     \boldsymbol{\mathcal Q}^\Transp(\tau)  (\mY(\tau) - \mB(\tau))
\\
\label{eq_trafo_in_extended_reference_system}
\widetilde{\mU}(\widetilde{\mY},\tau) &=   \boldsymbol{\mathcal Q}^\Transp(\tau)  
\left(
\mU
\left(    \boldsymbol{\mathcal Q}(\tau) \widetilde{\mY} + \mB(\tau)  \right) - \dot{  \boldsymbol{\mathcal Q}}(\tau)  \widetilde{\mY} -  \dot{\mB}(\tau) 
\right)
 \end{align}
where the new trajectory $\widetilde{\mY}(\tau)$ runs from $\widetilde{\mY}_0 = \widetilde{\mY}(\tau_0)$ to 
$\widetilde{\mY}_N = \widetilde{\mY}(\tau_N)$. Then,  $\mbox{TSE}$ in the moving reference system is
 \begin{eqnarray}
 \label{eq_define_TSEtilde}
\widetilde{\mbox{TSE}}_{t_0}^{t_N}(\vx_0,v_0) &=& 
\frac{1}{\Delta t} \ln \frac{ |  \widetilde{\mY}'(\tau_N)   | }{ | \widetilde{\mY}'(\tau_0) |  }
=
\frac{1}{\Delta t} \ln \frac{ | \widetilde{\mU}(\widetilde{\mY}_N, \tau_N) | }{ | \widetilde{\mU}(\widetilde{\mY}_0, \tau_0) |  }.
\\
 \label{eq_define_TRAtilde}
\widetilde{\mbox{TRA}}_{t_0}^{t_N}(\vx_0,v_0) &=& 
\frac{1}{\Delta t} \cos^{-1} \frac{ \widetilde{\mY}'(\tau_0)^\Transp \,  \widetilde{\mY}'(\tau_N)   }{ | \widetilde{\mY}'(\tau_0) | |  \widetilde{\mY}'(\tau_N)   |  }
 \end{eqnarray}
 To show objectivity of $\mbox{TSE}$ in the extended phase space, one has to prove $\mbox{TSE} = \widetilde{\mbox{TSE}}$ for any moving reference frame, as given by Eq.~\eqref{eq_define_moving_frame_1}.
To show quasi-objectivity of $\mbox{TRA}$ under averaged-vorticity condition, one has to prove  
$\mbox{TRA} = \widetilde{\mbox{TRA}}$ for all reference frames (\ref{eq_define_moving_frame_1}) in which the averaged-vorticity condition is fullfilled.

\subsection{A simple counter-example}

We show the non-objectivity of $\mbox{TSE}$ in the extended phase space by a simple counter-example. We set the 2D observed trajectory $\vx(t)$ and the underlying
velocity field $\vv(\vx,t)$ as
\begin{equation}
\label{eq_definetrajectory}
\vx(t) = 
\begin{pmatrix}
  e^t - 1 \\
   t \, (t+1)
\end{pmatrix}
\;\;\;,\;\;\;
\vv(\vx,t) = 
\begin{pmatrix}
  x+1 \\
   2 \, t+1
\end{pmatrix}
\end{equation}
for $t \in [t_0, t_N]=[0,1]$   and  $\vx = (x,y)^\Transp$. To calculate $\mbox{TSE}$ as in Eq.~\eqref{eq_define_TSE1}, we only need information at time $t_0$ and $t_N$. This gives
\begin{eqnarray}
\vx_0 = 
(0,0)^\Transp
&,&
\vx_N = 
(e-1,2)^\Transp
\\
\dot{\vx}(t_0) = \vv( \vx_0, t_0) = 
(1,1)^\Transp
&,&
\dot{\vx}(t_N) = \vv( \vx_N, t_N) = 
(e,3)^\Transp.
\end{eqnarray}
For the non-dimensionalization transformation, we set $\tau_0=0$, resulting in $\tau_N= \frac{1}{T}$. This gives with Eqs.~\eqref{eq:define_y} and \eqref{eq_define_u}
\begin{equation}
\vy(\tau) = \frac{1}{L}
\begin{pmatrix}
  e^{T \tau}  - 1\\
   T \tau (T \tau+1)
\end{pmatrix}
\;\;\;,\;\;\;
\vu(  \vy, \tau) = 
\frac{1}{v_0}  \begin{pmatrix}
 L  \bar{x}+1 \\
   2 \, T \tau+1
\end{pmatrix}
\end{equation}
with  $\vy = (\bar{x}, \bar{y})^\Transp$, and therefore we obtain at $\tau_0$ and $\tau_N$
\begin{eqnarray}
\vy_0  = 
\begin{pmatrix}
  0 \\
   0
\end{pmatrix}
&,&
\vy_N = \frac{1}{L}  
\begin{pmatrix}
  e-1 \\
   2
\end{pmatrix}
\\
\vu(\vy_0,\tau_0) =\frac{1}{v_0}
 \begin{pmatrix}
  1 \\
   1
\end{pmatrix}
&,&
\vu(\vy_N,\tau_N) =\frac{1}{v_0}
 \begin{pmatrix}
  e \\
   3
\end{pmatrix}.
\end{eqnarray}
Transforming to the extended phase space 
\begin{equation}
\mY = ( \bar{x}, \bar{y}, z )^\Transp
\end{equation}
using Eq.~\eqref{eq:define-extended-phase-space} gives 
 \begin{equation}
\mY(\tau) = 
\begin{pmatrix}
  \frac{1}{L} (e^{T \tau}  - 1)\\
   \frac{1}{L} ( T \tau (T \tau+1)) \\
   \tau
\end{pmatrix}
\;\;\;,\;\;\;
\mU(\mY) = 
\begin{pmatrix}
\frac{1}{v_0} ( L  \bar{x}+1) \\
\frac{1}{v_0} ( 2 \, T z +1)\\
1
\end{pmatrix}
\end{equation}
with the following position and tangent at the curve end points
\begin{eqnarray}
\mY_0  = 
(0,0,0)^\Transp
&,&
\mY_N =
\left(\frac{e-1}{L} , \frac{2}{L} , \frac{1}{T} \right)^\Transp
\\
\mU(\mY_0) =
\left(\frac{1}{v_0} , \frac{1}{v_0} , 1\right)^\Transp
&,&
\mU(\mY_N) =
\left(\frac{e}{v_0} , \frac{3}{v_0} , 1 \right)^\Transp.
\end{eqnarray}
Inserting into Eqs.~\eqref{eq_define_TSE2} and \eqref{eq_define_TRA2}, this results in \mbox{TSE} and \mbox{TRA}:
\begin{equation}
\label{eq_define_tse2}
\mbox{TSE} = \ln \sqrt{ \frac{ e^2 + 9 + v_0^2 }{ 2 + v_0^2 }  }
\;\;,\;\;
\mbox{TRA} = \cos{-1} \frac{ e + 3 + v_0^2 }{ \sqrt{2 + v_0^2}\sqrt{e^2 + 9 + v_0^2}  }.  
\end{equation}
For our counterexample, it is sufficient to choose a particular moving Euclidean reference system
 (\ref{eq_define_moving_frame_1}) by
 \begin{equation}
 \label{eq_Galilean_moving_frame}
   \boldsymbol{\mathcal Q}(\tau) = \mI \;\;\;,\;\;\; \mB(\tau) = \tau 
  \begin{pmatrix}
\vb_c \\
0
\end{pmatrix}
 \end{equation}
where $\mI$ is the identity matrix and $\vb_c = (x_c,y_c)^\Transp$ is a constant 2D vector. For this particular reference system, we get by 
(\ref{eq_trafo_in_extended_reference_system2}),
(\ref{eq_trafo_in_extended_reference_system}):
 \begin{eqnarray}
\widetilde{\mY}(\tau) &=& \mY(\tau) - \tau  \begin{pmatrix}
\vb_c \\
0
\end{pmatrix}
\\
\widetilde{\mU}(\widetilde{\mY},\tau) &=& \mU \left( \mY + \tau
\begin{pmatrix}
\vb_c \\
0
\end{pmatrix}
\right) 
-
\begin{pmatrix}
\vb_c \\
0
\end{pmatrix}.
\end{eqnarray} 
 This gives the following trajectory end points and tangents:
\begin{eqnarray}
\widetilde{\mY}_0  = 
(0,0,0)^\Transp
\;\;\;  ,\;\;\;
\widetilde{\mY}_N =
\left( \frac{e-1}{L}  -  \frac{x_c}{T} \,\,,\,\,
\frac{2}{L}  -  \frac{y_c}{T} \,\,,\,\, \frac{1}{T} \right)^\Transp
\\
\label{eq_YTilde0}
\widetilde{\mU}(\widetilde{\mY}_0,\tau_0) =
\widetilde{\mY}'(\tau_0)
=
\left( \frac{1}{v0}  - x_c \,\,,\,\,
\frac{1}{v0}  - y_c \,\,,\,\, 1 \right)^\Transp
\\
\label{eq_YTildeN}
\widetilde{\mU}(\widetilde{\mY}_N, \tau_N) =
 \widetilde{\mY}'(\tau_N)
=
\left( \frac{e}{v0}  - x_c \,\,,\,\,
\frac{3}{v0} - y_c \,\,,\,\, 1 \right)^\Transp
\end{eqnarray}
and finally by inserting into Eq.~\eqref{eq_define_TSEtilde}, we get $\widetilde{\mbox{TSE}}$:
\begin{equation}
\label{eq_tilde_tse}
\widetilde{\mbox{TSE}} = \ln \sqrt{ \frac{ (e-v_0 x_c)^2 + (3 - v_0 y_c)^2 + v_0^2 }{  (1 - v_0 x_c)^2 +  (1 - v_0 y_c)^2 + v_0^2 }  }
\end{equation}
Analogously, $\widetilde{\mbox{TRA}}$ follows by inserting (\ref{eq_YTilde0})--(\ref{eq_YTildeN}) into (\ref{eq_define_TRAtilde}).
Since there is no positive constant $v_0$, cf. \eqref{eq:non-dimension-params}, that makes $\mbox{TSE}$   in (\ref{eq_define_tse2})   and   $\widetilde{\mbox{TSE}} $ in (\ref{eq_tilde_tse}) identical for any 
$\vb_c = (x_c,y_c)^\Transp$, non-objectivity of $\mbox{TSE}$ in the extended phase space is shown.
Since in our example both $\mU(\mY)$ and
$\widetilde{\mU}(\widetilde{\mY},\tau)$ have zero vorticity, the average-vorticity condition in  
\citep{Haller21singletrajecory} is trivially fulfilled. Thus, the difference of $\mbox{TRA}$ and  $\widetilde{\mbox{TRA}}$ gives that $\mbox{TRA}$ is not quasi-objective in the extended phase under the average-vorticity condition.

\subsection{Where is the error?}

\cite{Haller21singletrajecory} considered a non-zero vector $\vxi_0$ at $(\vx_0,t_0)$ that is advected with $\vv$ along $\vx(t)$, resulting in
\begin{equation}
\label{eq_pde_general}
\dot{\vxi}(t) = \nabla \vv(\vx(t),t) \; \vxi(t) 
\;\;\; ,\; \; \;
 \vxi(t_0) =   \vxi_0.
\end{equation}
Then, $\vxi(t)$ is observed under a moving reference system (\ref{eq_define_movingsystem}).
Objectivity of $\vxi$ is deduced from  (\ref{eq_pde_general}), (\ref{eq_define_movingsystem}):
\begin{equation}
\label{eq_defineobjectovpy_xi}
\widetilde{ \vxi }(t) =  \mQ^\Transp(t)  \; \vxi(t) 
\end{equation}
where $\widetilde{ \vxi }$ is the observation of $\vxi$ under the moving reference system (\ref{eq_define_movingsystem}). From 
(\ref{eq_defineobjectovpy_xi}) follows the objectivity of 
$\frac{1}{\Delta t} \ln \frac{| \vxi(t_N) | }{| \vxi_0 |}$.
We note that (\ref{eq_defineobjectovpy_xi}) follows from (\ref{eq_pde_general}) and (\ref{eq_define_movingsystem}) only if another implicit assumption holds: objectivity of the seeding vector  $\vxi_0$, i.e.,  $\widetilde{ \vxi }_0 =  \mQ^\Transp(t_0)  \;   \vxi_0$.

The approach of \cite{Haller21singletrajecory} is to set $\vxi_0 = \vv_0 = \vv(\vx_0, t_0)$. With this, additional conditions are  necessary  to ensure
\begin{eqnarray}
\label{eq_cond_obj_1}
\dot{\vv}(t) &=& \nabla \vv(\vx(t),t) \; \vv( \vx, t) 
\\
\label{eq_cond_obj_2}
\widetilde{\vv}( \widetilde{\vx} , t) &=&  \mQ^\Transp(t)  \; \vv( \vx , t) 
\end{eqnarray}
where (\ref{eq_cond_obj_1}) corresponds to  (\ref{eq_pde_general}) and (\ref{eq_cond_obj_2}) corresponds to  (\ref{eq_defineobjectovpy_xi}).
 To ensure (\ref{eq_cond_obj_1}), \cite{Haller21singletrajecory} introduced the condition
\begin{equation}
\nonumber
\mbox{(A1)}  \;\;\;\;\;\;  \;\;\;\;\;\; \;\;\;\;\;\;  \;\;\;\;\;\; 
 \;\;\;\;\;\;  \delta_t \vv(\vx,t) = \vNull
\end{equation}
in the current observation frame. However, condition (A1) does not ensure  (\ref{eq_cond_obj_2}) because $\vxi_0=\vv_0$ is not objective. Since the observation of 
$\vv$ under the moving reference system (\ref{eq_define_movingsystem}) is \citep{Haller20:can}
\begin{equation}
\widetilde{\vv}(\widetilde{\vx},t) = \mQ^\Transp(t) \left(  \vv(\vx,t)  - \dot{\mQ}(t) \; \widetilde{\vx} - \dot{\vb}(t) \right),
\end{equation}
Eq.~\eqref{eq_cond_obj_2} is only fulfilled for $ \dot{\mQ}=\vNull, \dot{\vb}=\vNull$, i.e., the reference frame is not moving but static, resulting in demanding that $\widetilde{\vv}(\widetilde{\vx},t)$ is steady. This means that the condition for the quasi-objectivity of TSE is the steadiness of both 
$\vv$ and $\widetilde{\vv} $ in all considered reference frames. We remark that this is a rather strong condition for quasi-objectivity: it excludes the consideration of all moving reference frames.

The transformation to the extended reference system transforms $\vv$ to the steady vector field $\mU$, making the condition (A1) for (\ref{eq_cond_obj_1}) in the extended reference frame obsolete. However, the observation  $\widetilde{\mU}$  of $\mU$ under a moving reference system
 (\ref{eq_define_moving_frame_1}) is \emph{not} a steady vector field anymore, as shown in
 (\ref{eq_trafo_in_extended_reference_system}). This means that
 \begin{equation}
  \widetilde{\mU}( \widetilde{\mY} , \tau ) =
   \boldsymbol{\mathcal Q(\tau)} \;
   \mU(\mY) 
 \end{equation} 
 does not hold in general but only for particular steady reference frames. Because of this, TSE is in the extended phase space not objective but only quasi-objective under restriction to a static reference system.
 
\paragraph*{Summary:}
The error was to assume that the observation of an autonomous system (steady vector field) in the extended phase space under a moving reference frame remains an autonomous system.

\paragraph*{Remarks:}
A similar argumentation gives that $\overline{\mbox{TSE}} $ is not objective in the extended phase space, and and that
$\overline{\mbox{TRA}}$ is is not quasi-objective in the extended phase space under the averaged-vorticity-based condition.
$\mbox{TSE}$, $\overline{\mbox{TSE}} $, $\mbox{TRA}$ and $\overline{\mbox{TRA}}$ are not even Galilean invariant because the moving reference system 
 (\ref{eq_Galilean_moving_frame}) in the counterexample was performing a Galilean transformation.

\subsection{Further Analysis of $\mbox{TSE}$ and $\mbox{TRA}$}

\begin{figure}[t]
    \centering
    \includegraphics[width=0.31\linewidth]{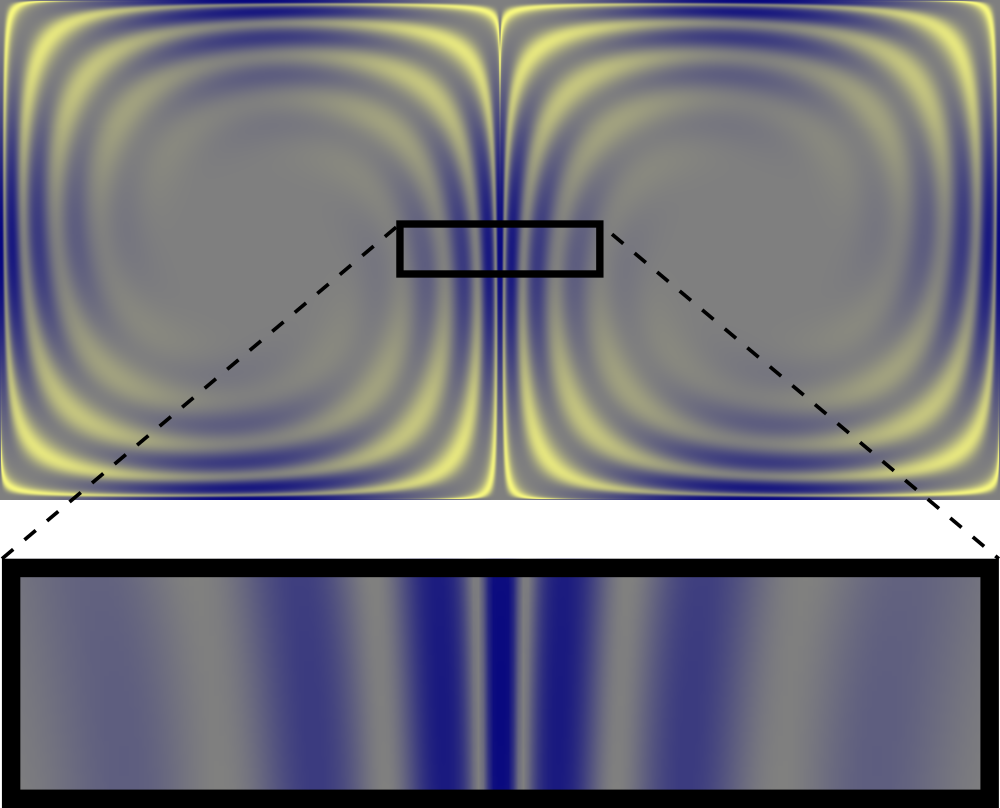}\hfill%
    \includegraphics[width=0.31\linewidth]{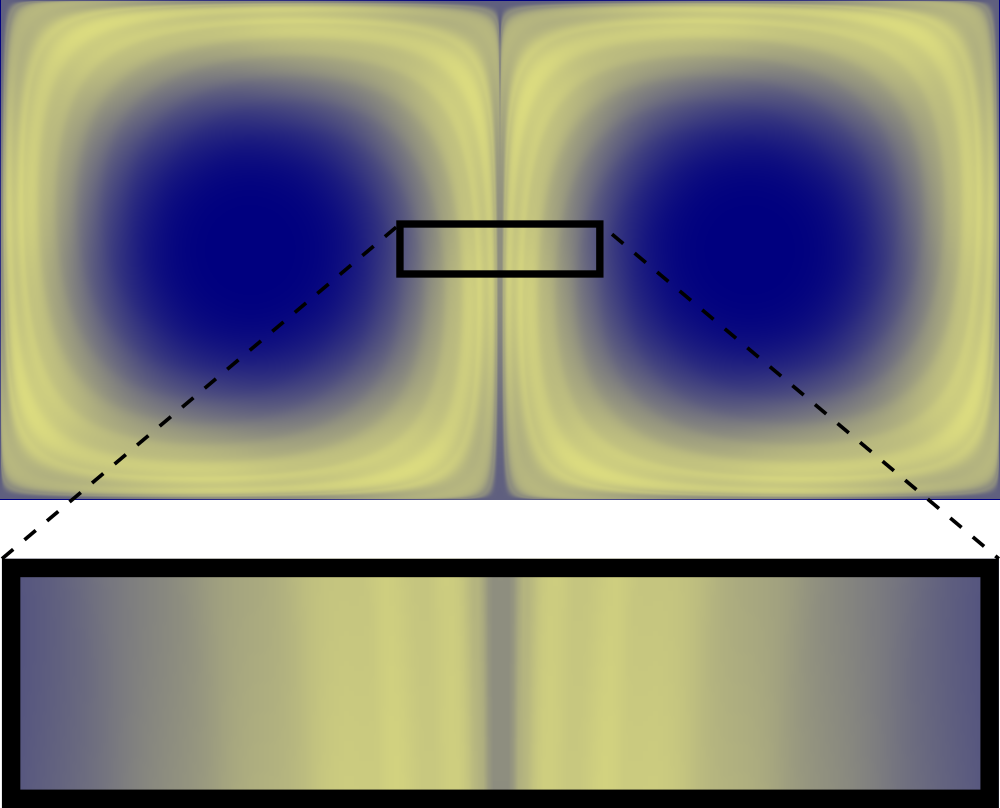}\hfill%
    \includegraphics[width=0.31\linewidth]{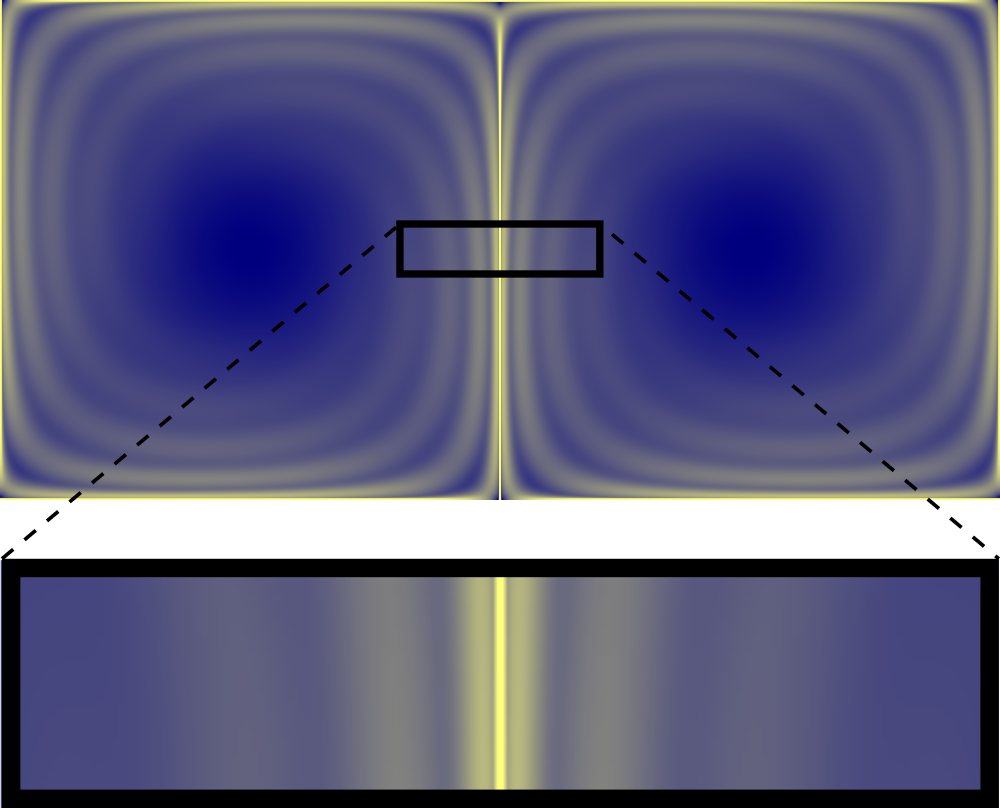}%
    
\begin{minipage}{0.31\linewidth}\centering%
    \begin{tikzpicture}
      \node[inner sep=0] (image) at (0,0)
      {\includegraphics[height=0.9\linewidth, width=0.3cm, angle=270]{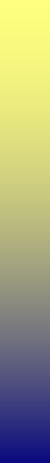}};
      \node[anchor=north east, black, overlay, inner xsep=0] at (image.south east) {\scriptsize $0.005$};
      \node[anchor=north west, black, overlay, inner xsep=0] at (image.south west) {\scriptsize $-0.005$};
      \node[anchor=north, black, overlay] at (image.south) {\scriptsize $\mbox{TSE}$ \vphantom{$\overline{\mbox{TRA}}$}};
      \draw [black] (image.north west) -- +(270:0.4cm);
      \draw [black] (image.north east) -- +(270:0.4cm);
    \end{tikzpicture}%
\end{minipage}
\hfill%
\begin{minipage}{0.31\linewidth}\centering%
    \begin{tikzpicture}
      \node[inner sep=0] (image) at (0,0)
      {\includegraphics[height=0.9\linewidth, width=0.3cm, angle=270]{images/color-gradient-sqr.png}};
            \node[anchor=north east, black, overlay, inner xsep=0] at (image.south east) {\scriptsize $0.025$};
      \node[anchor=north west, black, overlay, inner xsep=0] at (image.south west) {\scriptsize $0$};
      \node[anchor=north, black, overlay] at (image.south) {\scriptsize $\overline{\mbox{TSE}}$};
      \draw [black] (image.north west) -- +(270:0.4cm);
      \draw [black] (image.north east) -- +(270:0.4cm);
    \end{tikzpicture}%
\end{minipage}
\hfill%
\begin{minipage}{0.31\linewidth}\centering%
    \begin{tikzpicture}
      \node[inner sep=0] (image) at (0,0)
      {\includegraphics[height=0.9\linewidth, width=0.3cm, angle=270]{images/color-gradient-sqr.png}};
      \node[anchor=north east, black, overlay, inner xsep=0] at (image.south east) {\scriptsize $0.6$};
      \node[anchor=north west, black, overlay, inner xsep=0] at (image.south west) {\scriptsize $0$};
      \node[anchor=north, black, overlay] at (image.south) {\scriptsize $\mbox{FTLE}$ \vphantom{$\overline{\mbox{TRA}}$}};
      \draw [black] (image.north west) -- +(270:0.4cm);
      \draw [black] (image.north east) -- +(270:0.4cm);
    \end{tikzpicture}%
\end{minipage}

\vspace{0.2cm}
    \caption{$\mbox{TSE}$ (left), $\overline{\mbox{TSE}}$ (center) and $\mbox{FTLE}$ (right) calculated on a \textsc{Steady Double Gyre} flow for integration duration $\tau=10$:
        $\vv(x,y) = (
            -0.1 \pi \sin(x\pi)\cos(y\pi),
            0.1 \pi \cos(x\pi)\sin(y\pi)
        )^\Transp$.
    The centerline is a strongly separating structure, as can be seen in the FTLE image. However, both $\mbox{TSE}$ and $\overline{\mbox{TSE}}$ exhibit high values not on this line, but rather a "camelback" around it.
    This result is similar for different choices of $v_0$, shown here for $v_0 = 1$.}
    \label{fig:tse}
\end{figure}

\begin{figure}[t]
    \centering
    \raisebox{4.2em}{\rotatebox[origin=c]{90}{\small{${\mbox{TRA}}$\vphantom{$\overline{\mbox{TRA}}$}}}}~%
    \begin{tikzpicture}%
	\node[anchor=south west,inner sep=0] (image) at (0,0) {\includegraphics[width=0.9\linewidth]{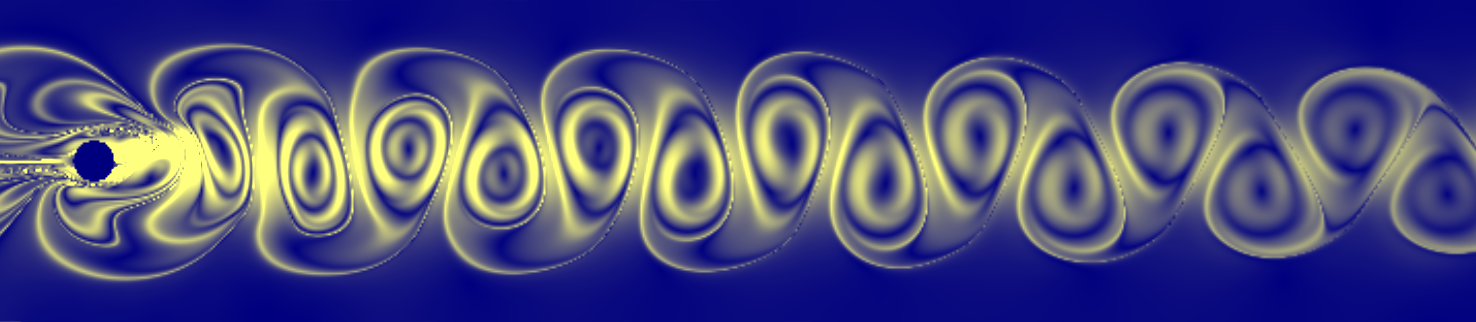}};%
	\begin{scope}[node distance=-1.8mm and -1.2mm, x={(image.south east)},y={(image.north west)}]%
	\node[draw=none] (label) at (1.03,0.5) {\includegraphics[width=0.015\linewidth]{images/color-gradient-sqr.png}}; %
	\node[draw=none,overlay,below= of label] {\scriptsize $0.0$};%
	\node[draw=none,overlay,above= of label] {\scriptsize $0.12$};%
	\end{scope}%
\end{tikzpicture}\\%
    \raisebox{4.2em}{\rotatebox[origin=c]{90}{\small{$\overline{\mbox{TRA}}$\vphantom{$\overline{\mbox{TRA}}$}}}}~%
    \begin{tikzpicture}%
	\node[anchor=south west,inner sep=0] (image) at (0,0) {\includegraphics[width=0.9\linewidth]{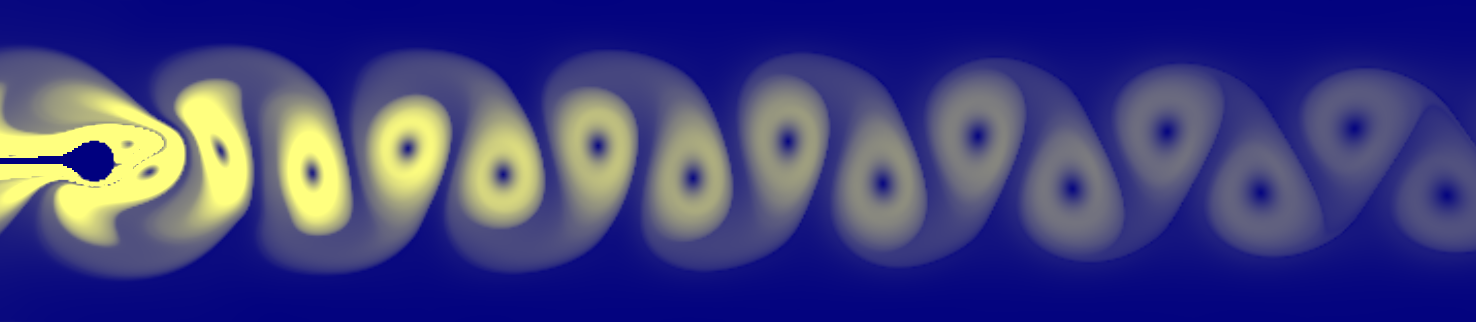}};%
	\begin{scope}[node distance=-1.8mm and -1.2mm, x={(image.south east)},y={(image.north west)}]%
	\node[draw=none] (label) at (1.03,0.5) {\includegraphics[width=0.015\linewidth]{images/color-gradient-sqr.png}}; %
	\node[draw=none,overlay,below= of label] {\scriptsize $0.0$};%
	\node[draw=none,overlay,above= of label] {\scriptsize $0.7$};%
	\end{scope}%
\end{tikzpicture}\\%
    \raisebox{4.4em}{\rotatebox[origin=c]{90}{\small{LAVD\vphantom{$\overline{\mbox{TRA}}$}}}}~%
    \begin{tikzpicture}%
	\node[anchor=south west,inner sep=0] (image) at (0,0) {\includegraphics[width=0.9\linewidth]{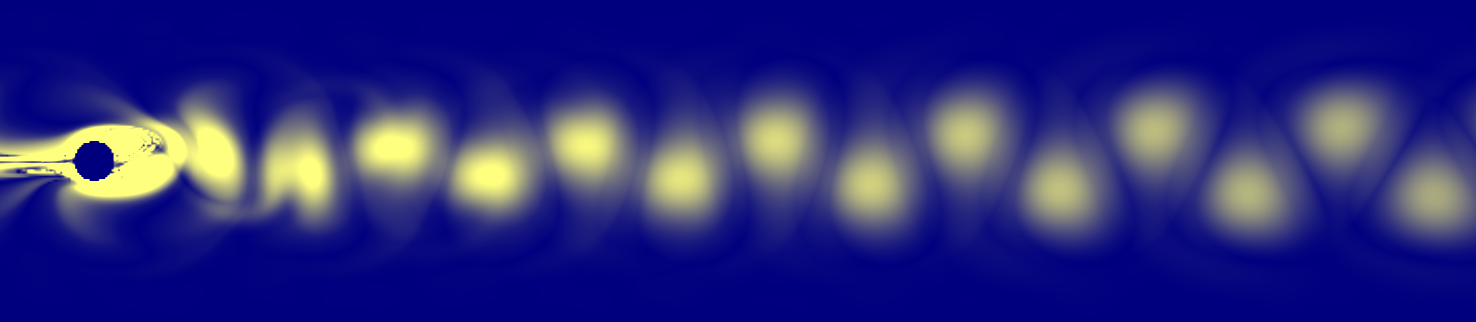}};%
	\begin{scope}[node distance=-1.8mm and -1.2mm, x={(image.south east)},y={(image.north west)}]%
	\node[draw=none] (label) at (1.03,0.5) {\includegraphics[width=0.015\linewidth]{images/color-gradient-sqr.png}}; %
	\node[draw=none,overlay,below= of label] {\scriptsize $0.0$};%
	\node[draw=none,overlay,above= of label] {\scriptsize $8.0$};%
	\end{scope}%
\end{tikzpicture}%
    \caption{Comparison of ${\mbox{TRA}}$, $\overline{\mbox{TRA}}$, and LAVD for an integration duration of $\tau=3$ in the \textsc{Cylinder} flow. Note the radial camelback effects in ${\mbox{TRA}}$ and $\overline{\mbox{TRA}}$, both having low values in the interior of vortices. In LAVD, the avg. vorticity was taken from the full domain.}
    \label{fig:tra}
\end{figure}

Being not objective (neither in Euclidean nor in extended phase space) does not necessarily mean that $\mbox{TSE}$ and $\mbox{TRA}$ are not useful. In fact, \cite{Haller21singletrajecory} and 
\cite{bartos2021quasiobjective}
show a number of successful applications. Because of this, we further analyze $\mbox{TSE}$ and $\mbox{TRA}$ on a dense field of trajectories. We observe a "camelback effect" of TSE that can be seen in Figure~\ref{fig:tse}: in a flow, hyperbolic separators are usually lines in 2D and surfaces in 3D (one may think of FTLE ridges). TSE tends to become large in areas close to hyperbolic separators, but small again exactly on the separators. This means, a low TSE can indicate either absence of hyperbolic separators, or an exact hit of a  hyperbolic separator.    
For $\mbox{TRA}$ and $\overline{\mbox{TRA}}$, we observe a "radial camelback effect" in Figure~\ref{fig:tra}: in a neighborhood of a vortical area $\mbox{TRA}$ and $\overline{\mbox{TRA}}$ tend to get large, but towards the center of rotation (vortex core) both measures exhibit smaller values.  Also this makes the interpretation of low TRA values ambiguous, limiting the applicability of TRA.
For both images, we set $v_0=1$. For reference, we visualized the vortices with Lagrangian averaged vorticity deviation (LAVD)~\citep{haller2016}, where the vorticity average was computed for the entire domain.

 \section{Trajectory Vorticity }

 Once we have seen that single-trajectory measures are not objective (neither in the Euclidean nor in the extended phase space), we search for objective measures that are based on more than one but still few trajectories. For this, we assume that the trajectories are in coherent (hyperbolic or elliptic) areas and therefore driven by similar phenomena.
 For hyperbolic regions, such an objective measure is relative dispersion, cf. Eqs.~\eqref{eq_RD1}--\eqref{eq_RD2}, which is computed from (at least) two trajectories. 
In the following, we introduce the -- to the best of our knowledge -- first objective measure of elliptic flow behavior that is based on very few trajectories only, called {\em Trajectory Vorticity} (TRV).
We begin with the formal definition in Section~\ref{sec:definition-trv}. Afterwards, we explain the derivation and properties in Section \ref{sec:properties-trv}.

 \subsection{Definition of TRV}
\label{sec:definition-trv}

In 2D/3D, we consider three/four distinct $C^2$ continuous trajectories $\vx_1=\vx_1(t)$,  $\vx_2=\vx_2(t)$, $\vx_3=\vx_3(t)$, [$\vx_4=\vx_4(t)$] with first derivatives $\dot{\vx}_1, \dot{\vx}_2, \dot{\vx}_3, [\dot{\vx}_4]$ and second derivatives $\ddot{\vx}_1, \ddot{\vx}_2, \ddot{\vx}_3,  [\ddot{\vx}_4]$. (Note that content in brackets $[\;]$ refers to additional content present in 3D but not in 2D.)
We introduce an $n$-dimensional  anti-symmetric matrix function
\begin{equation}
\mathbf{trv} = \mathbf{trv}_{\vx_1,\vx_2,\vx_3 [,\vx_4]}(t),
\end{equation}
based on this, we define the Lagrangian Trajectory Vorticity \mbox{TRV}  by integrating $\mathbf{trv}$  as
\begin{eqnarray}
\mbox{TRV}_{\vx_1,\vx_2,\vx_3 [,\vx_4]}^{t_0,t_N} &=& 
\frac{1}{\Delta t }
\frac{\sqrt{2}}{2}
\left|
\int_{t_0}^{t_N}  \mathbf{trv} \; dt
\right|_{Fr}
\\
\overline{\mbox{TRV}}_{\vx_1,\vx_2,\vx_3 [,\vx_4]}^{t_0,t_N} &=& 
\frac{1}{\Delta t}
\frac{\sqrt{2}}{2}
\int_{t_0}^{t_N} \left| \mathbf{trv} \right|_{Fr} \, dt
\end{eqnarray}
with $\Delta t = t_N-t_0$ and $_{Fr}$ denoting the Frobenius norm of a matrix.
To define $\mathbf{trv} $,  we introduce the time-dependent matrices
\begin{eqnarray}
\label{eq_trv_1}
\mX &=& \mX(t)  = \begin{pmatrix}
\vx_1 & \vx_2 & \vx_3 & [\vx_4]\\
1 & 1 & 1 & [1]
\end{pmatrix}
\\
\dot{\mX}  &=&
\dot{\mX}(t)  = \begin{pmatrix}
\dot{\vx}_1 & \dot{\vx}_2 & \dot{\vx}_3 & [\dot{\vx}_4]\\
0 & 0 & 0& [0]
\end{pmatrix}
\\
\label{eq_trv_1a}
\ddot{\mX} &=& 
\ddot{\mX}(t)  = \begin{pmatrix}
\ddot{\vx}_1 & \ddot{\vx}_2 & \ddot{\vx}_3 & [\ddot{\vx}_4]\\
0 & 0 & 0& [0]
\end{pmatrix}
\end{eqnarray}
from which we compute
\begin{equation}
\label{eq_defineH}
\mH = \mH(t) = \dot{\mX}  \;  \mX^{-1} \;\;\;,\;\;\;
\dot{\mH} = \dot{\mH}(t) = (\ddot{\mX} - \mH \dot{\mX}) \; \mX^{-1}.
\end{equation}
Setting $\mI_z = (\mI, \vNull)$ with $\mI$ being the identity matrix and $\vNull$ being the zero column-vector, we compute
\begin{equation}
\label{eq_defineJ}
\mJ = \mJ(t) = \mI_z \; \mH \; {\mI_z}^\Transp \;\;\;,\;\;\;
\dot{\mJ} = \dot{\mJ}(t) = \mI_z \; \dot{\mH} \; {\mI_z}^\Transp.
\end{equation}
Further, we consider the symmetric and anti-symmetric parts
\begin{equation}
\label{eq_define_S_W}
\mS = \frac{\mJ + \mJ^\Transp}{2}\;\;\;,\;\;\;
\dot{\mS} = \frac{\dot{\mJ} + \dot{\mJ}^\Transp}{2}
\;\;\;,\;\;\;
\mW = \frac{\mJ - \mJ^\Transp}{2}.
\end{equation}
Let $\mE$ be the rotational matrix containing the (normalized) eigenvectors of $\mS$ as columns, i.e., the transformation
\begin{equation}
\label{eq_define_E}
\overline{\mS} = \mE^\Transp \; \mS \; \mE\;\;\;,\;\;\;
\overline{\dot{\mS}} = \mE^\Transp \; \dot{\mS} \; \mE
\end{equation}
yields a diagonal matrix $\overline{\mS}$. From this, we compute 
\begin{equation}
\overline{\mW}_s = 
\begin{pmatrix}
0 &  -u_3 & [u_2]\\
u_3 & 0 & [-u_1] \\
[-u_2] & [u_1] & [0]
\end{pmatrix}
\end{equation}
with
\begin{equation}
\label{eq_define_u1u2u3}
([u_1, u_2,] u_3) = \left(
\left[
 \frac{ \overline{\dot{\mS}}_{3,2} }  { \overline{\mS}_{2,2} - \overline{\mS}_{3,3}  } ,
 \frac{ \overline{\dot{\mS}}_{1,3} }  { \overline{\mS}_{3,3} - \overline{\mS}_{1,1}  } ,
\right] 
 \frac{ \overline{\dot{\mS}}_{2,1} }  { \overline{\mS}_{1,1} - \overline{\mS}_{2,2}  } 
 \right)
\end{equation}
where  $\overline{\mS}_{i,j}$ denotes the entry at $[i,j]$ of the matrix $\overline{\mS}$.
Then, the back transformation 
\begin{equation}
\label{eq_define_Ws_1}
\mW_s = \mE \; \overline{\mW}_s \; \mE^\Transp
\end{equation}
gives 
\begin{equation}
\label{eq_trv_2}
\mathbf{trv}_{\vx_1,\vx_2,\vx_3 [,\vx_4]}(t) = 
\left\{
\begin{array}{cc}
\mW - \mW_s  & \mbox{if $\mW_s$ is computable } \\
\vNull & \mbox{else}
\end{array}
\right.
\end{equation}
where $\mW_s$ is computable if $\mX$ is invertible and $\dot{\mS}$ has distinct eigenvalues.
Here, $\vNull$ denotes the zero matrix.

TRV can also be computed for more than 3 (in 2D) or 4 (in 3D) trajectories. In this case, $\mX, \ddot{\mX}, \ddot{\mX}$
in (\ref{eq_trv_1})--(\ref{eq_trv_1a}) receive more columns, and $\mX^{-1}$ in (\ref{eq_defineH}) denotes the \emph{right} Moore-Penrose pseudo-inverse $\mX^\Transp(\mX\mX^\Transp)^{-1}$ instead of the matrix inverse.


\subsection{Properties and equivalent definitions of TRV}
\label{sec:properties-trv}

\begin{theorem}
\label{thm_1}
$\mathbf{trv}_{\vx_1,\vx_2,\vx_3 [,\vx_4]}^{t_0,t_N}(t)$ is objective.
\end{theorem}
The formal proof of this theorem is in the appendix. From theorem \ref{thm_1} follows directly that $\mbox{TRV}$ and $\overline{\mbox{TRV}}$ are objective as well. In addition to this, we give further information about the interpretation of $\mbox{TRV}$ and $\overline{\mbox{TRV}}$ in the following. 

The main idea for the introduction of $\mathbf{trv}$ is to consider a time-dependent vector field $\vv(\vx,t)$ that is fitted locally to the given trajectories, and to apply existing approaches for the objectivization of $\vv$. In fact, the vector field $\vv(\vx,t)$ given by
\begin{equation}
\label{eq_define_v}
\begin{pmatrix}
\vv(\vx,t)\\
0
\end{pmatrix}
=  \mH(t) \; 
\begin{pmatrix}
\vx\\
1
\end{pmatrix}
\end{equation}
fits  $\vx_1(t), \vx_2(t), \vx_3(t) [, \vx_4(t)]$ in the sense that all trajectories $\vx_1(t), \vx_2(t), \vx_3(t) [, \vx_4(t)]$ are path lines of $\vv$. Note that $\vv$ is linear in space but non-linear in time: fixing $t$ results in a linear vector field. Then, the Jacobian and the time partial derivative of $\vv$ are
\begin{equation}
\vv_t(\vx,t)
=  \dot{\mH}(t) \; 
\begin{pmatrix}
\vx\\
1
\end{pmatrix}
\;\;\;,\;\;\;
\begin{pmatrix}
\mJ(t) &   \va(t)\\
\vNull^\Transp & 0
\end{pmatrix} =  \mH(t). 
\end{equation}
Note that $\vv_t$ is linear in space, and $\mJ$ and $\va$ are constant in space. Also note that $\mS$ and $\mW$ in (\ref{eq_define_S_W}) denote the rate-of-strain tensor and the spin tensor of  $\vv$, respectively. 

Now, $\mathbf{trv}$ is obtained by applying standard objectivization approaches to $\vv$. In fact, $\mathbf{trv}$ is obtained by replacing $\mW$ with the relative spin tensor $\mW_r  = \mW - \mW_s$ using the strain rotation rate tensor 
\begin{equation}
\label{eq_define_Ws_2}
\mW_s = - \mE \, {\mE_t}^\Transp
\end{equation}
where $\mE$ is defined in (\ref{eq_define_E}) and $\mE_t = \frac{\partial \, \mE }{ \partial \, t}$,
as done by \cite{Drouot76,Astarita79}.
The identity of (\ref{eq_define_Ws_1}) and (\ref{eq_define_Ws_2})
follows directly from (\ref{eq_define_E})--(\ref{eq_define_u1u2u3}).

Interestingly, $\mathbf{trv}$ can also be obtained in a different way: by unsteadiness minimization following \cite{Guenther:2017:Siggraph}. Observing $\vv$ defined in (\ref{eq_define_v}) in a moving reference frame $\widetilde{\vx} = \mR(t) \, \vx + \vc(t)$ gives for the time-derivative of $\vv$ in the new reference frame \citep{Guenther:2017:Siggraph}
\begin{equation}
\widetilde{\vv}_t = \mR \,( \vv_t - \mM \, \vu)
\end{equation}
with 
$\mM = ( -\mJ \, \vx_p + \vv_p \,,\, \mJ \,,\, \vx_p \,,\, \mI)$, 
$\vx_p = 
\begin{pmatrix}
0 & -1\\ 1 & 0
\end{pmatrix}
\vx
$,
$\vv_p = 
\begin{pmatrix}
0 & -1\\ 1 & 0
\end{pmatrix}
\vv
$ in 2D, and
$\mM = ( -\mJ \, \mX + \mV \,,\, \mJ \,,\, \mX \,,\, \mI)$, $\mX=sk(\vx)$, $\mV=sk(\vv)$ in 3D, and $\vu$ is a 6-vector in 2D and 12-vector in 3D:
\begin{equation}
\vu =
\begin{pmatrix}
\vu_1\\ \vu_2\\ \vu_3\\ \vu_4
\end{pmatrix} =
\begin{pmatrix}
ap( \mR^\Transp   \dot{\mR}  )\\ 
\mR^\Transp  \dot{\vc}\\
ap( \mR^\Transp   \ddot{\mR}  - (\mR^\Transp   \dot{\mR})^2 )   \\ 
- ( \mR^\Transp  \ddot{\vc} -   \mR^\Transp  \dot{\mR} \mR^\Transp \dot{\vc}    )
\end{pmatrix}
\end{equation}
where $ap$ transforms the anti-symmetric part of a matrix to a scalar/vector: $ap(\mM)= \frac{1}{2}( \mM_{1,2}-\mM_{2,1})$ in 2D and $ap(\mM)= \frac{1}{2}( \mM_{3,2}-\mM_{2,3} 
\,,\,
\mM_{1,3}-\mM_{3,1}\,,\,\mM_{2,1}-\mM_{1,2}
)^\Transp$ in 3D. Conversely, $sk$ is the inverse function transforming a scalar/vector to an anti-symmetric matrix, here for 2D/3D:
\begin{equation}
sk(\alpha)=
\begin{pmatrix}
0 & \alpha\\ -\alpha & 0
\end{pmatrix}
\;\;\;/\;\;\;
sk\begin{pmatrix}
\alpha\\ \beta\\ \gamma
\end{pmatrix}=
\begin{pmatrix}
0 & -\gamma & \beta  \\ 
\gamma & 0 & -\alpha  \\ 
-\beta & \alpha & 0
\end{pmatrix}
\end{equation}
%
Note that due to the spatial linearity of $\vv$, both $\vv_t$ and $\widetilde{\vv}_t$ are spatially linear as well. Then, searching for an unsteadiness minimizing observation frame results in searching an unknown $\vu$ fulfilling
\begin{equation}
\label{eq_setminimizationproblem}
\int_U \|\widetilde{\vv}_t\|^2 \; d V \to \min
\end{equation}
where $U$ is a certain 2D/3D cube. 
Note that due to the spatial linearity of $\vv$, the problem in Eq.~\eqref{eq_setminimizationproblem} is under-determined in $\vu$, i.e., it has a whole family of solutions $\vu$. However, all solutions of (\ref{eq_setminimizationproblem}) have the same  component $\vu_1$, that is, component $\vu_1$ 
is independent of the size and location of $U$. With this, we get
\begin{equation}
\label{eq_u_1}
\mW_s = - sk(\vu_1)
\end{equation}
The proof of the equivalence of 
(\ref{eq_define_Ws_1}) and
(\ref{eq_u_1}) is a straight computation  for which we provide a Maple sheet in the accompanying material. Eq.~\eqref{eq_u_1} gives that $\mathbf{trv}$ can be computed by observing $\mW$ in an unsteadiness minimizing reference frame following \cite{Guenther:2017:Siggraph}.

\paragraph*{Remarks:}
The equivalence of unsteadiness minimization and relative spin tensor consideration shown here does not hold for general vector fields but only for spatially linear ones as considered here.

Another popular approach to objectivize flow measures is to replace $\mW$ by the spin-deviation tensor 
\begin{equation}
\mW - \frac{1}{vol(U)} \int_U \mW \, d V.
\end{equation}
For $\vv$ defined in (\ref{eq_define_v}), this gives a perfectly objective but trivial solution: it is zero everywhere.

 \cite{Haller20:can} raises concerns against considering the  relative spin tensor by defining a compatibility condition and showing that general  relative spin tensor consideration do not fulfill them. For our approach, this is not an issue because due to the spatial linearity of $\vv$ the compatibility condition of \cite{Haller20:can} is always fulfilled.

\section{Results}
In the following, we apply our approach to four data sets.
We begin with a synthetic example to demonstrate the capability of our approach to separate rotating motion from reference frame rotation.

\subsection{Three trajectories}

\begin{figure}[t]
    \centering
    \hspace{1em}%
    \begin{minipage}{0.24\linewidth}\centering%
    $(p, q) = (6.75, -2.25)$
    \end{minipage}\hfill%
    \begin{minipage}{0.24\linewidth}\centering%
    $(p, q) = (3.25, 1.25)$
    \end{minipage}\hfill%
    \begin{minipage}{0.24\linewidth}\centering%
    $(p, q) = (1.25, 3.25)$
    \end{minipage}\hfill%
    \begin{minipage}{0.24\linewidth}\centering%
    $(p, q) = (-2.25, 6.75)$
    \end{minipage}\hspace{0.34cm}$~$\\%
    
    \raisebox{5em}{\rotatebox[origin=c]{90}{\small{Original}}}%
    \includegraphics[width=0.2\linewidth]{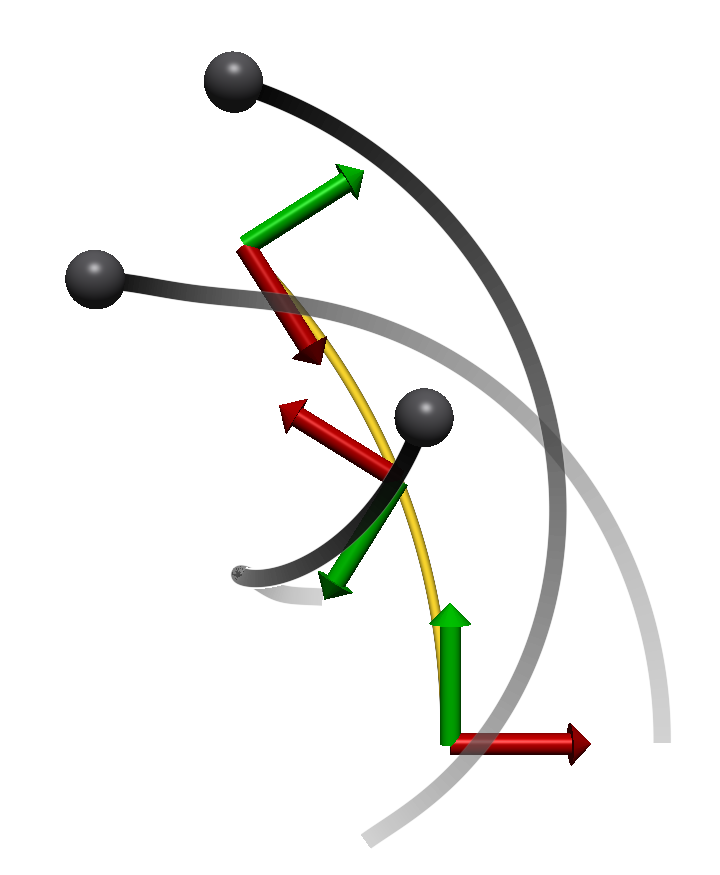}\hfill%
    \includegraphics[width=0.2\linewidth]{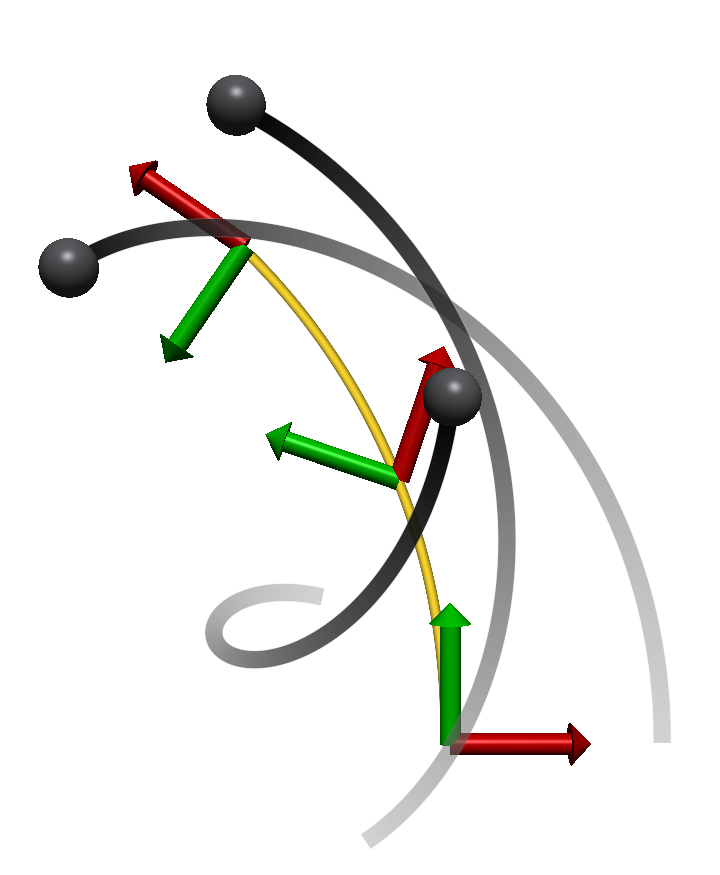}\hfill%
    \includegraphics[width=0.2\linewidth]{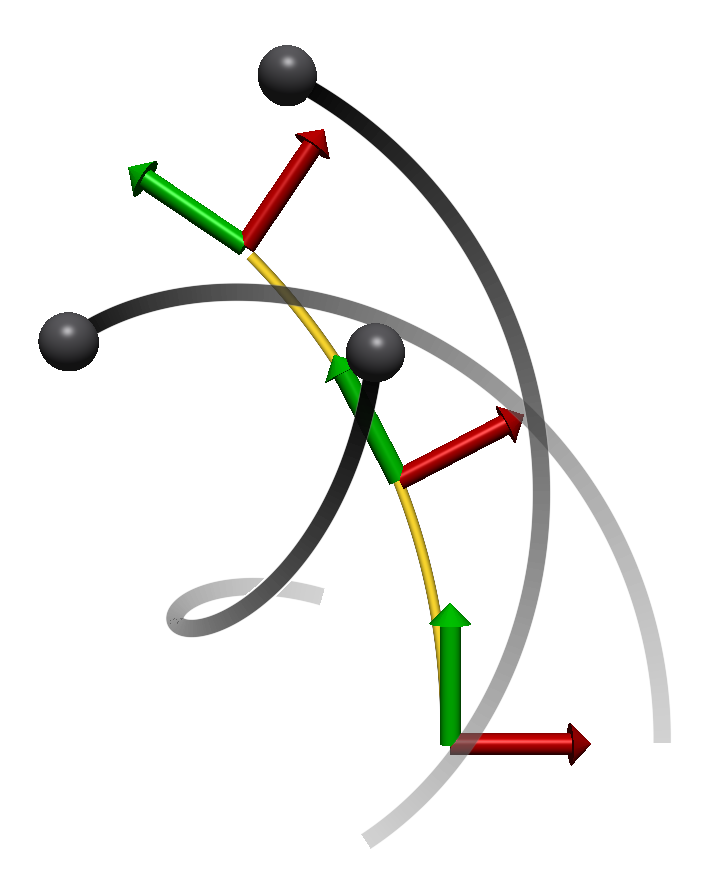}\hfill%
    \includegraphics[width=0.2\linewidth]{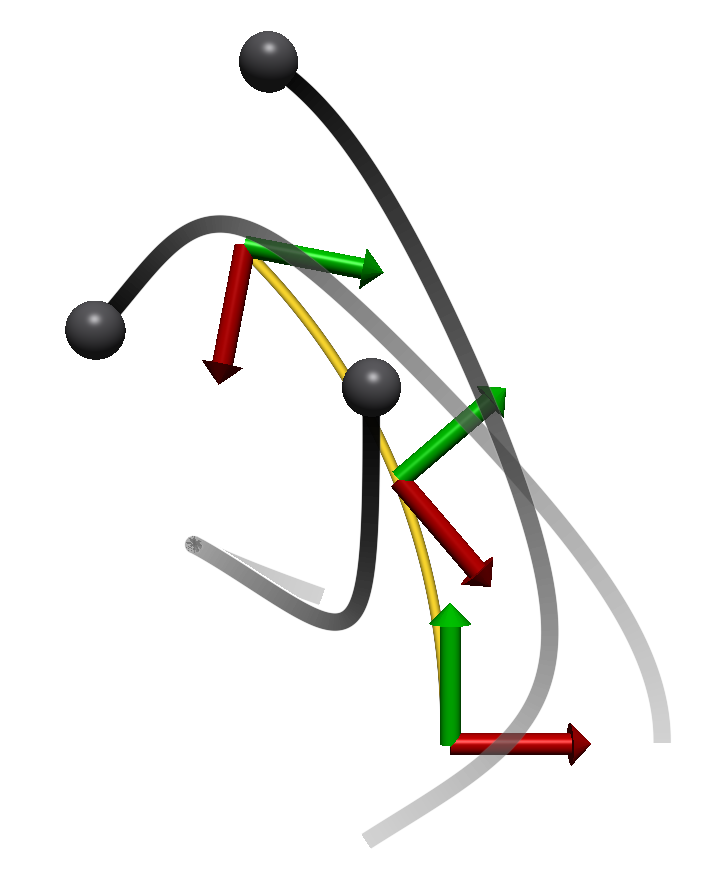}%
    
    \raisebox{4em}{\rotatebox[origin=c]{90}{\small{Correct system}}}%
    \includegraphics[width=0.2\linewidth]{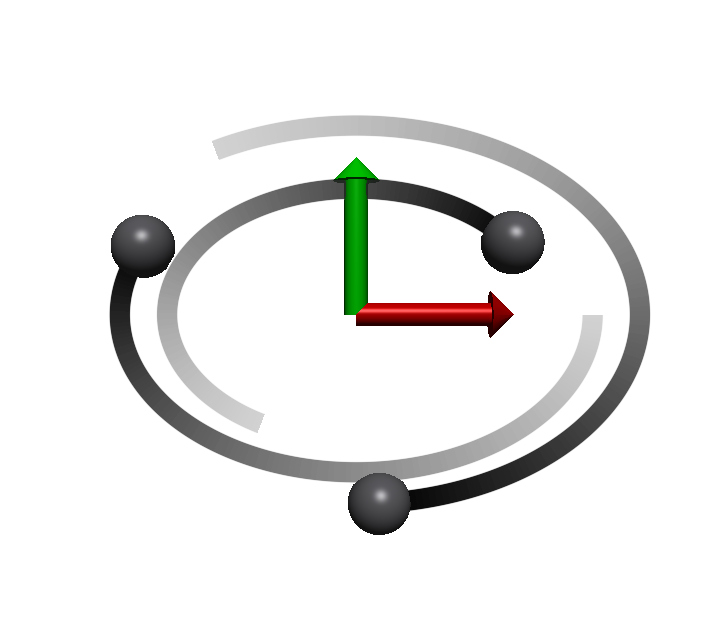}\hfill%
    \includegraphics[width=0.2\linewidth]{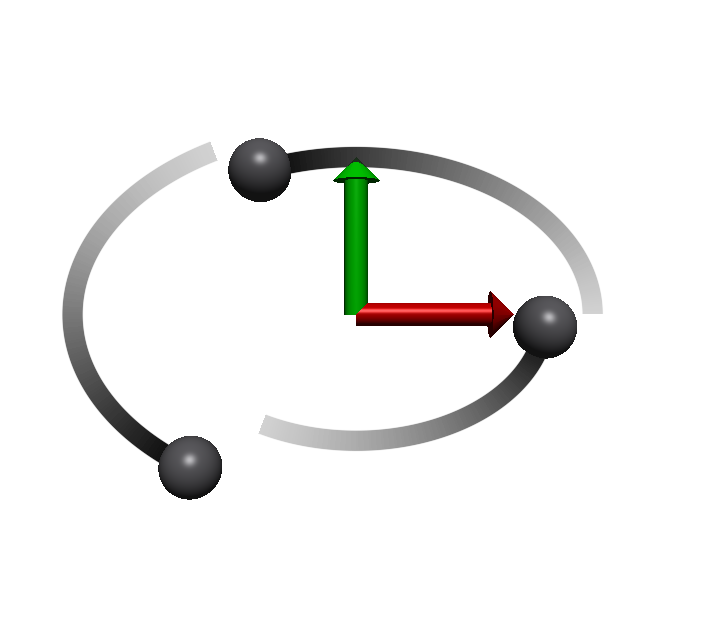}\hfill%
    \includegraphics[width=0.2\linewidth]{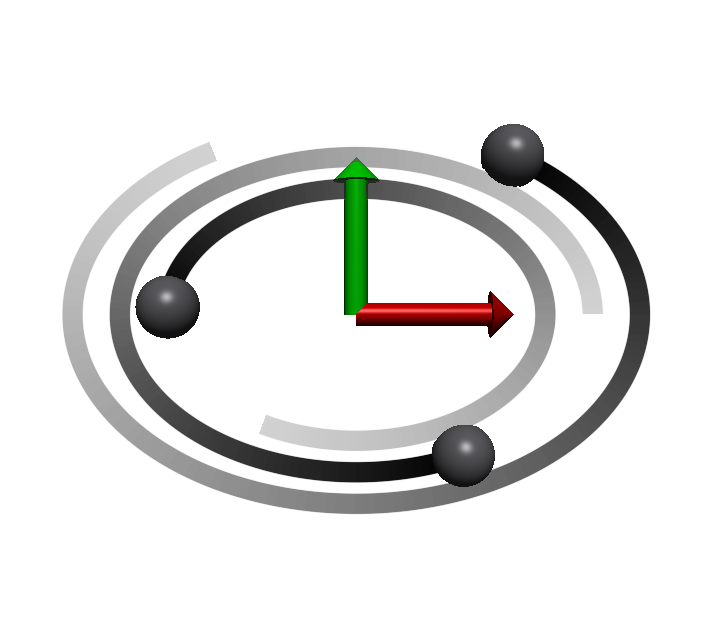}\hfill%
    \includegraphics[width=0.2\linewidth]{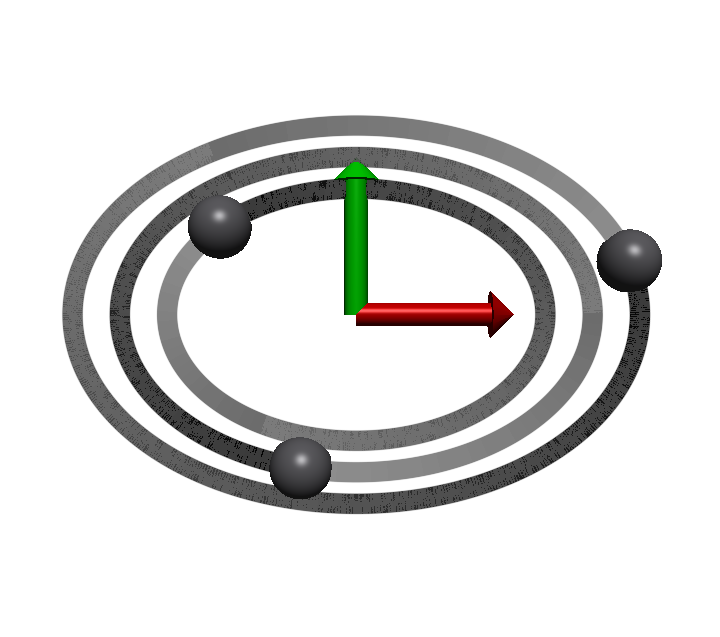}%
    
    \raisebox{3.6em}{\rotatebox[origin=c]{90}{\small{Wrong system}}}%
    \includegraphics[width=0.2\linewidth]{images/three-lines-lines_rf_1.png}\hfill%
    \includegraphics[width=0.2\linewidth]{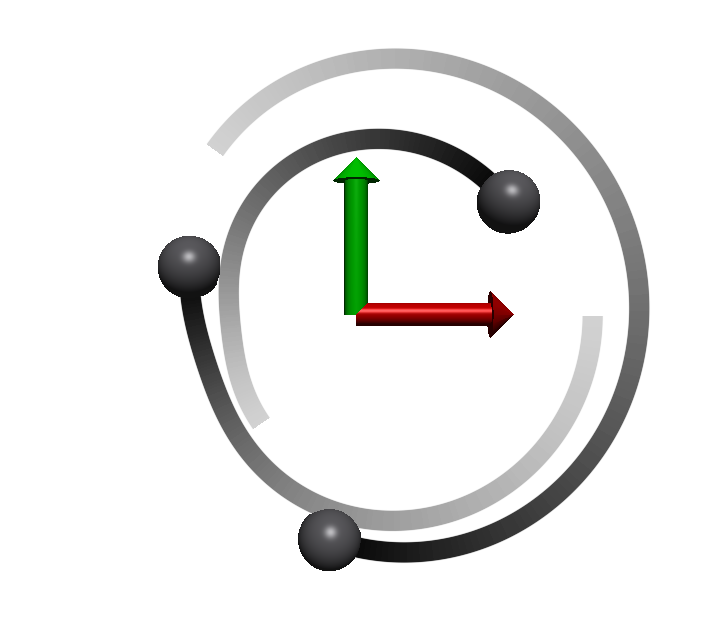}\hfill%
    \includegraphics[width=0.2\linewidth]{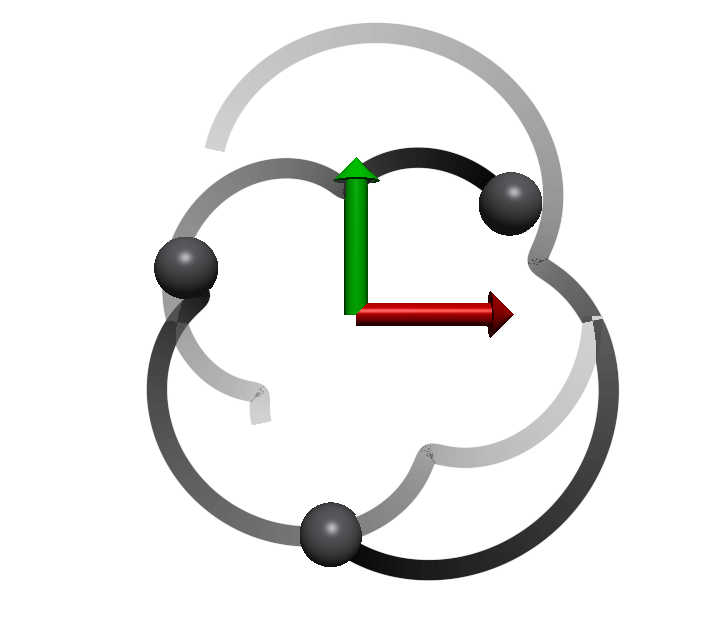}\hfill%
    \includegraphics[width=0.2\linewidth]{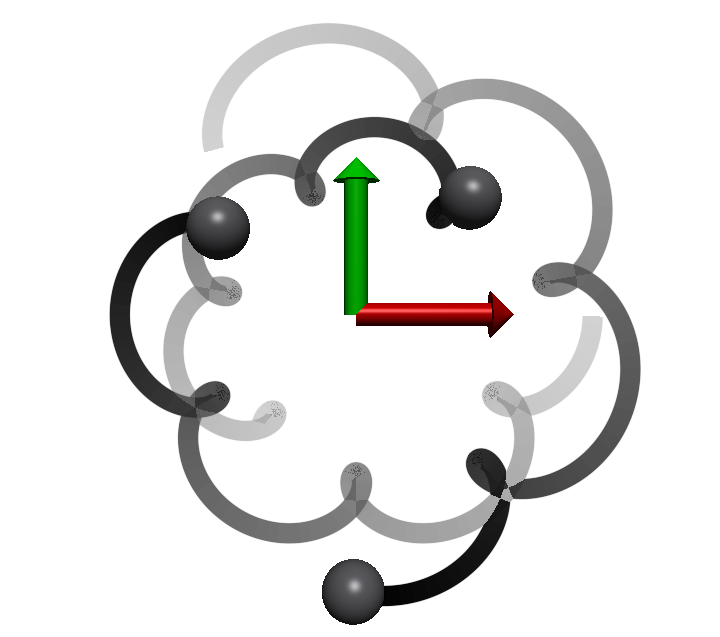}%
    \caption{Example of three trajectories rotating around a point on a circle at different speeds.
    In the first row, the trajectories and reference frame are shown over $\pi/4$.
    Removing the respective reference frame gives the result in the second row, rendered over $\pi/2$: the particles are moving in an ellipse around the origin, with their speed and direction depending on the choice of $q$.
    Removing the wrong reference frame (here taken from the first line set) leads to non-stationary behavior, as shown in the third row.
    }
    \label{fig:three-lines}
\end{figure}

We consider a simple data set consisting of  
the three 2D trajectories
\begin{eqnarray}
\vx_1(t) &=& \vo + \frac{3}{10} \cos(q\, t) \, \vr_1 +  \frac{1}{5} \sin(q\, t) \, \vr_2
\\ 
\nonumber
\vx_2(t) &=& \vo + \frac{4}{5} \left( \frac{3}{10} \cos\left(q\, t -  \frac{2}{3}\pi  \right) \, \vr_1 +  \frac{1}{5} \sin\left(q\, t -  \frac{2}{3}\pi  \right) \, \vr_2\right)
\\ 
\nonumber
\vx_3(t) &=& \vo + \frac{6}{5} \left( \frac{3}{10} \cos\left(q\, t +  \frac{2}{3}\pi  \right) \, \vr_1 +  \frac{1}{5} \sin\left(q\, t +  \frac{2}{3}\pi  \right) \, \vr_2\right)
\end{eqnarray} 
with 
\begin{equation}
\label{eq_optimalreferencesystem}
\vo = 
\begin{pmatrix}
\cos(t)\\ 
\sin(t)
\end{pmatrix} 
\;,\;
(\vr_1,\vr_2) = 
 \begin{pmatrix}
\cos(p \ t) & -\sin(p \ t)\\ 
\sin(p \ t) & \cos(p \ t)
\end{pmatrix}. 
\end{equation}
The trajectories are the result of a superposition of three rotational movements: a rotation around the origin with the angular speed 1, 
a rotation of the local reference system with the angular speed $p$, and a rotation of the particles in this local reference system with the angular speed $q$.

The objectivity of TRV ensures that we can separate the movement of the reference system from the movement of the particles in it.  In fact, applying our approach gives 
\begin{equation}
\mathbf{trv} = 
\begin{pmatrix}
0 & -\frac{13}{12} q\\ 
\frac{13}{12} q & 0
\end{pmatrix}, 
\end{equation}
as shown in the supplemental Maple sheet. Then, the optimal moving coordinate system is given by Eq.~\eqref{eq_optimalreferencesystem}
where $\vo$ is the origin and $\vr_1,\vr_2$ are the coordinate axes.
We illustrate the trajectories and the corresponding coordinate axes for the combinations 
$(p,q) = (6.75, -2.25), (3.25, 1.25), (1.25, 3.25),  (-2.25, 6.75)$ in Figure~\ref{fig:three-lines} in three different reference systems each. Note that 
the sum of the angular speed of the reference system and the particles therein is constant for all 4 instances: $p+q=4.5$. The upper row of Figure~\ref{fig:three-lines} shows the motion of the particles in a fixed global reference system as well as the motion of the optimal moving reference systems. From the particle motion in the fixed global system it is hard to infer the rotation behavior of the trajectories around each other. This changes when switching to the optimal local moving reference system (middle rows): here we can clearly observe clockwise rotation in the first column and a counterclockwise rotation of different angular speed in the remaining columns. For reference, the lower row shows the observation in the reference frame of the first column, showing a non-stationary particle behavior.

\newcommand{\trajcylfig}[1]{\begin{tikzpicture}%
	\node[anchor=south west,inner sep=0] (image) at (0,0) {\includegraphics[width=0.30\linewidth]{{#1}}};%
	\begin{scope}[node distance=-1.8mm and -1.2mm, x={(image.south east)},y={(image.north west)}]%
	\node[draw=none,overlay,text=red!80!black] at (0.93,0.15) {\scriptsize $x$};%
	\node[draw=none,overlay,text=green!80!black] at (0.06,0.05) {\scriptsize $y$};%
	\node[draw=none,overlay,text=blue!80!black] at (0.04,0.95) {\scriptsize $t$};%
	\end{scope}%
\end{tikzpicture}}

\begin{figure*}[t]
    \centering
    \hspace{1em}%
    \begin{minipage}{0.30\linewidth}\centering%
    Near-steady frame
    \end{minipage}\hfill%
    \begin{minipage}{0.30\linewidth}\centering%
    Original frame
    \end{minipage}\hfill%
    \begin{minipage}{0.30\linewidth}\centering%
    Fast-moving frame
    \end{minipage}\hspace{0.44cm}$~$\\%
    \raisebox{3em}{\rotatebox[origin=c]{90}{\small{input\vphantom{$\overline{\mbox{TRA}}$}}}}%
    \trajcylfig{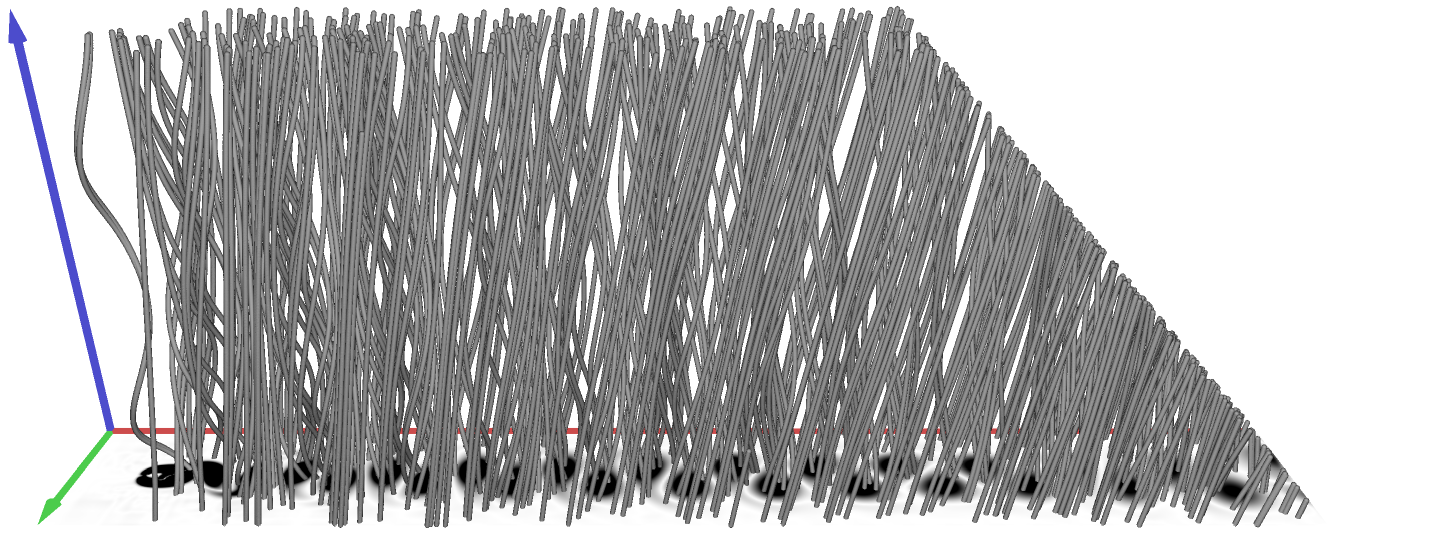}\hfill%
    \trajcylfig{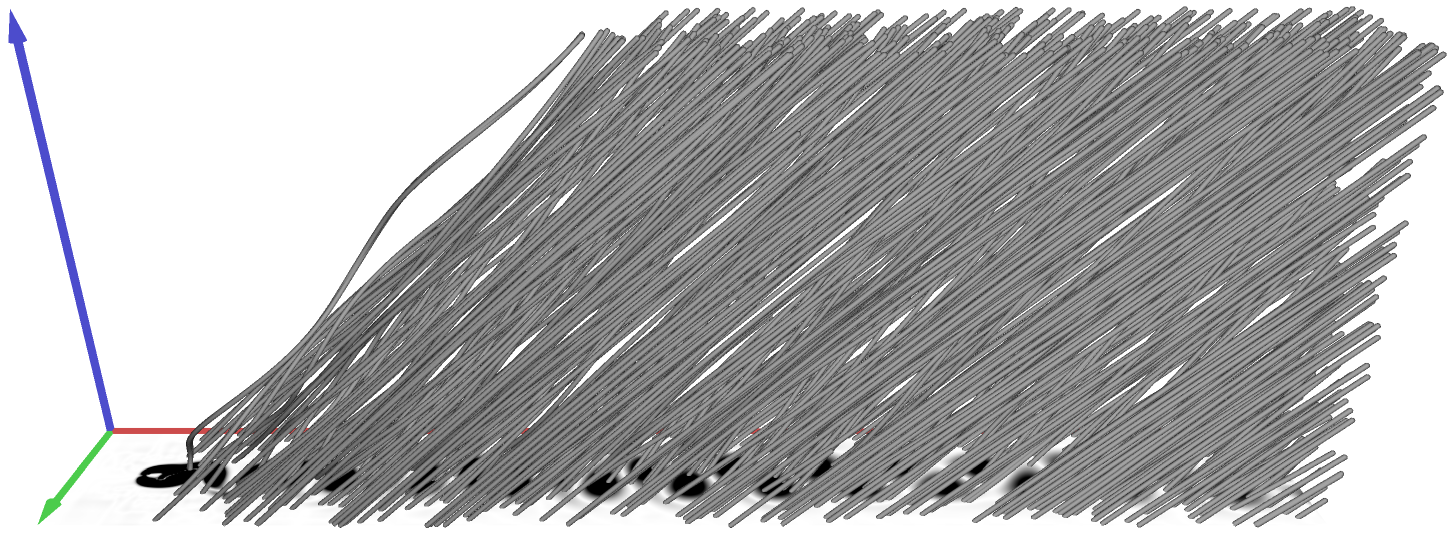}\hfill%
    \trajcylfig{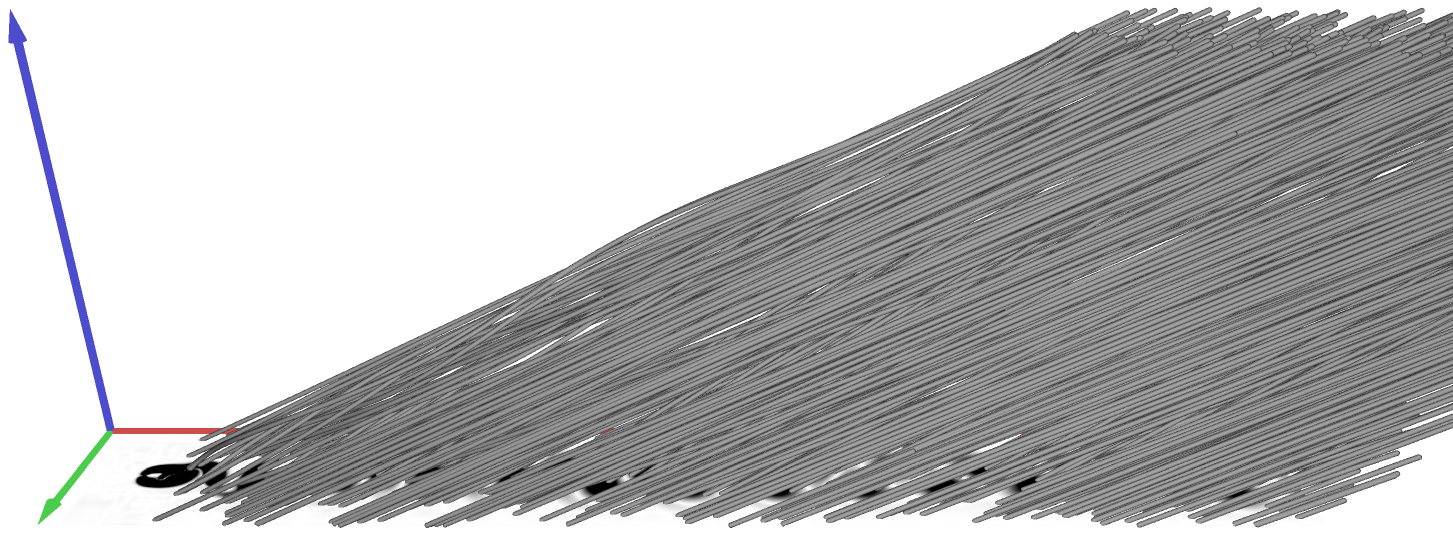}\hspace{0.44cm}$~$\\%
    \raisebox{3em}{\rotatebox[origin=c]{90}{\small{$\mbox{TRA}$\vphantom{p$\overline{\mbox{TRA}}$}}}}%
    \trajcylfig{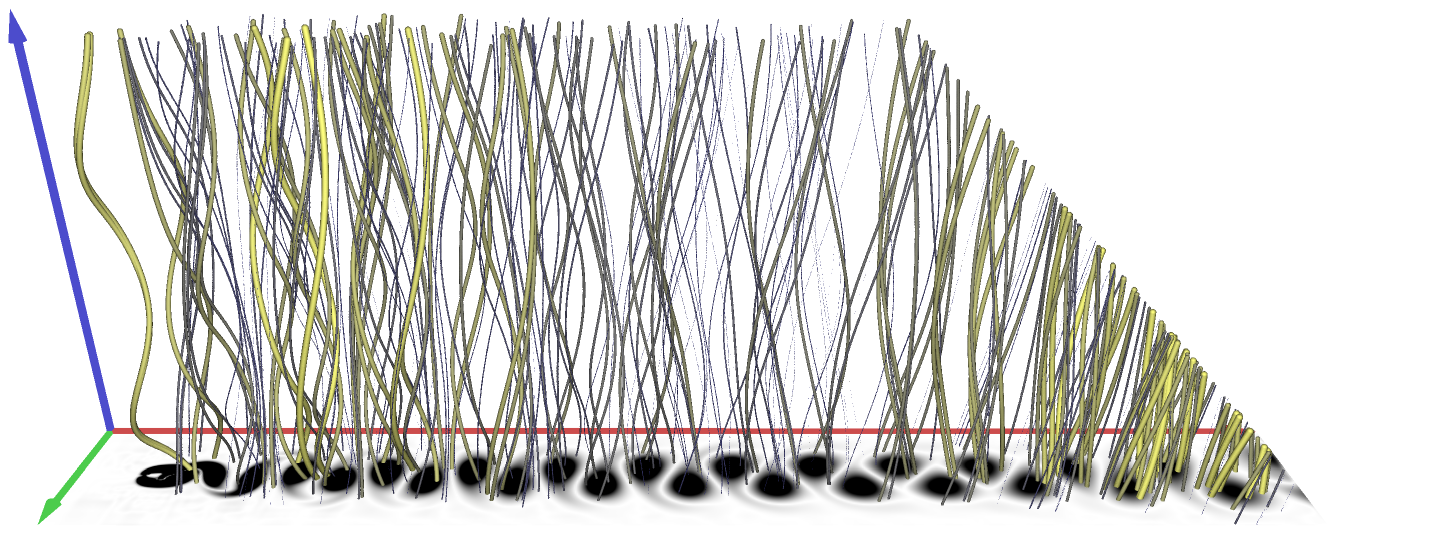}\hfill%
    \trajcylfig{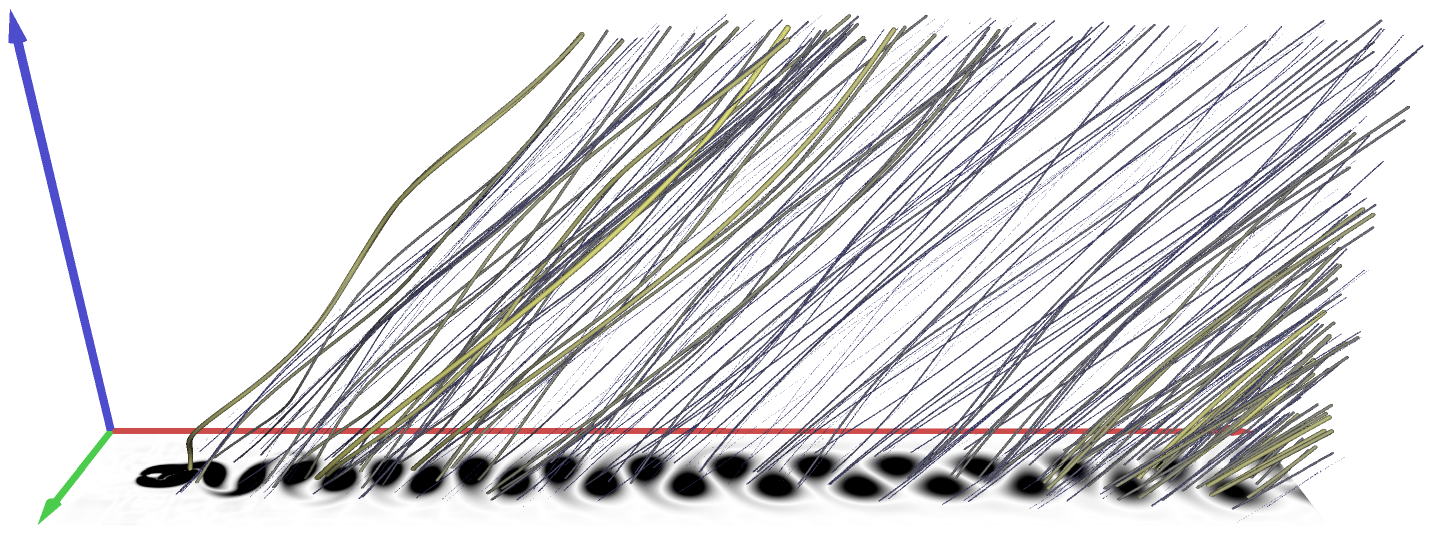}\hfill%
    \trajcylfig{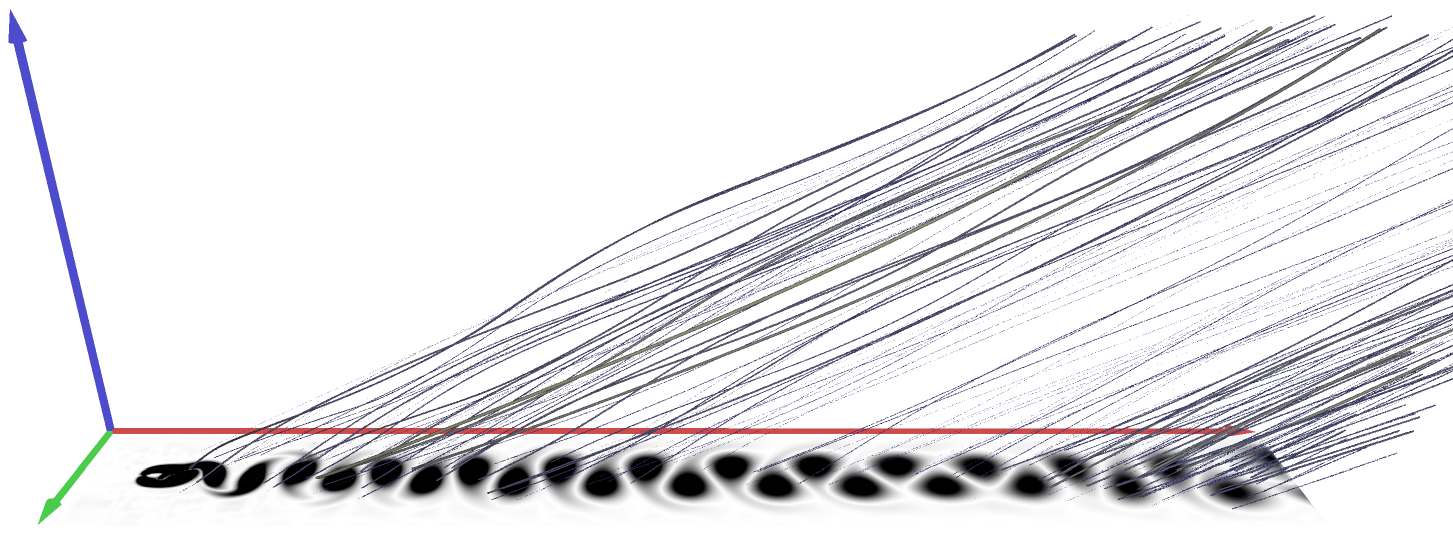}\hfill%
    \raisebox{2.5mm}{\begin{tikzpicture}%
	\node[anchor=south west,inner sep=0] (image) at (0,0) {\includegraphics[width=0.008\linewidth]{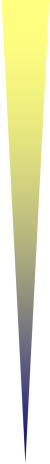}};%
	\begin{scope}[node distance=-0.8mm and -0.2mm, x={(image.south east)},y={(image.north west)}]%
	\node[draw=none,overlay,below= of image] {\scriptsize $0.04$};%
	\node[draw=none,overlay,above= of image] {\scriptsize $0.25$};%
	\end{scope}%
\end{tikzpicture}}\\%
    \raisebox{3em}{\rotatebox[origin=c]{90}{\small{$\overline{\mbox{TRA}}$\vphantom{p$\overline{\mbox{TRA}}$}}}}%
    \trajcylfig{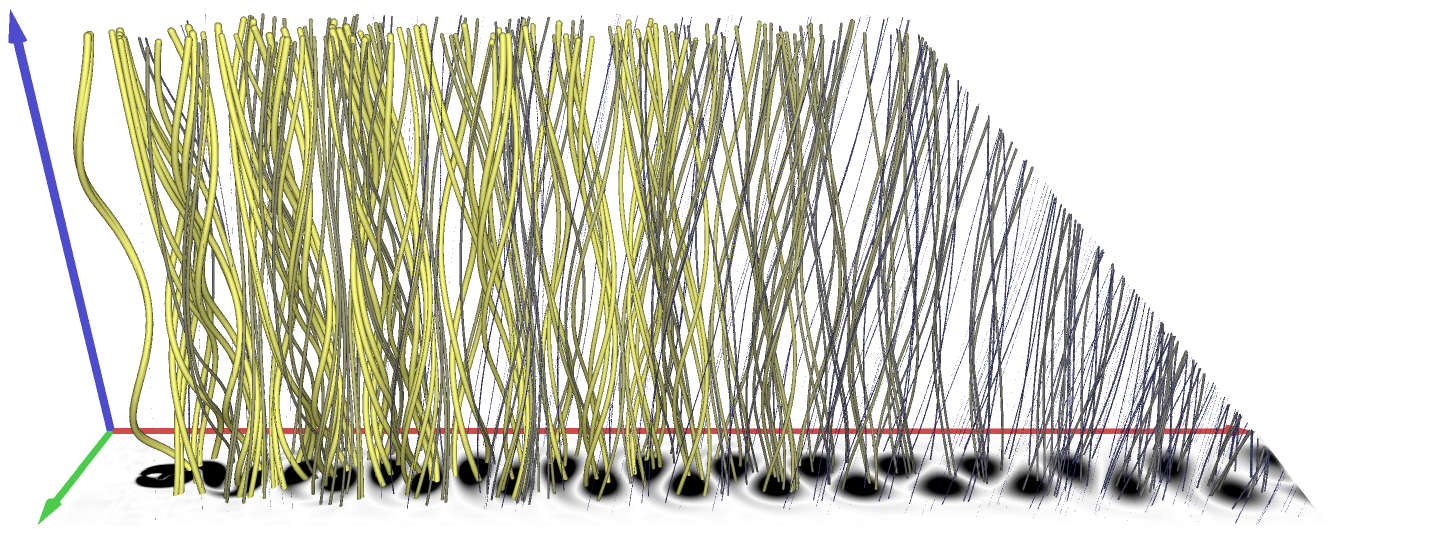}\hfill%
    \trajcylfig{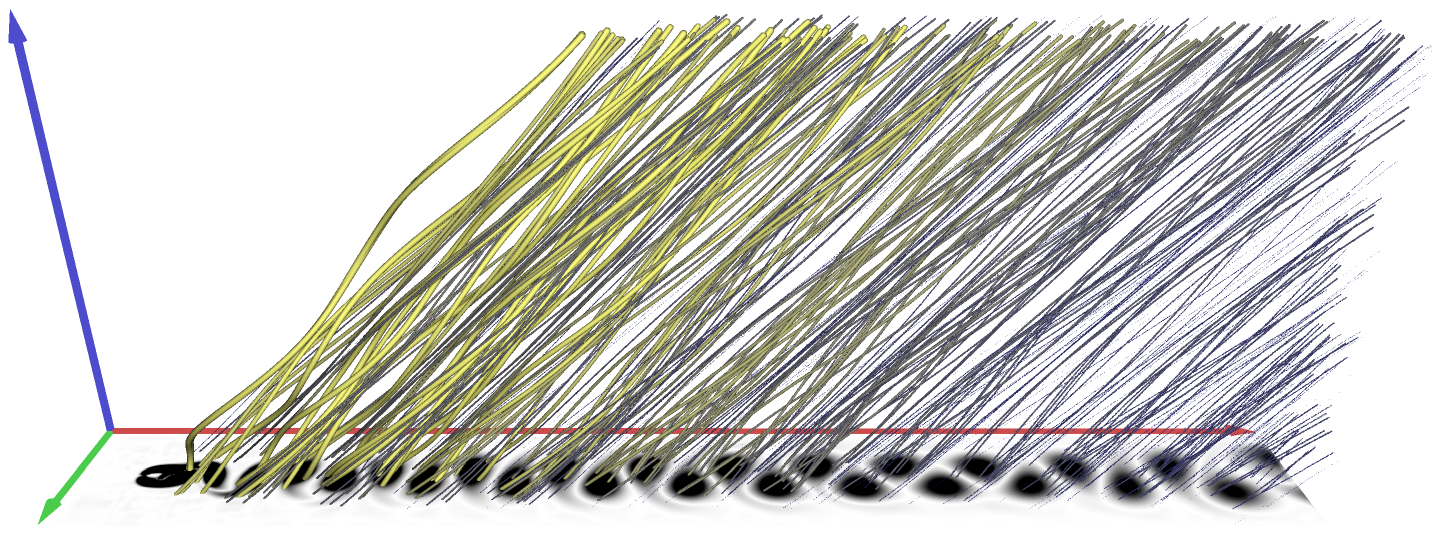}\hfill%
    \trajcylfig{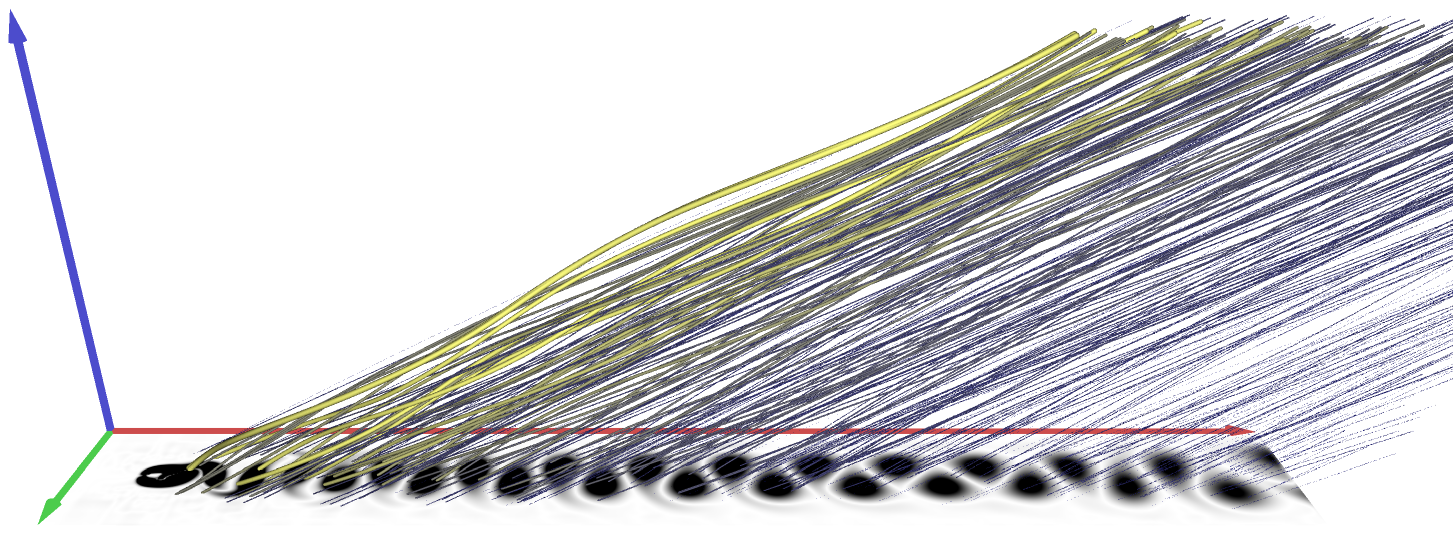}\hfill%
    \raisebox{2.5mm}{\begin{tikzpicture}%
	\node[anchor=south west,inner sep=0] (image) at (0,0) {\includegraphics[width=0.008\linewidth]{images/color-gradient.png}};%
	\begin{scope}[node distance=-0.8mm and -0.2mm, x={(image.south east)},y={(image.north west)}]%
	\node[draw=none,overlay,below= of image] {\scriptsize $0.04$};%
	\node[draw=none,overlay,above= of image] {\scriptsize $0.60$};%
	\end{scope}%
\end{tikzpicture}}\\%
    \raisebox{3em}{\rotatebox[origin=c]{90}{\small{$\mbox{TRV}$\vphantom{p$\overline{\mbox{TRA}}$}}}}%
    \trajcylfig{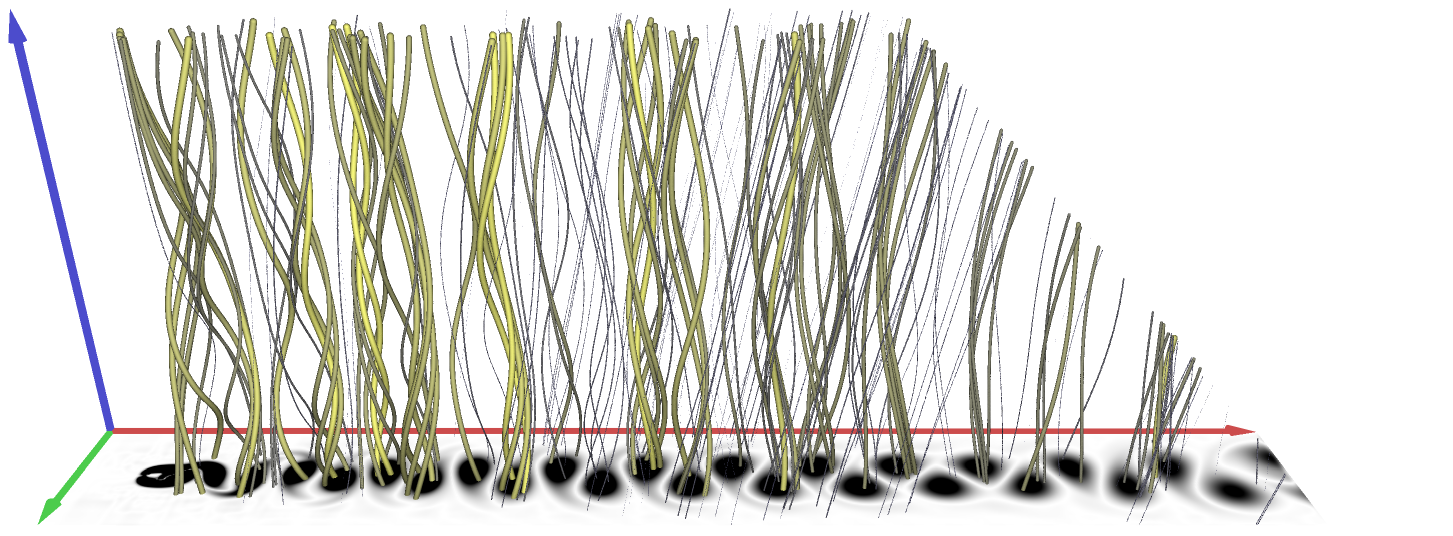}\hfill%
    \trajcylfig{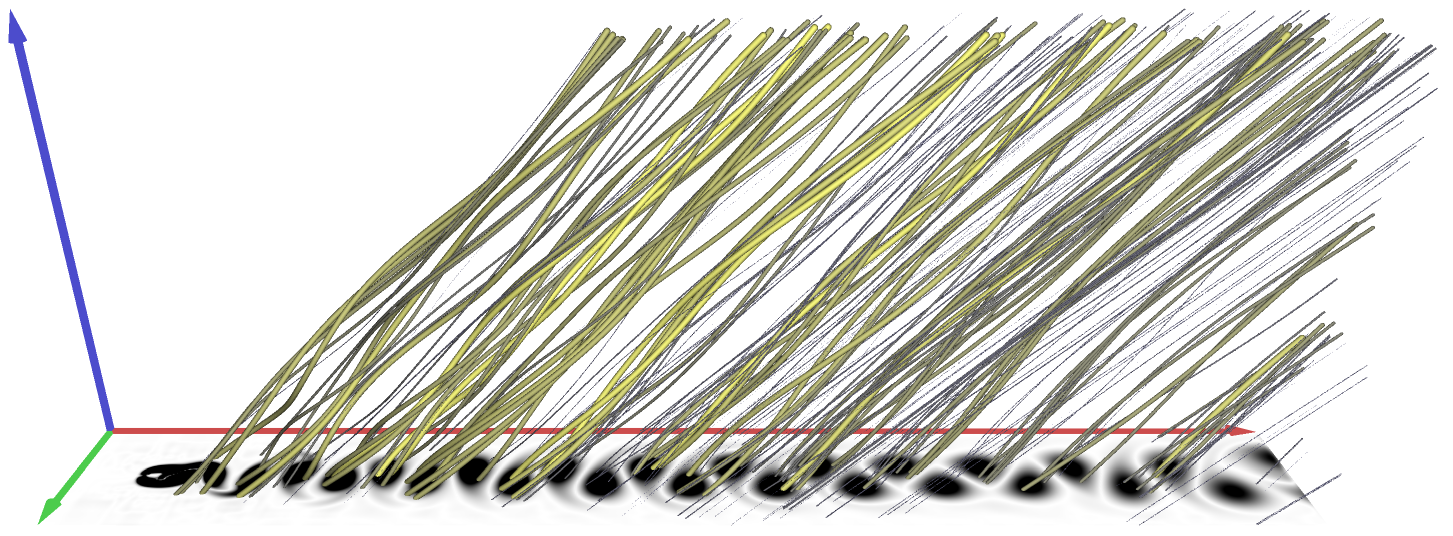}\hfill%
    \trajcylfig{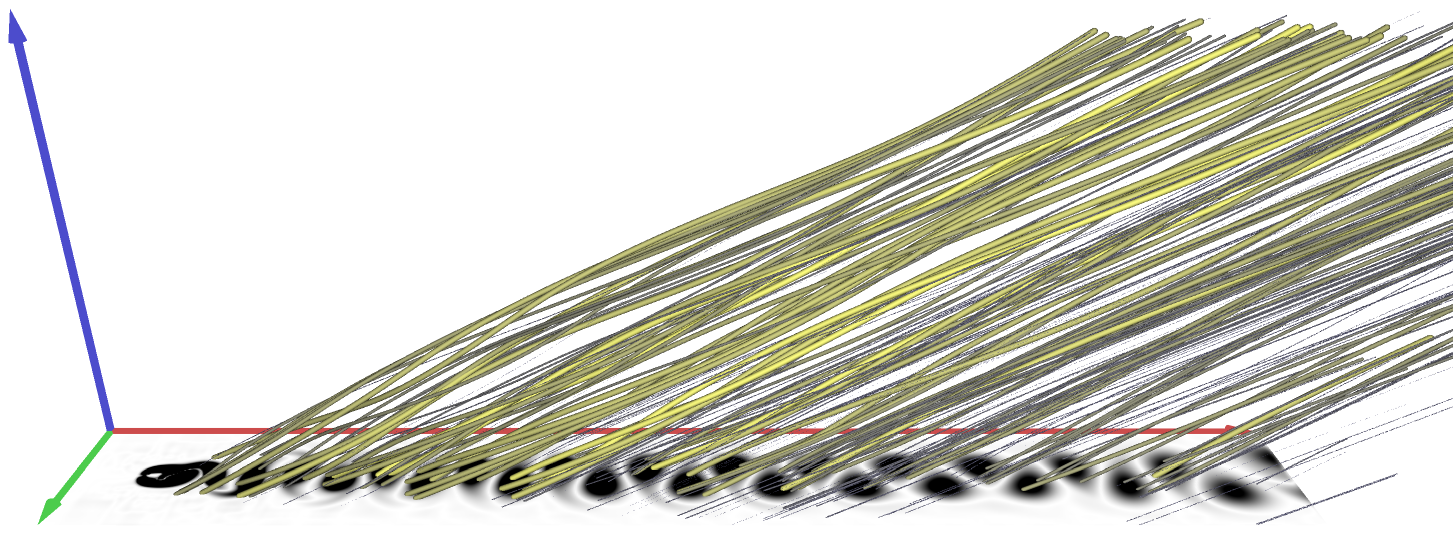}\hfill%
    \raisebox{2.5mm}{\begin{tikzpicture}%
	\node[anchor=south west,inner sep=0] (image) at (0,0) {\includegraphics[width=0.008\linewidth]{images/color-gradient.png}};%
	\begin{scope}[node distance=-0.8mm and -0.2mm, x={(image.south east)},y={(image.north west)}]%
	\node[draw=none,overlay,below= of image] {\scriptsize $0.01$};%
	\node[draw=none,overlay,above= of image] {\scriptsize $0.03$};%
	\end{scope}%
\end{tikzpicture}}\\%
    \raisebox{3em}{\rotatebox[origin=c]{90}{\small{$\overline{\mbox{TRV}}$\vphantom{p$\overline{\mbox{TRA}}$}}}}%
    \trajcylfig{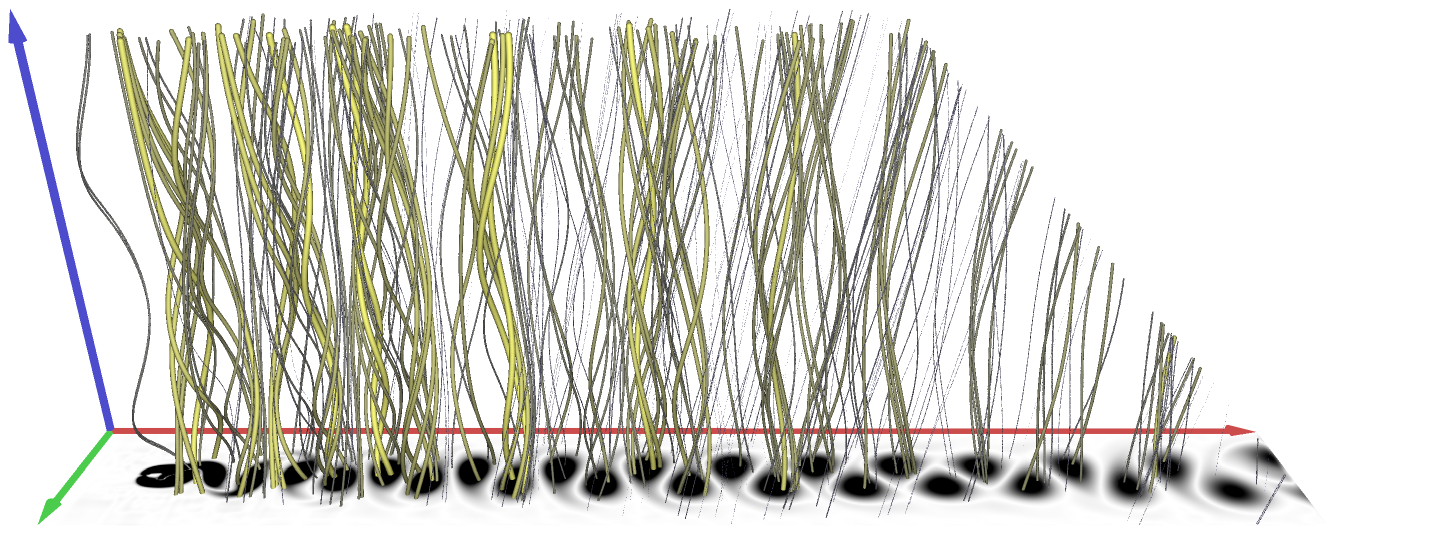}\hfill%
    \trajcylfig{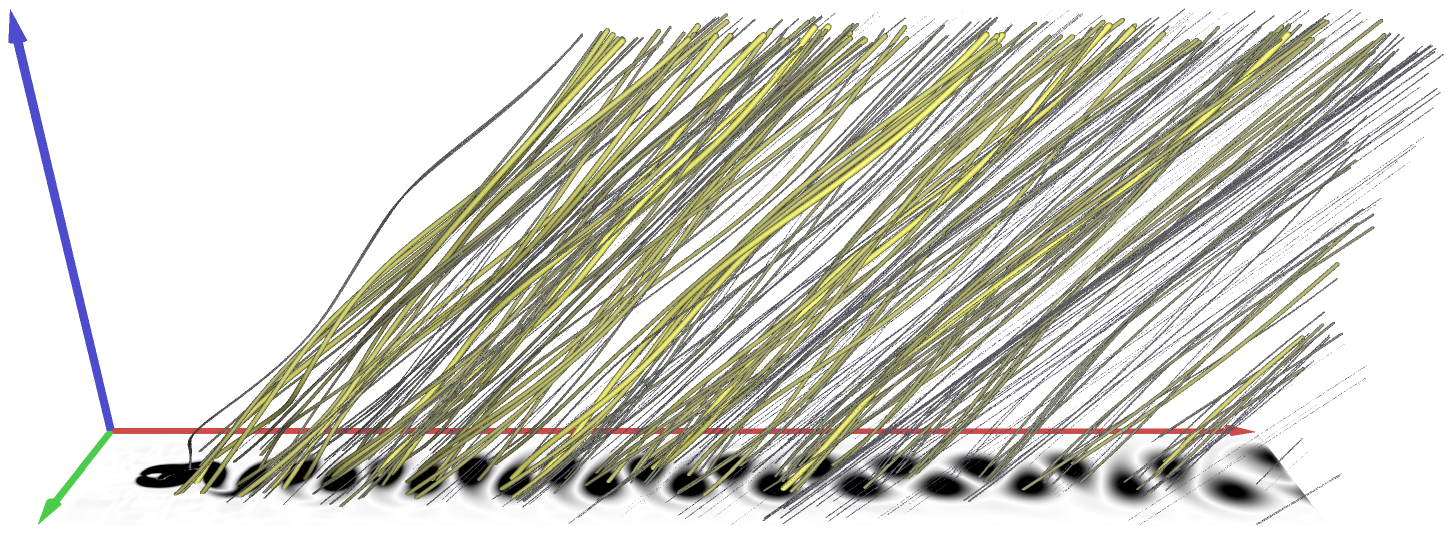}\hfill%
    \trajcylfig{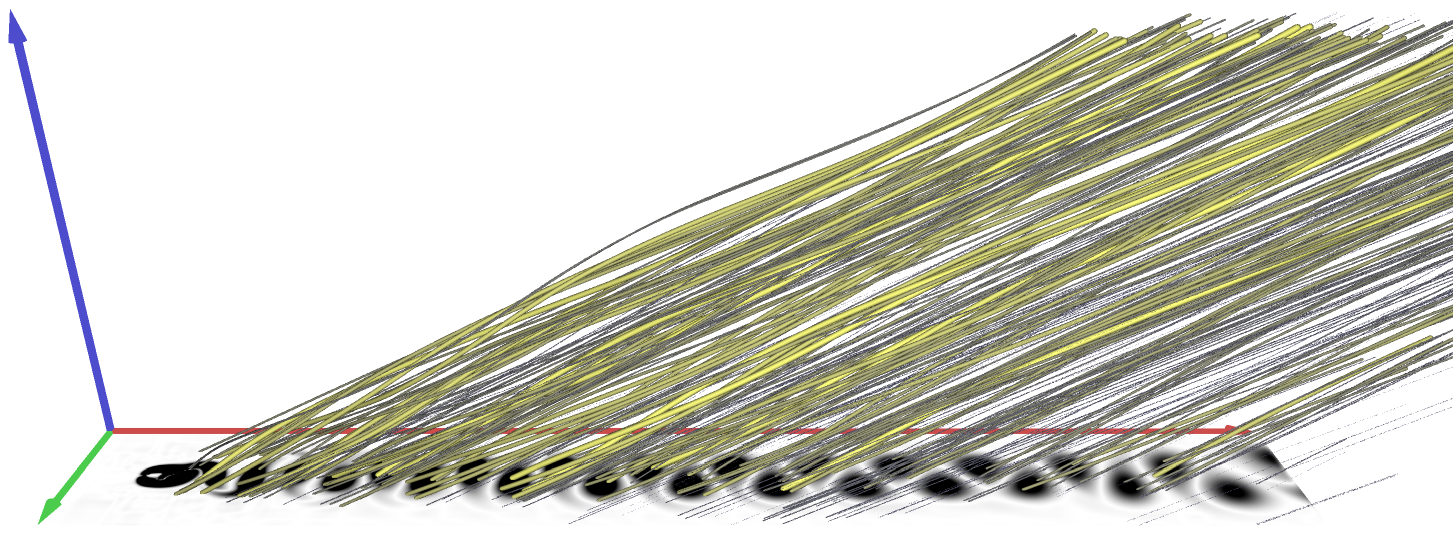}\hfill%
    \raisebox{2.5mm}{\begin{tikzpicture}%
	\node[anchor=south west,inner sep=0] (image) at (0,0) {\includegraphics[width=0.008\linewidth]{images/color-gradient.png}};%
	\begin{scope}[node distance=-0.8mm and -0.2mm, x={(image.south east)},y={(image.north west)}]%
	\node[draw=none,overlay,below= of image] {\scriptsize $0.01$};%
	\node[draw=none,overlay,above= of image] {\scriptsize $0.03$};%
	\end{scope}%
\end{tikzpicture}}%
    \vspace{-0.5em}
    \caption{Trajectory vortex measures calculated from 500 randomly placed pathlines in the \textsc{Cylinder} flow for three Galilean observers, here shown in 2D space-time. Left to right: observer moving approximately relative to vortices, the original observer, and observer moving faster. For reference, vorticity magnitude is visualized in the first time slice. Note that the value of $\mbox{TRA}$ and $\overline{\mbox{TRA}}$ changes for different observers, while $\mbox{TRV}$ and $\overline{\mbox{TRV}}$ give consistent results. Pathlines were transformed from the original via Eq.~\eqref{eq_define_movingsystem} with $\mQ(t)=\mI$ and $\vb(t)=(\pm 0.9\,t,0)^\Transp$ to the near-steady and fast-moving frame, respectively.}
    \label{fig:cyl-transform}
\end{figure*}

\subsection{Cylinder Flow}

We apply our approach to the numerically simulated \textsc{Cylinder} data set, which was simulated using Gerris flow solver~\citep{gerrisflowsolver} and was published by~\cite{Guenther:2017:Siggraph}. Such a data set is not the main target of our approach because here the underlying velocity field is available. We use it as test data set since we can compute an arbitrary number of trajectories and can compare measures based on them with "ground truth" measures form the underlying velocity field.

\begin{figure}[t]
    \centering
    \includegraphics[width=0.47\linewidth]{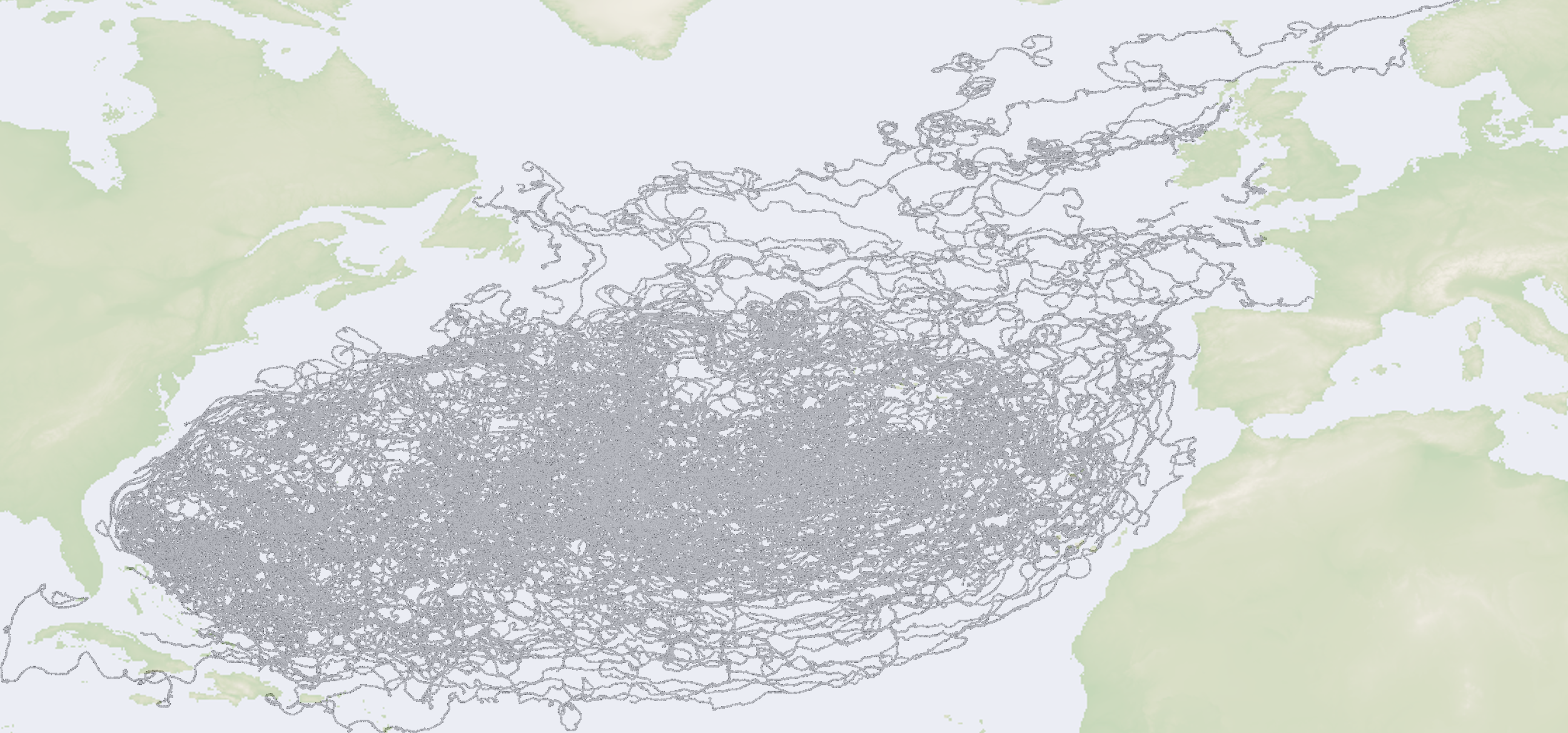}\hfill%
    \begin{tikzpicture}%
	\node[anchor=south west,inner sep=0] (image) at (0,0) {\includegraphics[width=0.47\linewidth]{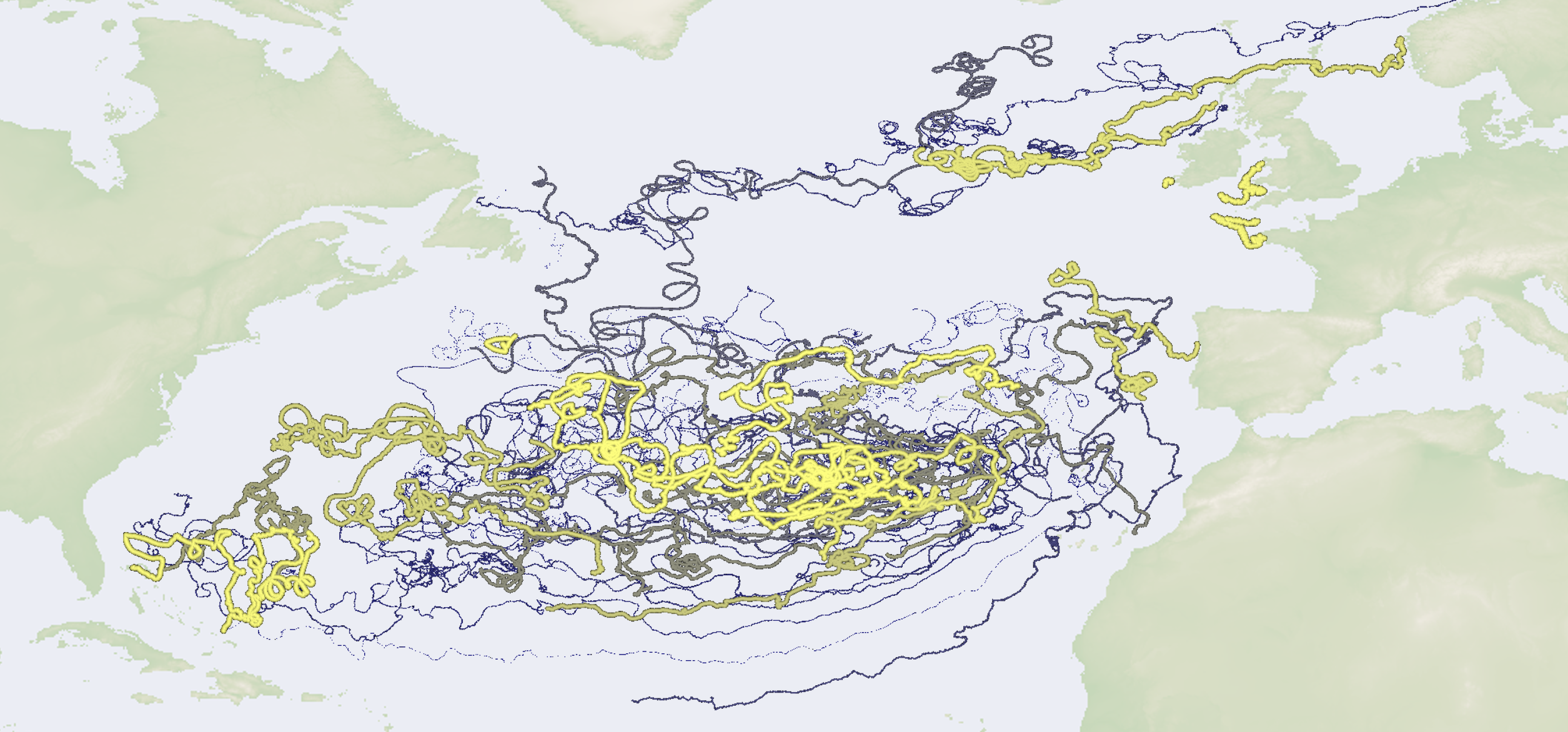}};%
	\begin{scope}[node distance=-1.8mm and -1.2mm, x={(image.south east)},y={(image.north west)}]%
	\node[draw=none] (label) at (1.03,0.5) {\includegraphics[width=0.015\linewidth]{images/color-gradient.png}}; %
	\node[draw=none,overlay,below= of label] {\scriptsize $0.6$};%
	\node[draw=none,overlay,above= of label] {\scriptsize $1.0$};%
	\end{scope}%
\end{tikzpicture}%
    \caption{
    On the left, input trajectories of drifting buoys in the Atlantic are shown. On the right, our new trajectory vortex measure $\overline{\mbox{TRV}}$ is visualized, revealing rotational particle behavior. The $\overline{\mbox{TRV}}$ value is mapped to color and line radius. Lines with too small $\overline{\mbox{TRV}}$ are removed to avoid visual clutter.}
    \label{fig:drifter-compare}
\end{figure}

\begin{figure*}[t]
    \centering
    \hfill\includegraphics[width=0.45\linewidth]{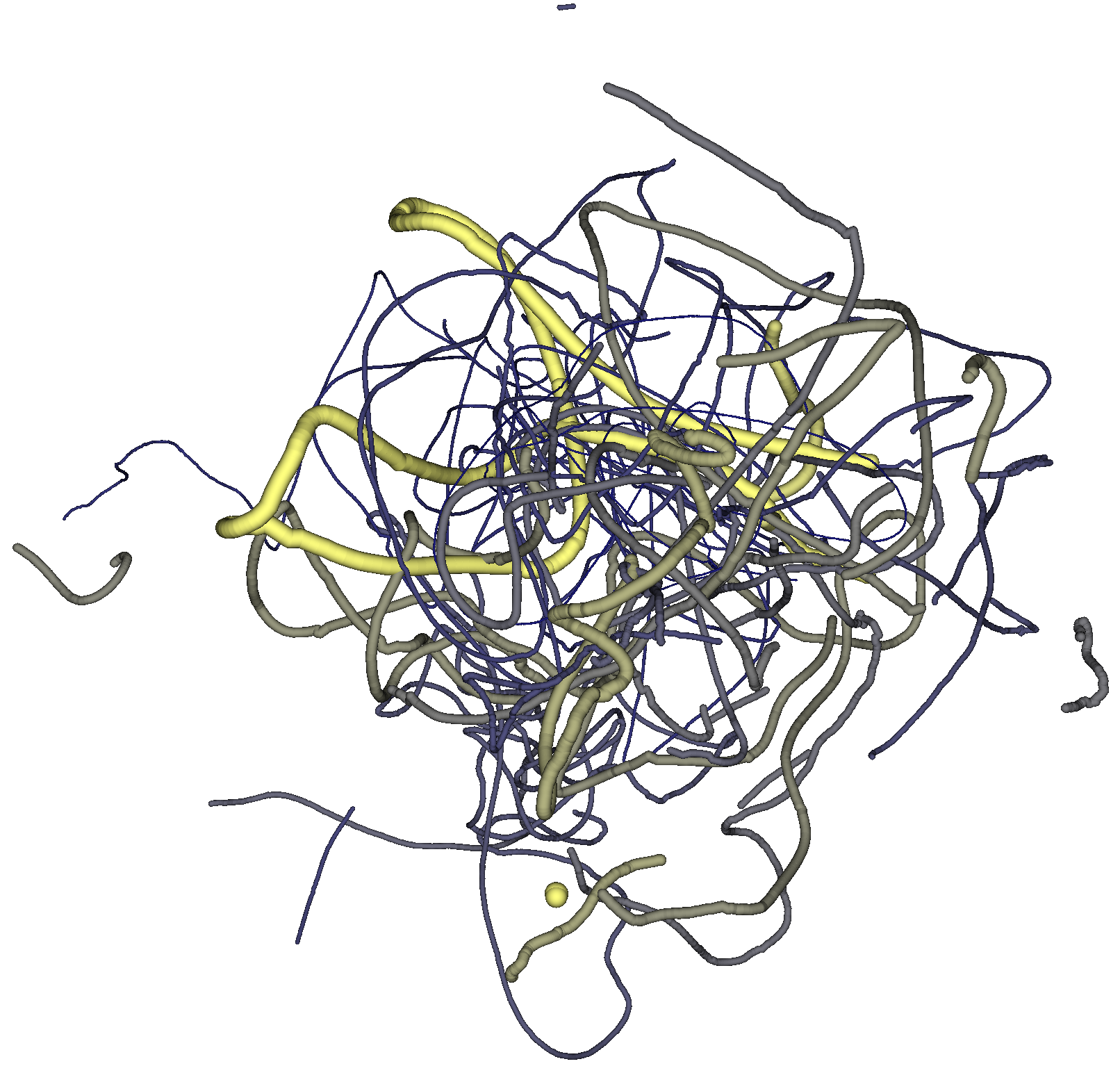}\hfill%
    \begin{tikzpicture}%
	\node[anchor=south west,inner sep=0] (image) at (0,0) {\includegraphics[width=0.45\linewidth]{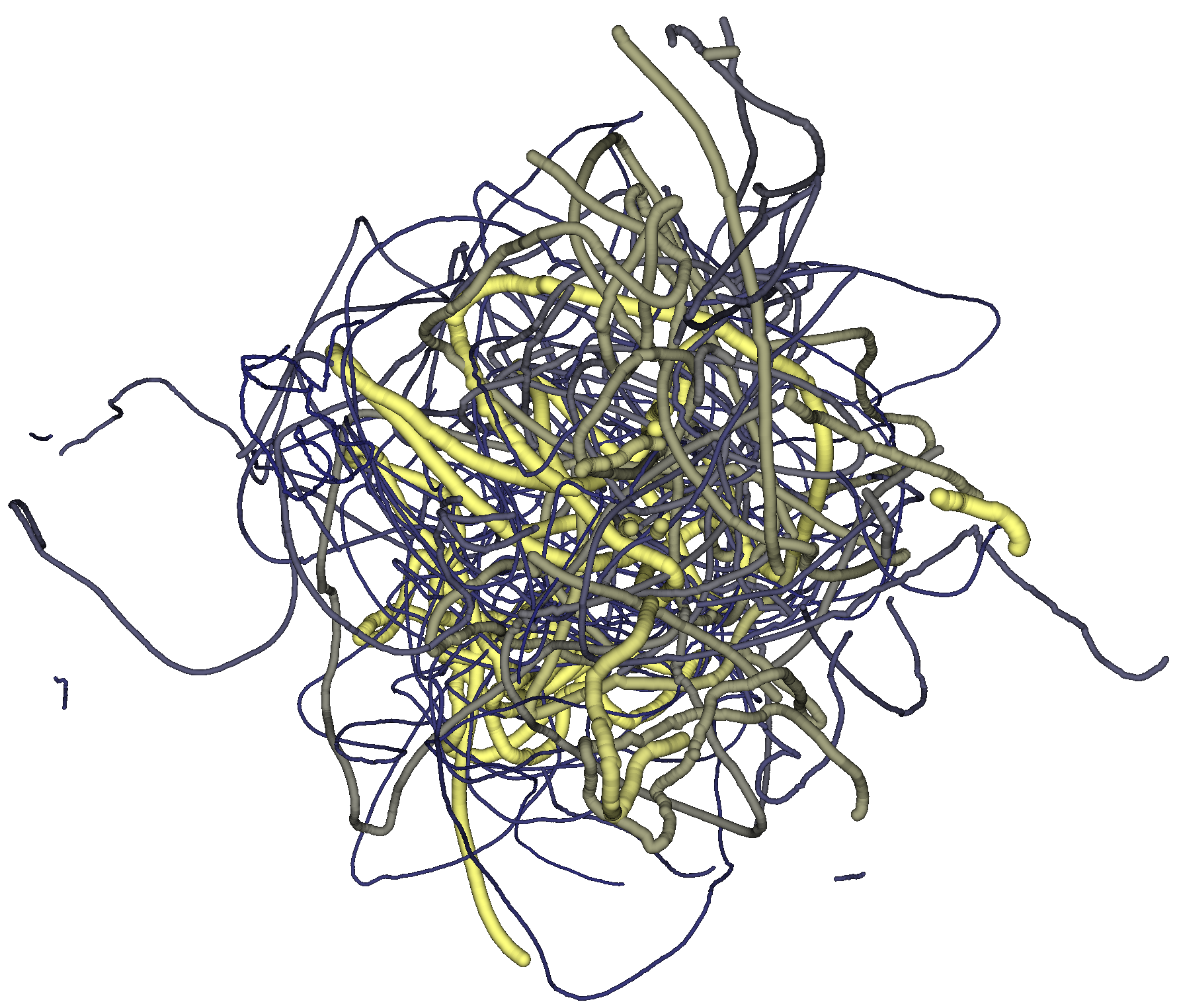}};%
	\begin{scope}[node distance=-1.8mm and -1.2mm, x={(image.south east)},y={(image.north west)}]%
	\node[draw=none] (label) at (1.03,0.5) {\includegraphics[width=0.015\linewidth]{images/color-gradient.png}}; %
	\node[draw=none,overlay,below= of label] {\scriptsize $0.0$};%
	\node[draw=none,overlay,above= of label] {\scriptsize $1.4$};%
	\end{scope}%
\end{tikzpicture}\hfill\hfill%
    \caption{$\overline{\mbox{TRV}}$ calculated for two different seeding times in the measured \textsc{Midge} data set. For this 3D data set, four neighboring lines are considered. The total number of input trajectories in the data sets are 38 (left) and 56 (right).}
    \label{fig:mosquito}
\end{figure*}


Figure \ref{fig:cyl-transform} shows trajectory measures for rotating behavior from 500 randomly seeded trajectories (pathlines) observed in three different reference systems: a system moving with the approximate speed of the vortices  (left column), the original references system (middle column), and a system moving faster (right column). For reference, we color code the vorticity magnitude of the underlying velocity field in the first time slice, giving a reliable indicator where to expect rotation behavior. To compute $\mbox{TRV}$ and $\overline{\mbox{TRV}}$, for a trajectory, the 4 nearest neighboring curves were selected from the sampled set of trajectories based on the squared distance:
\begin{align}
    d(\xx_1(t),\xx_2(t))=\int_{t_0}^{t_N}|\xx_1(\tau)-\xx_2(\tau)|^2 \;\mathrm{d}\tau.
    \label{eq:distance-metric}
\end{align} 
Since the trajectories were in temporal correspondence, more general distance metrics, such as the reduced mean closest point distance~\citep{Oeltze2014blood}, were not necessary.
Figure \ref{fig:cyl-transform} illustrates again that $\mbox{TRA}$ and $\overline{\mbox{TRA}}$ are not objective: corresponding trajectories for different observations frames (columns in Figure \ref{fig:cyl-transform}) have different colors. For $\mbox{TRV}$ and $\overline{\mbox{TRV}}$, we observe the same colors for different frames, confirming objectivity. We also note that $\mbox{TRV}$ and $\overline{\mbox{TRV}}$ tend to have high values in regions of high vorticity magnitude, confirming the detection of rotation trajectory behavior.

\subsection{Ocean Drifter Trajectories}

One instrument for measuring oceanic flow are drifting buoys, which get released into the ocean and are tracked by satellites.
The result is a time series per buoy encompassing their position, speed, and potentially other measurements from equipment attached to the drifters.
Several hundred such drifters are currently deployed by the National Oceanic and Atmospheric Administration (NOAA) of the USA, with tracking data freely available~\citep{GDP:buoys}.
We applied our $\overline{\mbox{TRV}}$ measure with \eqref{eq:distance-metric} and 2 nearest neighbors per curve on a subset of 203 drifters in the North Atlantic ocean, tracked over two years, to identify vortical behavior.
The results are shown in Figure~\ref{fig:drifter-compare}:
strong rotational behavior can be seen in the center of the North Atlantic Gyre, a region where water gets trapped by the surrounding currents.
Other regions highlighted by $\overline{\mbox{TRV}}$ include the beginning of the Gulf stream west of Florida, where eddy vortices are known to shear off, and areas close to the European coastline.

\subsection{Midge Trajectories}

We analyze trajectories of tracked swarms of 
{\em Chironomus riparius}. The data set is described and provided by \cite{sinhuber2019three}.
{\em Chironomus riparius} are
a midge species that consistently and predictably forms mating swarms over visual cues \citep{Downe1973}. To create the data set, the midges were bred in a laboratory environment, including a constant temperature and humidity and day/night sequences by artificial illumination.  The observed swarms describe male midges, mostly observed in artificial dusk. The tracking was done by an optical 3-camera system.
Swarms of {\em Chironomus riparius} are known to nucleate over visual features on the ground, such as tree stumps or stream banks \citep{Downe1973}. In the experiment, this was simulated by adding "swarm markers" to the setup.
We apply $\overline{\mbox{TRV}}$ to further analyze the movement around visual features. In particular, we analyze if a common objective rotation  behavior can be observed. Figure \ref{fig:mosquito} shows the trajectories for two different seeding times. While the pure shape of the trajectories does not reveal any patters, we found a few trajectories with high $\overline{\mbox{TRV}}$ values, mostly in the inner parts of the data set. Our approach can confirm a swirling behavior of a few trajectories, while for the majority of the particles, no objective rotation behavior is detected.

\setlength{\tabcolsep}{2mm}
\begin{table}[b]
    \centering
    \begin{tabular}{c|r|r|r|r|r}
        Data set & dist. (ms) & $\overline{\mbox{TRV}}$ (ms) & lines & vertices & fit \\\hline
    \textsc{Drifter} & 26.41 & 161.98 & 203 & 125,472 & 4 \\
    \textsc{Midge} & 0.17 & 8.81 & 38 & 1,558 & 3 \\
    \textsc{Midge} & 0.36 & 15.19 & 56 & 2,728 & 3 \\
    \textsc{Cylinder} & 25.93 & 162.54 & 500 & 115,241 & 5 \\
    \textsc{Cylinder} & 25.99 & 184.01 & 500 & 112,014 & 10 \\
    \textsc{Cylinder} & 27.49 & 229.90 & 500 & 109,815 & 15 \\
    \textsc{Cylinder} & 354.35 & 482.14 & 2,000 & 471,906 & 5 \\
    \textsc{Cylinder} & 357.94 & 571.12 & 2,000 & 462,457 & 10 \\
    \textsc{Cylinder} & 360.12 & 758.48 & 2,000 & 455,018 & 15 \\
    \textsc{Cylinder} & 1,945.26 & 1,204.98 & 5,000 & 1,199,686 & 5 \\
    \textsc{Cylinder} & 1,930.10 & 1,440.55 & 5,000 & 1,183,928 & 10 \\
    \textsc{Cylinder} & 1,933.80 & 1,948.60 & 5,000 & 1,171,023 & 15
    \end{tabular}
    \caption{Runtime measurements for the computation of the full distance matrix (in millisec.), the $\overline{\mbox{TRV}}$ computation (in millisec.), the number of trajectories in the set, the total number of vertices in the set, and the number of neighboring lines used for the fit.}
    \label{tab:performance}
\end{table}

\section{Discussion}

\newcommand{\trajcylcmpfig}[2]{\begin{tikzpicture}%
	\node[anchor=south west,inner sep=0] (image) at (0,0) {\includegraphics[width=0.31\linewidth]{{#1}}};%
	\begin{scope}[node distance=-1.8mm and -1.2mm, x={(image.south east)},y={(image.north west)}]%
	\node[draw=none,overlay,text=red!80!black] at (0.89,0.14) {\scriptsize $x$};%
	\node[draw=none,overlay,text=green!80!black] at (0.02,0.76) {\scriptsize $y$};%
	\node[draw=none,overlay,text=blue!80!black] at (0.04,0.96) {\scriptsize $t$};%
	\draw[draw=black] (0.2067,0.7737) rectangle (0.4519,0.9710);
	\node[anchor=south west,inner sep=0] (image) at (0,0) {\includegraphics[width=0.14\linewidth]{{#2}}};%
	\begin{scope}[node distance=-0.8mm and -0.2mm, x={(image.south east)},y={(image.north west)}]%
	\draw[draw=black] (0,0) rectangle (1,1);
	\end{scope}%
	\end{scope}%
\end{tikzpicture}}

\begin{figure*}[t]
    \centering
    \hspace{1em}%
    \begin{minipage}{0.31\linewidth}\centering%
    fit 5 nearest lines
    \end{minipage}\hfill%
    \begin{minipage}{0.31\linewidth}\centering%
    fit 10 nearest lines
    \end{minipage}\hfill%
    \begin{minipage}{0.31\linewidth}\centering%
    fit 15 nearest lines
    \end{minipage}\hspace{0.34cm}$~$\\%
    \raisebox{7em}{\rotatebox[origin=c]{90}{\small{$500$ trajectories}}}~%
    \trajcylcmpfig{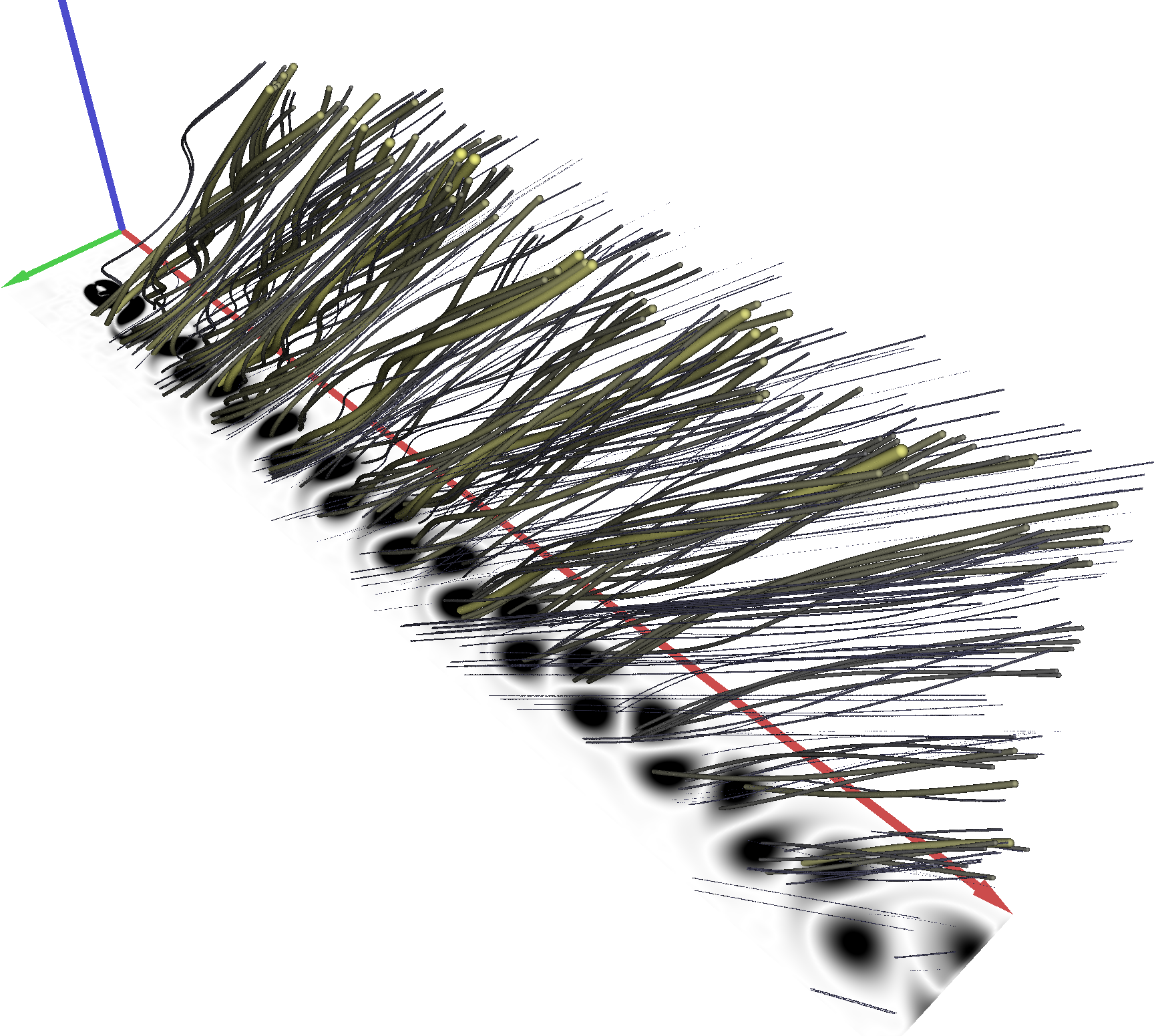}{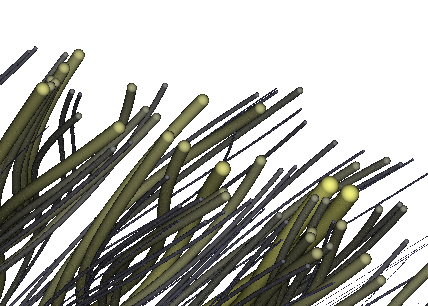}\hfill%
    \trajcylcmpfig{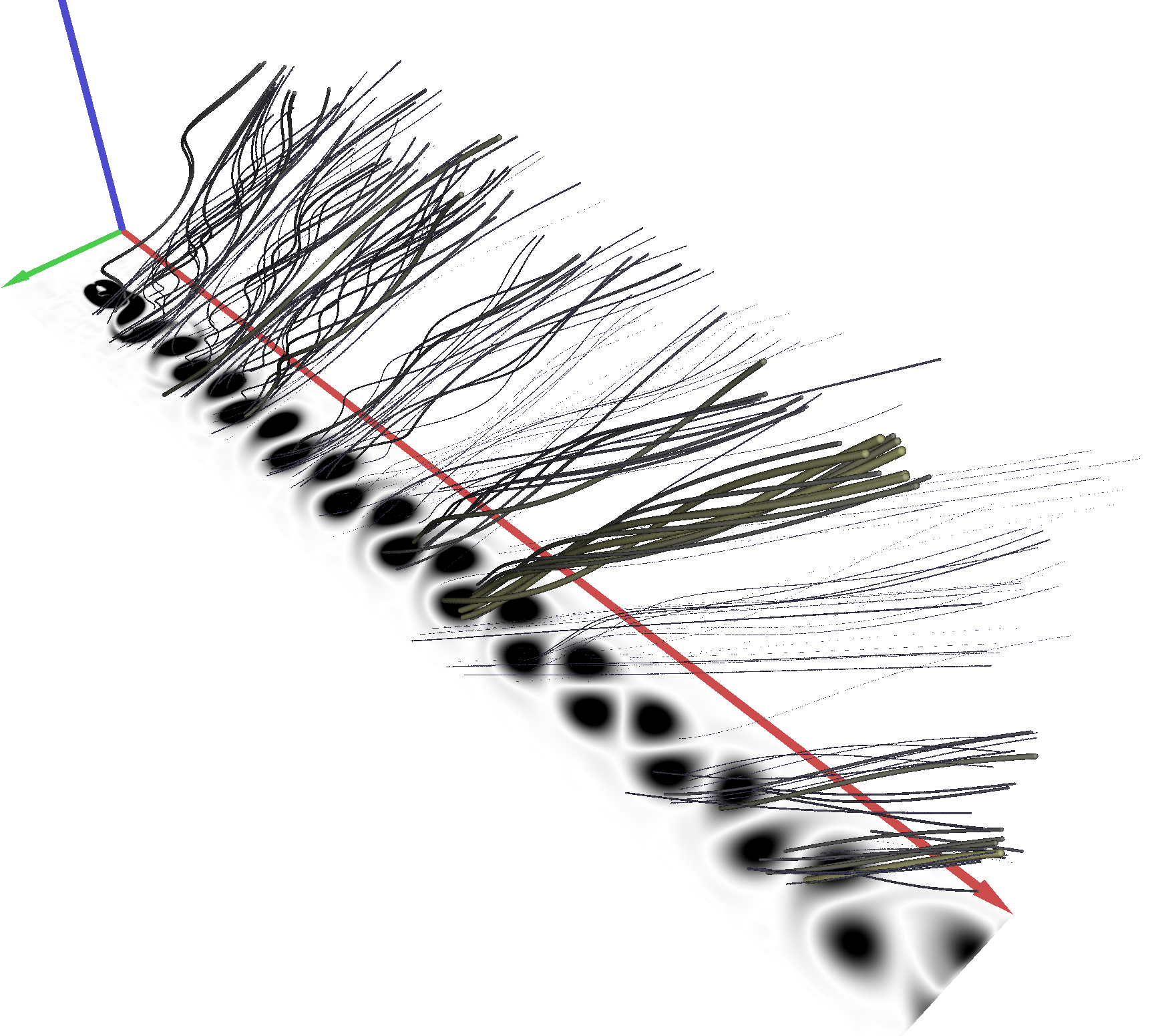}{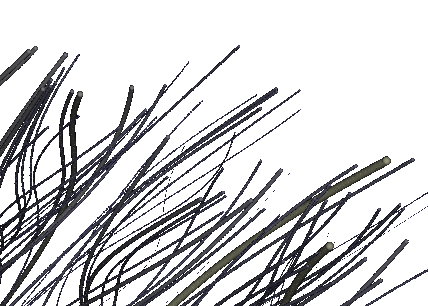}\hfill%
    \trajcylcmpfig{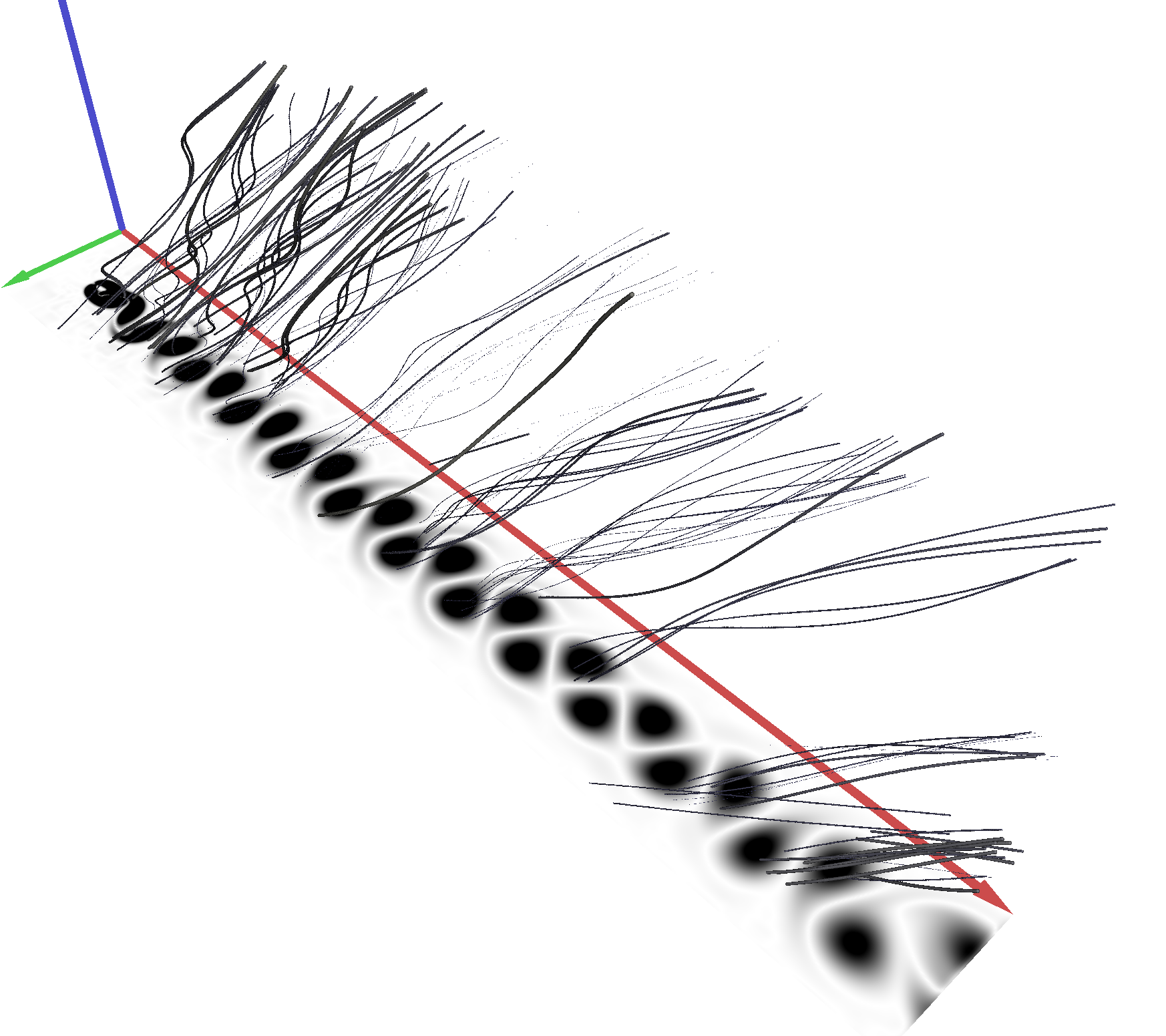}{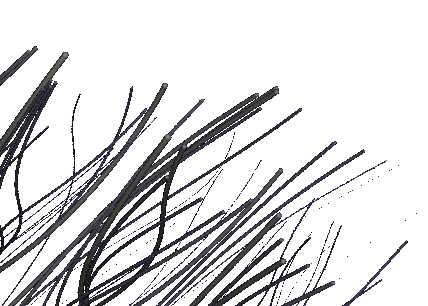}\hfill%
    \raisebox{12.5mm}{\begin{tikzpicture}%
	\node[anchor=south west,inner sep=0] (image) at (0,0) {\includegraphics[width=0.013\linewidth]{images/color-gradient.png}};%
	\begin{scope}[node distance=-0.8mm and -0.2mm, x={(image.south east)},y={(image.north west)}]%
	\node[draw=none,overlay,below= of image] {\scriptsize $0.01$};%
	\node[draw=none,overlay,above= of image] {\scriptsize $0.035$};%
	\end{scope}%
\end{tikzpicture}}\\%
    \raisebox{7em}{\rotatebox[origin=c]{90}{\small{$2,000$ trajectories}}}~%
    \trajcylcmpfig{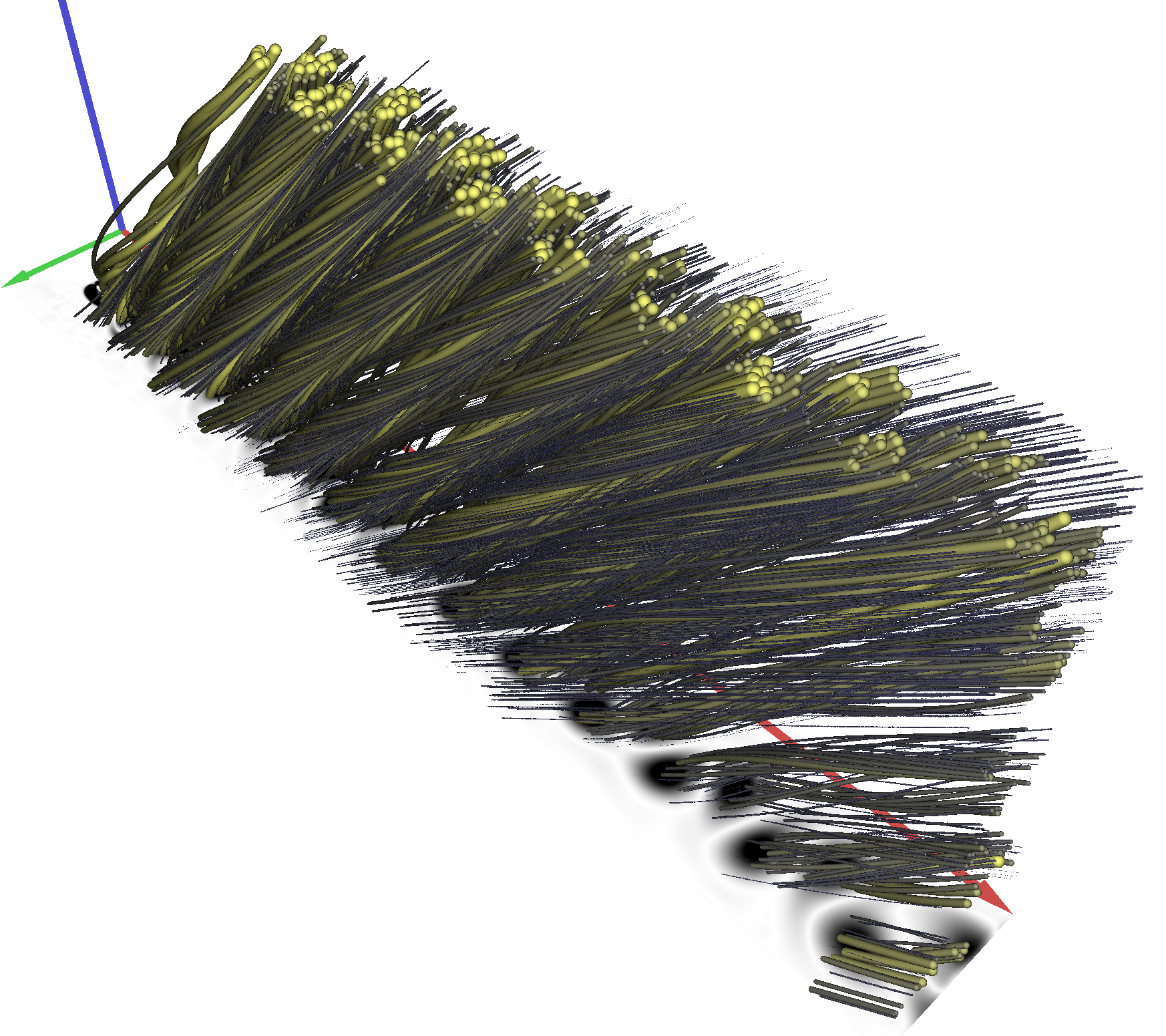}{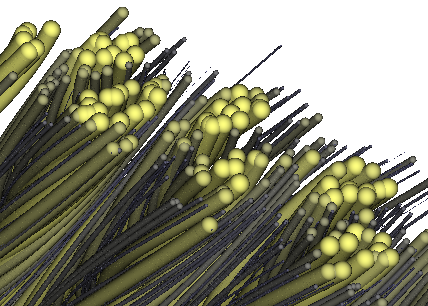}\hfill%
    \trajcylcmpfig{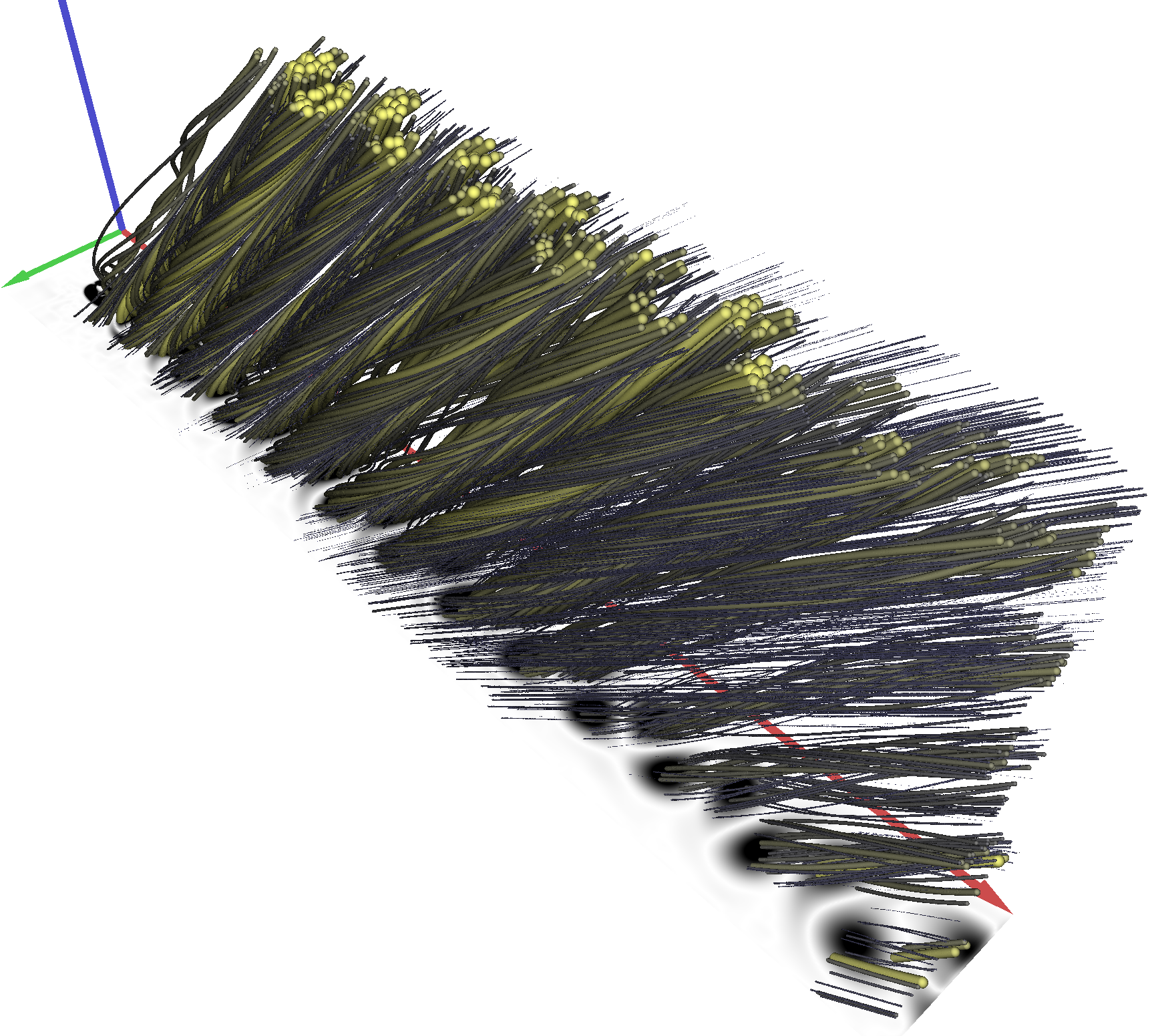}{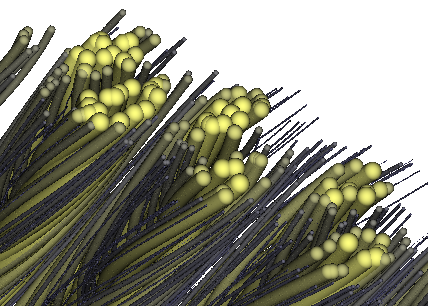}\hfill%
    \trajcylcmpfig{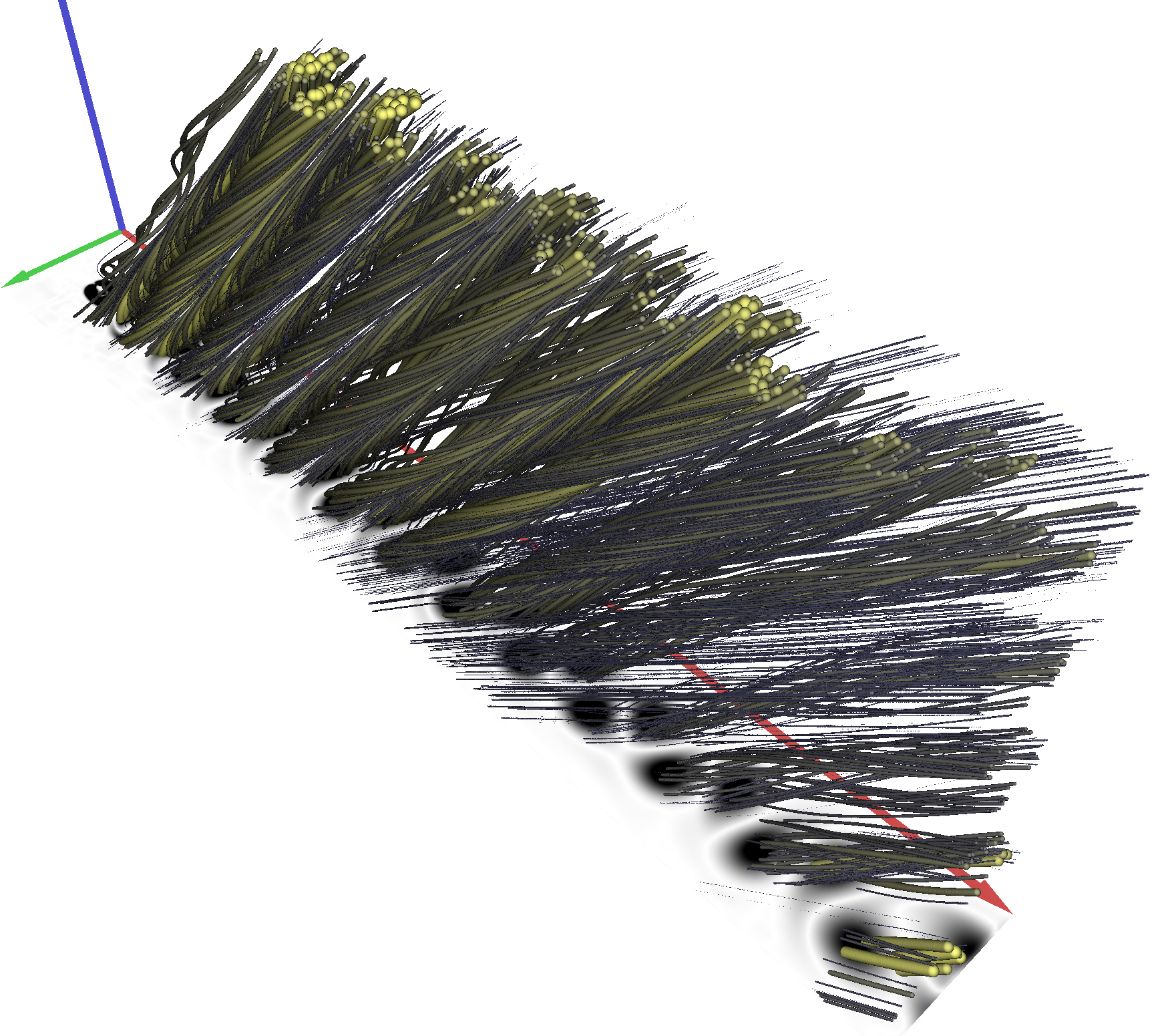}{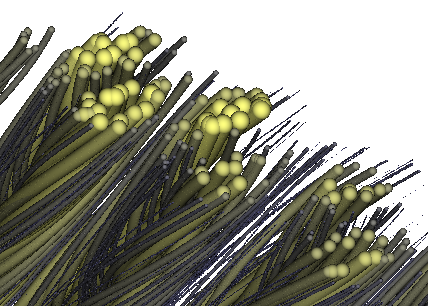}\hfill%
    \raisebox{12.5mm}{\begin{tikzpicture}%
	\node[anchor=south west,inner sep=0] (image) at (0,0) {\includegraphics[width=0.013\linewidth]{images/color-gradient.png}};%
	\begin{scope}[node distance=-0.8mm and -0.2mm, x={(image.south east)},y={(image.north west)}]%
	\node[draw=none,overlay,below= of image] {\scriptsize $0.01$};%
	\node[draw=none,overlay,above= of image] {\scriptsize $0.035$};%
	\end{scope}%
\end{tikzpicture}}\\%
    \raisebox{7em}{\rotatebox[origin=c]{90}{\small{$5,000$ trajectories}}}~%
    \trajcylcmpfig{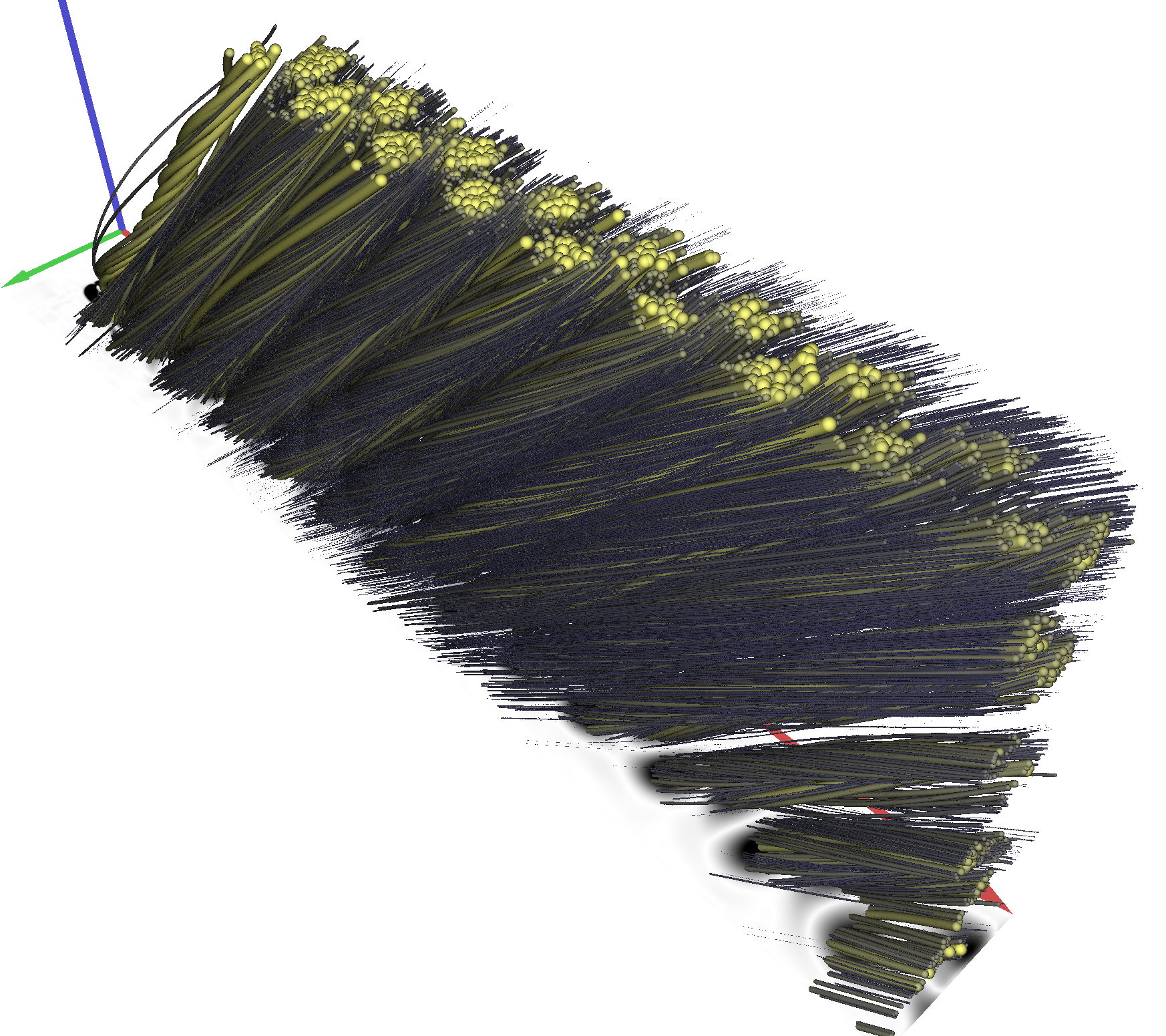}{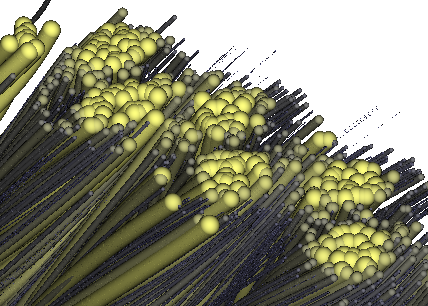}\hfill%
    \trajcylcmpfig{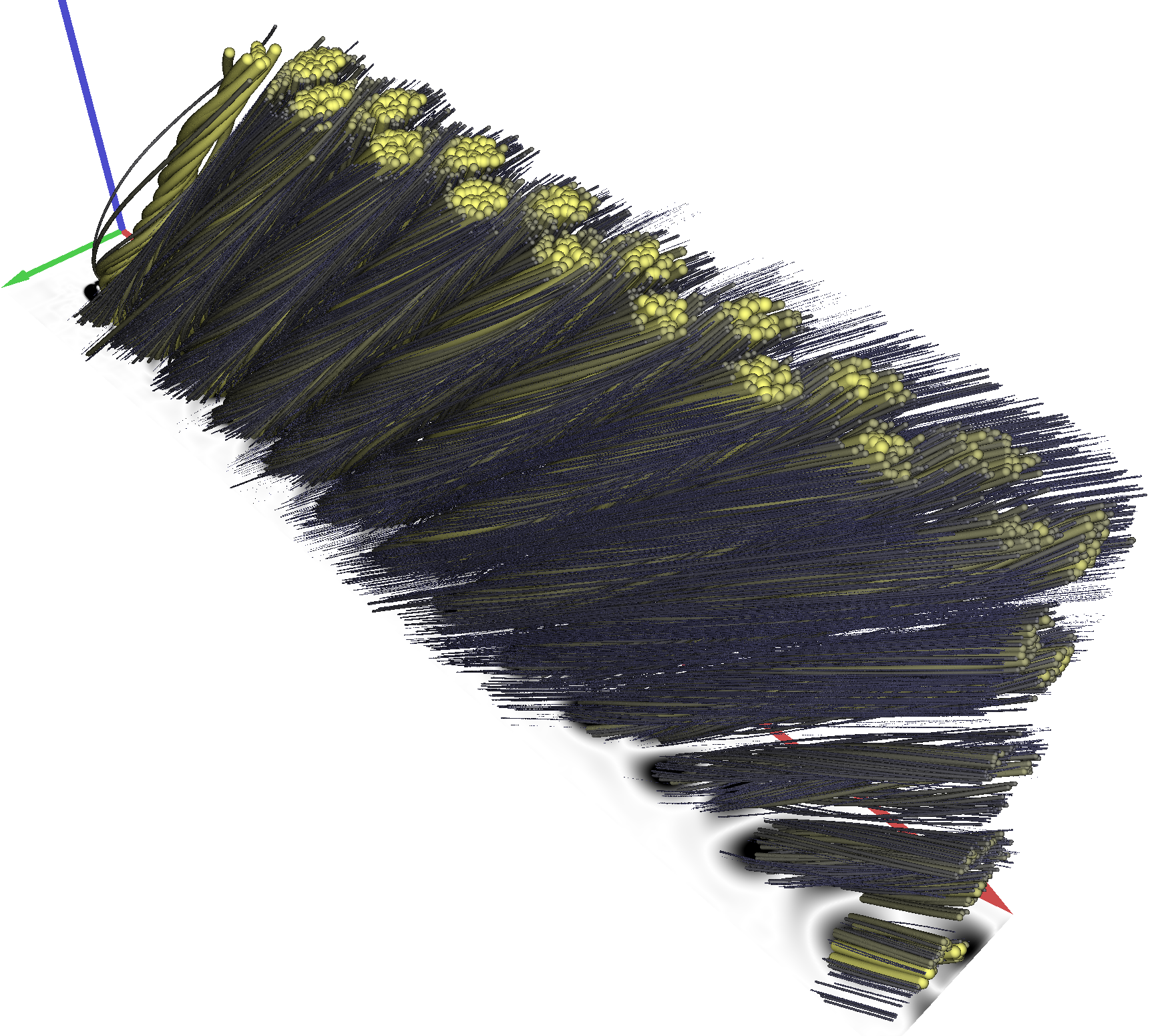}{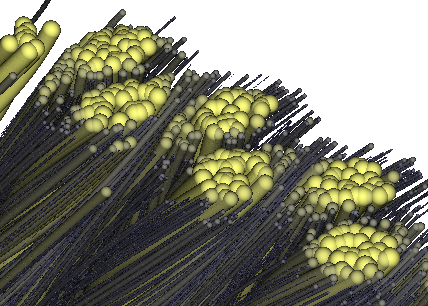}\hfill%
    \trajcylcmpfig{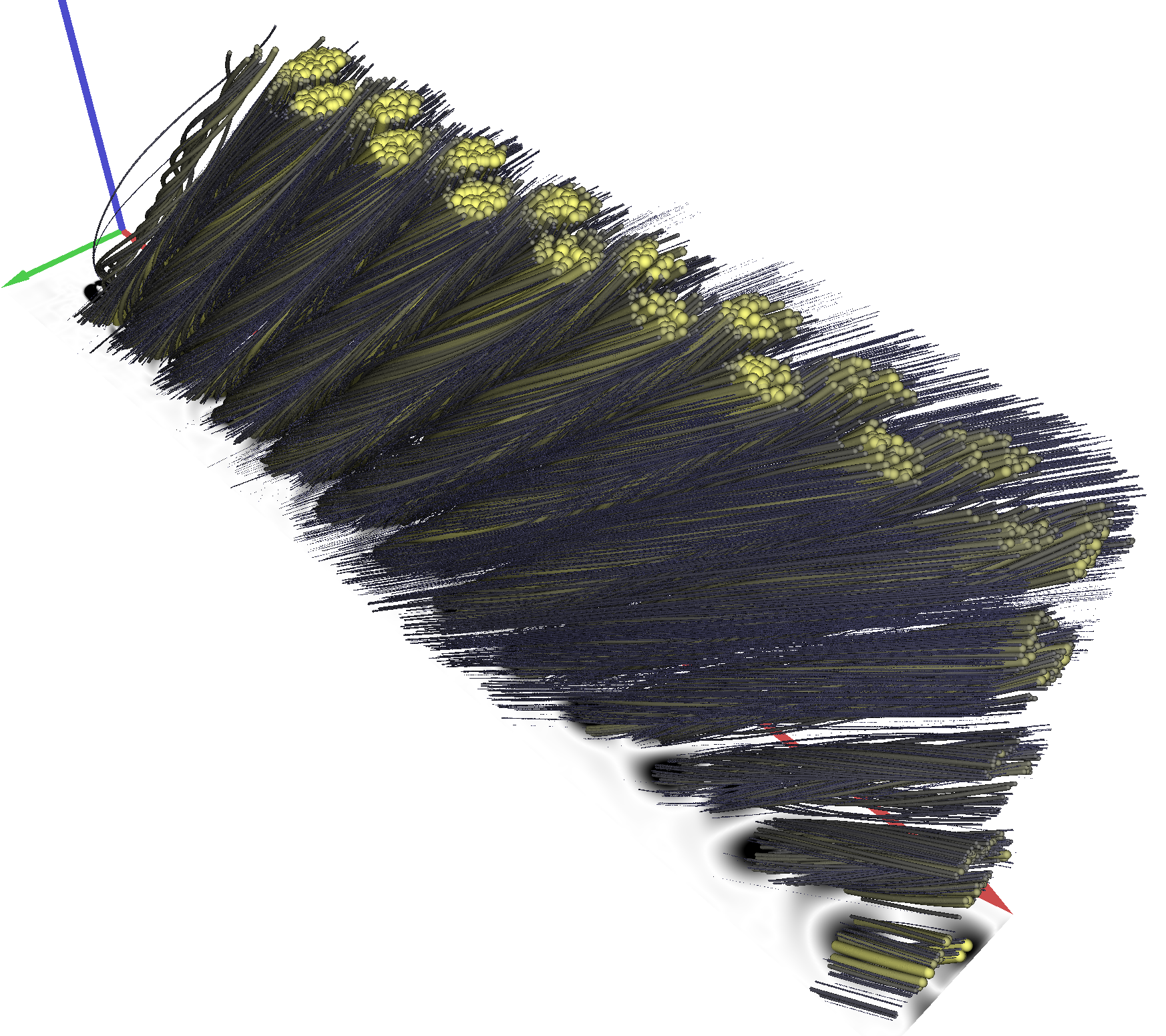}{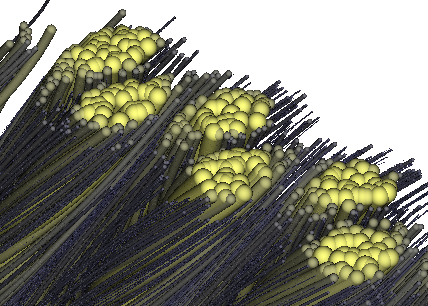}\hfill%
    \raisebox{12.5mm}{\begin{tikzpicture}%
	\node[anchor=south west,inner sep=0] (image) at (0,0) {\includegraphics[width=0.013\linewidth]{images/color-gradient.png}};%
	\begin{scope}[node distance=-0.8mm and -0.2mm, x={(image.south east)},y={(image.north west)}]%
	\node[draw=none,overlay,below= of image] {\scriptsize $0.01$};%
	\node[draw=none,overlay,above= of image] {\scriptsize $0.035$};%
	\end{scope}%
\end{tikzpicture}}%
    \caption{Parameter studies for the number of input trajectories (rows) and the number of neighboring lines used for fitting $\mH$ (columns), here for $\overline{\mbox{TRV}}$. For a low number of input trajectories (top row), adding too many neighbors includes lines that might not be part of a vortex. For a large number of input trajectories (bottom row), more neighboring lines result in a more stable estimation of rotating motion. Note that with increasing number of input trajectories, vortices are estimated more accurately, as the continuous field is sampled more densely.}
    \label{fig:cyl-param}
\end{figure*}
Since the input of $\mbox{TRV}$, $\overline{\mbox{TRV}}$ is a finite (low) number of trajectories, the quality of the results depends on the input trajectories. We analyze how the results depend on the  density of the input trajectories, and how  $\mbox{TRV}$ and $\overline{\mbox{TRV}}$  behave if the input is "garbage" (i.e., trajectories far away from each other, showing a different behavior driven by different phenomena). We do the analysis on the cylinder data set because for this an underlying velocity field as "ground truth" is available. 
Figure \ref{fig:cyl-param} shows the result for different amounts of input trajectories (rows) and different sample sizes (columns) for computing  $\overline{\mbox{TRV}}$.  For a low number of input trajectories, sampling many lines tends to include more lines from different regions. This results in fewer detected high  $\overline{\mbox{TRV}}$ (Figure \ref{fig:cyl-param} upper right). This confirms a desired behavior: "garbage input" leads to low $\overline{\mbox{TRV}}$ values. On the other hand, a larger number of input trajectories (lower row) give a more stable estimation of the rotating motion.


Table \ref{tab:performance} lists performance measurements for all considered data sets, computed on an Intel Core i9-10980XE CPU with 3.00 GHz.
For the real-world data, computations were in the order of milliseconds, while the largest test set took in total about 4 seconds to compute $\overline{\mbox{TRV}}$ for all 5,000 trajectories.
In our implementation, we compute the full distance matrix between all trajectories first. Afterwards, the distance matrix is reused when finding the $k$-nearest trajectories for each of the curves, by iterating the corresponding row in the distance matrix and collecting the $k$ smallest items in a max heap. To calculate the tangents and accelerations numerically, we use a sixth-order accurate finite-difference scheme~\citep{Fornberg1988:FD}.


\section{Conclusion}

We have introduced Trajectory Vorticity (TRV), the -- to the best of our knowledge -- first approach to analyze rotation behavior based on only few trajectories in an objective way. We proved objectivity of TRV and showed that TRV can be carried over from two independent established objectivization methods for velocity field data. We also analyzed and corrected statements about objectivity of previous trajectory based techniques from the literature.

\newcommand{\sqz}{\hspace{0mm}}
\appendix
\section{Appendix}

What follows is the proof that $\mathbf{trv}$ as defined in (\ref{eq_trv_1})--(\ref{eq_trv_2}) is objective. 
We have to show that the computability condition in
(\ref{eq_trv_2}) is objective, and that $\mW-\mW_s$
is objective. The first condition holds because 
$\dot{\mS}$ is objective \citep{Astarita79}. 
To show the objectivity of   
$\mW-\mW_s$, we
consider the observation of the trajectories in a moving reference system performing a Euclidean transformation of the form as in Eq.~\eqref{eq_define_movingsystem}.
We denote the observed measures from    (\ref{eq_trv_1})--(\ref{eq_trv_2}) with a tilde. This gives for the observations of $\mX, \dot{\mX}, \ddot{\mX}$ in the new reference system:
\begin{align}
\label{eq_Xtilde}
\widetilde{\mX} &=
\boldsymbol{\mathcal R}^\Transp ( \mX - \mA)
\\
\widetilde{\dot{\mX}} &=
\dot{\boldsymbol{\mathcal R}}^\Transp ( \mX - \mA) + \boldsymbol{\mathcal R}^\Transp ( \dot{\mX} - \dot{\mA})
\\
\widetilde{\ddot{\mX}} &=
\ddot{\boldsymbol{\mathcal R}}^\Transp ( \mX - \mA) 
+ 2 \dot{\boldsymbol{\mathcal R}}^\Transp ( \dot{\mX} - \dot{\mA})
+ \boldsymbol{\mathcal R}^\Transp ( \ddot{\mX} - \ddot{\mA})
\end{align}
with
\begin{align}
\nonumber
\mA &= \begin{pmatrix}
\va & \sqz\va & \sqz\va & \sqz[\va]\\
0  & \sqz 0  &\sqz 0  &\sqz[0 ]  
\end{pmatrix}
,&
\dot{\mA} &= \begin{pmatrix}
\dot{\va} &\sqz\dot{\va}  & \sqz\dot{\va}  & \sqz[\dot{\va} ]\\
0  & \sqz 0  &\sqz 0  &\sqz[0 ]  
\end{pmatrix}
,&
\ddot{\mA} &= \begin{pmatrix}
\ddot{\va} &\sqz\ddot{\va}  & \sqz\ddot{\va}  & \sqz[\ddot{\va} ]\\
0  & \sqz 0  &\sqz 0  &\sqz[0 ]  
\end{pmatrix}
\\
\nonumber
\boldsymbol{\mathcal R} &= 
\begin{pmatrix}
\mR & \vNull \\
\vNull^\Transp & 1 
\end{pmatrix}
,&
\dot{\boldsymbol{\mathcal R}} &= 
\begin{pmatrix}
\dot{\mR} & \vNull \\
\vNull^\Transp & 0 
\end{pmatrix}
,&
\ddot{\boldsymbol{\mathcal R}} &= 
\begin{pmatrix}
\ddot{\mR} & \vNull \\
\vNull^\Transp & 0 
\end{pmatrix}.
\end{align}
Eqs.~\eqref{eq_defineH} and \eqref{eq_Xtilde} give
\begin{align}
\widetilde{\mX}^{-1} &= 
\mX^{-1} \,
\begin{pmatrix}
\mI & \va \\
\vNull^\Transp & 1 
\end{pmatrix}
\,
\boldsymbol{\mathcal R}
\\
\label{eq_defineHtilde}
\widetilde{\mH} &= 
\boldsymbol{\mathcal R}^\Transp \, \mH \, \boldsymbol{\mathcal R} + \dot{\boldsymbol{\mathcal R}}^\Transp \boldsymbol{\mathcal R}
+
\left( \vNull  \;\;,\;\;  \boldsymbol{\mathcal R}^\Transp 
\left( 
 \dot{\mX} \mX^{-1}  
\begin{pmatrix}
\va \\
0
\end{pmatrix}
- 
\begin{pmatrix}
\dot{\va} \\
0
\end{pmatrix}
 \right)
 \right)
\end{align}
and from Eqs.~\eqref{eq_defineJ} and \eqref{eq_defineHtilde} follows
\begin{eqnarray}
\nonumber
\widetilde{\mJ} &=& \mI_z \, \widetilde{\mH} \,  \mI_z^\Transp
\,=\,
 \mR^\Transp \, \mJ \,\mR +  \dot{\mR}^\Transp  \, \mR 
\\
\nonumber
\widetilde{\dot{\mJ}} &=& 
    \mR^\Transp \, \dot{\mJ} \,\mR
+  \dot{\mR}^\Transp \, \mJ \,\mR
+  \mR^\Transp \, \mJ \, \dot{\mR}
+ \dot{\mR}^\Transp \, \dot{\mR} 
+ \ddot{\mR}^\Transp \,  \mR.
\end{eqnarray}
Since $ \dot{\mR}^\Transp  \, \mR $ and $\dot{\mR}^\Transp \, \dot{\mR} 
+ \ddot{\mR}^\Transp \,  \mR$ are anti-symmetric, we get
\begin{eqnarray}
\label{eq_defineWTilde}
\widetilde{\mW} &=&  \mR^\Transp \, \mW \,\mR +  \dot{\mR}^\Transp  \, \mR 
\\
\label{eq_defineSTilde}
\widetilde{\mS} &=& \mR^\Transp \, \mS \,\mR 
\\
\label{eq_defineSdotTilde}
\widetilde{\dot{\mS}} &=& 
    \mR^\Transp \, \dot{\mS} \,\mR
+  \dot{\mR}^\Transp \, \mS \,\mR
+  \mR^\Transp \, \mS \, \dot{\mR}.
\end{eqnarray}
Eq.~\eqref{eq_defineSTilde} gives: if $\ve$ is an eigenvector of $\mS$, then  $\mR^\Transp \, \ve$ is an eigenvector of $\widetilde{\mS}$. 
From this follows
\begin{equation}
\label{eq_define_ETilde}
\widetilde{\mE} = \mR^\Transp \, \mE
\end{equation}
which gives 
\begin{eqnarray}
\label{eq_sdot1}
\widetilde{\overline{\mS}} &=& \widetilde{\mE}^\Transp \, \widetilde{\mS} \, \widetilde{\mE}
= (\mE^\Transp \, \mR) \, ( \mR^\Transp \, \mS \,\mR  ) \, ( \mR^\Transp \, \mE) = \overline{\mS}
\\
\widetilde{\overline{\dot{\mS}}} &=& 
 \widetilde{\mE}^\Transp \, \widetilde{\dot{\mS}} \, \widetilde{\mE} \\
 &=&
 \overline{\dot{\mS}} + \mE^\Transp \, (
 \mR \, \dot{\mR}^\Transp \, \mS + \mS  \, \dot{\mR}  \, \mR^\Transp) \, \mE
 \\
 &=&
 \overline{\dot{\mS}} +  (\mE^\Transp  \, \mR \,  \dot{\mR}^\Transp   \mE)  \, \overline{\mS}
 + \overline{\mS}  \,   (\mE^\Transp  \, \dot{\mR}  \mR^\Transp \,    \mE)
 \\
 \label{eq_sdot2}
 &=&
 \overline{\dot{\mS}} +  (\mE^\Transp  \, \mR \,  \dot{\mR}^\Transp   \mE)  \, \overline{\mS}
 - \overline{\mS}  \,   (\mE^\Transp  \, \mR \,  \dot{\mR}^\Transp   \mE).
\end{eqnarray}
Eqs.~\eqref{eq_sdot1} and \eqref{eq_sdot2}, $\overline{\mS}$ being a diagonal matrix, and  $(\mE^\Transp  \, \mR \,  \dot{\mR}^\Transp   \mE)$ being anti-symmetric gives
\begin{equation}
\label{eq_define_Wsbar}
\widetilde{\overline{ \mW}}_s = \overline{ \mW}_s +  \mE^\Transp  \, \mR \,  \dot{\mR}^\Transp   \mE.
\end{equation}
Then, Eqs.~\eqref{eq_define_Ws_1}, \eqref{eq_define_ETilde}, and \eqref{eq_define_Wsbar} give
\begin{equation}
\label{eq_defineWTildes}
\widetilde{\mW}_s =  \widetilde{\mE} \; \widetilde{\overline{\mW}_s} \; \widetilde{\mE}^\Transp
=
\mR^\Transp \, \mW_s \, \mR + \dot{\mR}^\Transp  \, \mR.
\end{equation}
Finally, Eqs.~\eqref{eq_defineWTilde} and \eqref{eq_defineWTildes} give
$ \widetilde{\mathbf{trv}} = \widetilde{\mW} - \widetilde{\mW}_s = \mR^\Transp \,  \mathbf{trv}  \, \mR $ 
which proves the theorem.

\bibliographystyle{unsrtnat}
\bibliography{literature}  

\begin{thebibliography}{31}
\providecommand{\natexlab}[1]{#1}
\providecommand{\url}[1]{\texttt{#1}}
\expandafter\ifx\csname urlstyle\endcsname\relax
  \providecommand{\doi}[1]{doi: #1}\else
  \providecommand{\doi}{doi: \begingroup \urlstyle{rm}\Url}\fi

\bibitem[McLoughlin et~al.(2010)McLoughlin, Laramee, Peikert, Post, and
  Chen]{McLoughlin09}
Tony McLoughlin, Robert~S Laramee, Ronald Peikert, Frits~H Post, and Min Chen.
\newblock Over two decades of integration-based, geometric flow visualization.
\newblock \emph{Computer Graphics Forum}, 29\penalty0 (6):\penalty0 1807--1829,
  2010.

\bibitem[Edmunds et~al.(2012)Edmunds, Laramee, Chen, Max, Zhang, and
  Ware]{Edmunds12:SurfaceSTAR}
Matt Edmunds, Robert~S. Laramee, Guoning Chen, Nelson Max, Eugene Zhang, and
  Colin Ware.
\newblock Surface-based flow visualization.
\newblock \emph{Computers \& Graphics}, 36\penalty0 (8):\penalty0 974--990,
  2012.
\newblock ISSN 0097-8493.
\newblock \doi{http://dx.doi.org/10.1016/j.cag.2012.07.006}.

\bibitem[Bujack et~al.(2020)Bujack, Yan, Hotz, Garth, and
  Wang]{Bujack20:UnsteadyTopoSTAR}
Roxana Bujack, Lin Yan, Ingrid Hotz, Christoph Garth, and Bei Wang.
\newblock State of the art in time-dependent flow topology: Interpreting
  physical meaningfulness through mathematical properties.
\newblock \emph{Computer Graphics Forum}, 39\penalty0 (3):\penalty0 811--835,
  2020.
\newblock \doi{https://doi.org/10.1111/cgf.14037}.

\bibitem[Bujack and Joy(2015)]{Bujack15:LagrangianRepresentation}
Roxana Bujack and Kenneth~I. Joy.
\newblock Lagrangian representations of flow fields with parameter curves.
\newblock In \emph{2015 IEEE 5th Symposium on Large Data Analysis and
  Visualization (LDAV)}, pages 41--48, 2015.
\newblock \doi{10.1109/LDAV.2015.7348070}.

\bibitem[Sane et~al.(2018)Sane, Bujack, and Childs]{Sane18:InSituLagrangian}
Sudhanshu Sane, Roxana Bujack, and Hank Childs.
\newblock Revisiting the evaluation of in situ lagrangian analysis.
\newblock In \emph{Proceedings of the Symposium on Parallel Graphics and
  Visualization}, EGPGV '18, page 63–67, Goslar, DEU, 2018. Eurographics
  Association.

\bibitem[Lumpkin and Centurioni(2019, accessed 2021-11-15)]{GDP:buoys}
Rick Lumpkin and Luca Centurioni.
\newblock Noaa global drifter program quality-controlled 6-hour interpolated
  data from ocean surface drifting buoys. subset from 2019-06-30 to 2021-06-30,
  2019, accessed 2021-11-15.

\bibitem[Sinhuber et~al.(2019)Sinhuber, Van Der~Vaart, Ni, Puckett, Kelley, and
  Ouellette]{sinhuber2019three}
Michael Sinhuber, Kasper Van Der~Vaart, Rui Ni, James~G Puckett, Douglas~H
  Kelley, and Nicholas~T Ouellette.
\newblock Three-dimensional time-resolved trajectories from laboratory insect
  swarms.
\newblock \emph{Scientific Data}, 6\penalty0 (1):\penalty0 1--8, 2019.

\bibitem[Haller et~al.(2021)Haller, Aksamit, and
  Encinas-Bartos]{Haller21singletrajecory}
George Haller, Nikolas Aksamit, and Alex~P. Encinas-Bartos.
\newblock Quasi-objective coherent structure diagnostics from single
  trajectories.
\newblock \emph{Chaos: An Interdisciplinary Journal of Nonlinear Science},
  31\penalty0 (4):\penalty0 043131, 2021.
\newblock \doi{10.1063/5.0044151}.
\newblock URL \url{https://doi.org/10.1063/5.0044151}.

\bibitem[Truesdell and Noll(1965)]{truesdell_book}
Clifford Truesdell and Walter Noll.
\newblock \emph{The Nonlinear Field Theories of Mechanics}.
\newblock Springer, 1965.

\bibitem[Astarita(1979)]{Astarita79}
Gianni Astarita.
\newblock Objective and generally applicable criteria for flow classification.
\newblock \emph{Journal of Non-Newtonian Fluid Mechanics}, 6\penalty0
  (1):\penalty0 69--76, 1979.

\bibitem[Haller(2005)]{haller2005}
George Haller.
\newblock An objective definition of a vortex.
\newblock \emph{Journal of Fluid Mechanics}, 525:\penalty0 1--26, 2005.

\bibitem[Shadden et~al.(2005)Shadden, Lekien, and Marsden]{Shadden2005}
Shawn~C Shadden, Francois Lekien, and Jerrold~E Marsden.
\newblock Definition and properties of lagrangian coherent structures from
  finite-time lyapunov exponents in two-dimensional aperiodic flows.
\newblock \emph{Physica D: Nonlinear Phenomena}, 212\penalty0 (3):\penalty0
  271--304, 2005.

\bibitem[Drouot and Lucius(1976)]{Drouot76}
R~Drouot and M~Lucius.
\newblock Approximation du second ordre de la loi de comportement des fluides
  simples. lois classiques d{\'e}duites de l?introduction d?un nouveau tenseur
  objectif.
\newblock \emph{Archiwum Mechaniki Stosowanej}, 28\penalty0 (2):\penalty0
  189--198, 1976.

\bibitem[Haller et~al.(2016)Haller, Hadjighasem, Farazmand, and
  Huhn]{haller2016}
George Haller, Alireza Hadjighasem, Mohammad Farazmand, and Florian Huhn.
\newblock Defining coherent vortices objectively from the vorticity.
\newblock \emph{Journal of Fluid Mechanics}, 795:\penalty0 136--173, 2016.

\bibitem[Liu et~al.(2019)Liu, Gao, and Liu]{Liu19:SpinDeviation}
Jianming Liu, Yisheng Gao, and Chaoqun Liu.
\newblock An objective version of the rortex vector for vortex identification.
\newblock \emph{Physics of Fluids}, 31\penalty0 (6):\penalty0 065112, 2019.
\newblock \doi{10.1063/1.5095624}.
\newblock URL \url{https://doi.org/10.1063/1.5095624}.

\bibitem[G{\"u}nther et~al.(2017)G{\"u}nther, Gross, and
  Theisel]{Guenther:2017:Siggraph}
T.~G{\"u}nther, M.~Gross, and H.~Theisel.
\newblock Generic objective vortices for flow visualization.
\newblock \emph{ACM Transactions on Graphics (Proc. SIGGRAPH)}, 36\penalty0
  (4):\penalty0 141:1--141:11, 2017.

\bibitem[G{\"u}nther and Theisel(2019)]{Guenther:2019:VIS}
T.~G{\"u}nther and H.~Theisel.
\newblock Objective vortex corelines of finite-sized objects in fluid flows.
\newblock \emph{IEEE Transactions on Visualization and Computer Graphics (Proc.
  IEEE Scientific Visualization 2018)}, 25\penalty0 (1):\penalty0 956--966,
  2019.

\bibitem[{Hadwiger} et~al.(2019){Hadwiger}, {Mlejnek}, {Theußl}, and
  {Rautek}]{Hadwiger19}
M.~{Hadwiger}, M.~{Mlejnek}, T.~{Theußl}, and P.~{Rautek}.
\newblock Time-dependent flow seen through approximate observer killing fields.
\newblock \emph{IEEE Transactions on Visualization and Computer Graphics},
  25\penalty0 (1):\penalty0 1257--1266, 2019.
\newblock \doi{10.1109/TVCG.2018.2864839}.

\bibitem[Baeza~Rojo and G{\"u}nther(2020)]{Rojo20}
Irene Baeza~Rojo and Tobias G{\"u}nther.
\newblock Vector field topology of time-dependent flows in a steady reference
  frame.
\newblock \emph{IEEE Transactions on Visualization and Computer Graphics},
  26\penalty0 (1):\penalty0 280--290, 2020.
\newblock \doi{10.1109/TVCG.2019.2934375}.

\bibitem[G{\"u}nther and Theisel(2020)]{Guenther:2020:TVCG}
T.~G{\"u}nther and H.~Theisel.
\newblock Hyper-objective vortices.
\newblock \emph{IEEE Transactions on Visualization and Computer Graphics},
  26\penalty0 (3):\penalty0 1532--1547, 2020.

\bibitem[{Rautek} et~al.(2021){Rautek}, {Mlejnek}, {Beyer}, {Troidl},
  {Pfister}, {Theußl}, and {Hadwiger}]{Rautek21}
P.~{Rautek}, M.~{Mlejnek}, J.~{Beyer}, J.~{Troidl}, H.~{Pfister}, T.~{Theußl},
  and M.~{Hadwiger}.
\newblock Objective observer-relative flow visualization in curved spaces for
  unsteady 2d geophysical flows.
\newblock \emph{IEEE Transactions on Visualization and Computer Graphics},
  27\penalty0 (2):\penalty0 283--293, 2021.
\newblock \doi{10.1109/TVCG.2020.3030454}.

\bibitem[Zhang et~al.(2022)Zhang, Hadwiger, Theu{\ss}l, and
  Rautek]{Zhang2021KillingObserverInteraction}
Xingdi Zhang, Markus Hadwiger, Thomas Theu{\ss}l, and Peter Rautek.
\newblock Interactive exploration of physically-observable objective vortices
  in unsteady 2d flow.
\newblock \emph{IEEE Transactions on Visualization and Computer Graphics},
  28\penalty0 (2):\penalty0 to appear, 2022.

\bibitem[Haller(2021)]{Haller20:can}
George Haller.
\newblock Can vortex criteria be objectivized?
\newblock \emph{Journal of Fluid Mechanics}, 508:\penalty0 A25, 2021.
\newblock \doi{10.1017/jfm.2020.937}.

\bibitem[Theisel et~al.(2021)Theisel, Hadwiger, Rautek, Theußl, and
  Günther]{Theisel:2021:PhysFluids}
H.~Theisel, M.~Hadwiger, P.~Rautek, T.~Theußl, and T.~Günther.
\newblock Vortex criteria can be objectivized by unsteadiness minimization.
\newblock \emph{Physics of Fluids}, 33\penalty0 (10):\penalty0 107115, 2021.
\newblock \doi{10.1063/5.0063817}.
\newblock URL \url{https://aip.scitation.org/doi/abs/10.1063/5.0063817}.

\bibitem[Provenzale(1999)]{Provenzale_1999_review}
A.~Provenzale.
\newblock Transport by coherent barotropic vortices.
\newblock \emph{ANNUAL REVIEW OF FLUID MECHANICS}, 31:\penalty0 55--93, 1999.
\newblock URL
  \url{http://archipelago.uma.pt/pdf_library/Provenzale_1999_review.pdf}.

\bibitem[Haller and Yuan(2000)]{Haller2000}
G.~Haller and G.~Yuan.
\newblock Lagrangian coherent structures and mixing in two-dimensional
  turbulence.
\newblock \emph{Phys. D}, 147\penalty0 (3?4):\penalty0 352?370, December 2000.
\newblock ISSN 0167-2789.
\newblock \doi{10.1016/S0167-2789(00)00142-1}.
\newblock URL \url{https://doi.org/10.1016/S0167-2789(00)00142-1}.

\bibitem[Bartos et~al.(2021)Bartos, Aksamit, and
  Haller]{bartos2021quasiobjective}
Alex P.~Encinas Bartos, Nikolas~O. Aksamit, and George Haller.
\newblock Quasi-objective eddy visualization from sparse drifter data, 2021.

\bibitem[Popinet(2004)]{gerrisflowsolver}
S.~Popinet.
\newblock Free computational fluid dynamics.
\newblock \emph{ClusterWorld}, 2\penalty0 (6), 2004.
\newblock URL \url{http://gfs.sf.net/}.

\bibitem[Oeltze et~al.(2014)Oeltze, Lehmann, Kuhn, Janiga, Theisel, and
  Preim]{Oeltze2014blood}
Steffen Oeltze, Dirk~J Lehmann, Alexander Kuhn, G{\'a}bor Janiga, Holger
  Theisel, and Bernhard Preim.
\newblock Blood flow clustering and applications invirtual stenting of
  intracranial aneurysms.
\newblock \emph{IEEE transactions on visualization and computer graphics},
  20\penalty0 (5):\penalty0 686--701, 2014.

\bibitem[Downe and Caspary(1973)]{Downe1973}
A.~Downe and V.~Caspary.
\newblock The swarming behaviour of chironomus riparius (diptera: Chironomidae)
  in the laboratory.
\newblock \emph{Can. Entomol.}, 105:\penalty0 165–--171, 1973.

\bibitem[Fornberg(1988)]{Fornberg1988:FD}
Bengt Fornberg.
\newblock Generation of finite difference formulas on arbitrarily spaced grids.
\newblock \emph{Mathematics of computation}, 51\penalty0 (184):\penalty0
  699--706, 1988.

\end{thebibliography}






\end{document}